\documentclass[11pt]{article}
\usepackage{amsmath,amsfonts,epsfig,mathrsfs, verbatim}
\numberwithin{equation}{section}
\usepackage{lscape}

\ifx\ltimes\undefined
\newcommand{\ltimes}{{\kern3pt\hbox{\vrule width 0.4pt height 5.30pt
depth .0pt}\kern-1.76pt\times\kern1pt}} \fi

\ifx\lrtimes\undefined
\newcommand{\rtimes}{{\kern1pt\times\kern-4.76pt\kern3pt\hbox{\vrule width 0.4pt height 5.30pt
depth .0pt}}} \fi

\def\Z {\mathbb{Z}}
\def\R {\mathbb{R}}
\def\C {\mathbb{C}}

\def\bid{\hbox{1\hspace{-0.04in}I}} 

\def\s{\sigma}                                   

\def\x{\xi}

\def\G{\Gamma}

\def\cG{{\cal G}}
\def\cX{{\cal X}}

\newcommand{\fdh}[2]{H^{#1}_{\phantom{#1}#2}}

\newcommand{\p}{\partial}
\newcommand{\eps}{\epsilon}

\begin{document}

\begin{titlepage}

\vspace*{15mm}

\begin{center}

{\Large {${\cal N}=4$ Gauged Supergravity from Duality-Twist\\Compactifications of String Theory}} \\

\vspace*{20mm}

{ R A Reid-Edwards$^1$
and B Spanjaard$^2$} \\
\vspace*{10mm}

{\em II. Institute f\"{u}r Theoretische Physik }\\
{\em Universit\"{a}t Hamburg } \\ {\em DESY, Luruper Chaussee 149} \\
{\em D-22761 Hamburg, Germany} \\

\vspace*{30mm}

\end{center}

\begin{abstract}
\noindent We investigate the lifting of half-maximal four-dimensional gauged supergravities to compactifications of string theory. It is shown that a class of such supergravities can arise from compactifications of IIA string theory on manifolds of $SU(2)$-structure which may be thought of as $K3$ fibrations over $T^2$. Examples of these $SU(2)$-structure backgrounds, as smooth $K3$ bundles and as compactifications with $H$-flux, are given and we also find evidence for a class of non-geometric, Mirror-fold backgrounds. By applying the duality between IIA string theory on $K3$ and Heterotic string theory on $T^4$ fibrewise, we argue that these $SU(2)$-structure backgrounds are dual to Heterotic compactifications on a class $T^4$ fibrations over $T^2$. Examples of these fibrations as twisted tori, $H$-flux and T-fold compactifications are given. We also construct a new set of backgrounds, particular to Heterotic string theory, which includes a previously unknown class of Heterotic T-folds. A sigma model description of these backgrounds, from the Heterotic perspective, is presented in which we generalize the Bosonic doubled formalism to Heterotic string theory.
\end{abstract}

\vfill

\noindent {$^1$ronald.reid.edwards@desy.de\\ $^2$bastiaan.spanjaard@desy.de }
\date{}
\end{titlepage}

\newpage

\section{Introduction}
The study of gauged supergravities has received a new impetus in recent years. This has been due, in part, to the application of novel
techniques and concepts, such as generalized geometry, in constructing and analyzing new dimensional reduction scenarios. Generalized geometry allows us to construct internal manifolds that can be seen as generalizations of the well-known Calabi-Yau three-folds, $K3$ and tori. A dimensional reduction of ten-dimensional supergravity on such backgrounds leads to various lower-dimensional gauged supergravities. More specifically, one may argue from very general topological considerations, that a reduction on a manifold of reduced structure group explicitly breaks some supersymmetry \cite{Candelas:1985en,GMW,Gauntlett:2002sc,GKMW}. The amount of supersymmetry that is explicitly broken by the reduction depends on the structure group of the manifold. For example, six-dimensional $SU(3)$-structure manifolds explicitly break three-quarters of the supersymmetry, six-dimensional $SU(2)$-structure manifolds break half, while $\bid$-structure\footnote{Pronounced, `identity-structure'.} manifolds, of any dimension, do not explicitly break any supersymmetry. The surviving supersymmetry in the effective lower-dimensional theory may then be spontaneously broken for a given vacuum solution.

Reductions on six-dimensional manifolds with $SU(3)$-structure have been fruitfully employed in constructing new ${\cal N}=1$ and ${\cal N}=2$ gauged supergravities in four dimensions \cite{GLMW,GLW1,Grana:2006hr} which have minimally-coupled Abelian gauge groups and scalar potentials. The topological requirement that the six-dimensional reduction manifold has a certain structure group does not completely specify the internal space and, as such, there are many outstanding questions concerning the relationship between these four-dimensional ${\cal N}=1$ and ${\cal N}=2$ supergravities to ten-dimensional supergravity and string theory. In particular, an issue which has yet to achieve a satisfactory resolution is whether or not the effective four-dimensional theory correctly captures all of the low energy physics of string or supergravity theory compactified on such a manifold.

The standard procedure for producing a lower-dimensional supergravity from a higher-dimensional one is to propose a reduction ansatz for the higher-dimensional fields, which is substituted into the higher-dimensional action. One then integrates out all internal coordinate dependence in the higher-dimensional action. A judicious choice of reduction ansatz can lead to a lower dimensional supergravity with many attractive features, such as Yang-Mills gauge symmetries and scalar potentials. In many cases, it is possible to show that this procedure produces an action for a lower-dimensional supergravity which captures all of the low energy classical physics one would expect from a supergravity compactified on a compact manifold. In such cases, there is a clear relationship between the four-dimensional and ten-dimensional supergravities and one can interpret the lower-dimensional supergravity as a long wavelength limit of string theory with confidence. However, for many gauged supergravities constructed in this way, it is not clear how the reduction ansatz naturally arises from considering a compactification on a conventional manifold. It is important then to distinguish between those cases where the ansatz can be related to a compactification on a known manifold and the cases where the situation is not so clear. Moreover it is important to use terminology that makes this distinction clear. Below we briefly clarify the terminology pertaining to both cases as it will be used throughout this paper.

A \emph{compactification} will refer to a decomposition of the higher-dimensional fields in terms of the harmonics of a well-defined internal manifold followed by a truncation, which sets certain higher modes in the Kaluza-Klein tower of states to zero. In order to construct an effective theory, valid up to a certain energy scale, one usually chooses to truncate out modes with an effective mass above this energy scale and indeed this is what is done in conventional compactifications on tori and Calabi-Yau manifolds, where only the lowest modes are kept. So that we can be sure that all of the light modes are kept and none thrown away in the truncation, we therefore need to have a good understanding of the internal geometry.

By contrast, a dimensional \emph{reduction} is simply an algorithm that may be employed to obtain a lower-dimensional supergravity from a higher-dimensional one. In many cases, such as for tori and Calabi-Yau manifolds, the standard reduction algorithm is equivalent to the Kaluza-Klein compactification and truncation described above; however, generally one may have no idea of what the reduction corresponds to physically; it is merely a recipe to generate a lower-dimensional supergravity\footnote{A \emph{consistent reduction} is a reduction in which solutions of the reduced theory lift to solutions of the full, higher-dimensional, theory. Similarly one may speak of a \emph{consistent truncation} as a truncation such that the surviving modes solve the higher-dimensional equations of motion.}.

One might then say that many of the ${\cal N}=1$ and ${\cal N}=2$ gauged supergravities mentioned above are well understood at the level of a dimensional reduction - an algorithm to construct one supergravity from another, higher-dimensional, supergravity - but are yet to be fully understood as compactifications in the rigorous Kaluza-Klein sense. It is, in fact, quite difficult to construct explicit examples of $SU(n)$-structure manifolds of $2n$ real dimensions. It is considerably easier to construct an $SU(n)$-structure manifold of real dimension greater than $2n$. For example, one may construct a $(2n+k)$-dimensional $SU(n)$-structure manifold, as a smooth bundle where the fibres are ($2n$ real-dimensional) $SU(n)$-holonomy manifolds, examples of which are well known, over a $k$-dimensional base. Examples of seven-dimensional $SU(3)$-structure manifolds of this kind, constructed as Calabi-Yau bundles over $S^1$, were given in \cite{Aharony:2008rx}. Further examples, giving six-dimensional manifolds of $SU(2)$-structure, will be given in this paper (see also \cite{Cvetic:2007ju} for a related class of $SU(3)$-structure backgrounds constructed in a similar manner).

Whilst a central motivation of this growing field of generalized reductions has been to address the urgent need to construct credible moduli stabilization and supersymmetry-breaking scenarios from flux compactifications, these studies have also led to additional insights into the fundamental structure of string theory. In particular, there is mounting evidence that many gauged supergravities can not arise from a compactification of a higher dimensional supergravity on a conventional manifold, but can be lifted to string theory on a non-geometric background. Such backgrounds have no analogue in field theories such as General Relativity and shed light on possible string- or M-theoretic generalizations of spacetime \cite{Kachru:2002sk,Hellerman:2002ax,Hull ``A geometry for non-geometric string backgrounds''}. The need for an explicit constructions of the internal background is perhaps of greater importance in the non-geometric case, where our experience is more limited. Many non-geometric $\bid$-structure backgrounds can be thought of as conventional manifolds locally, with transition functions between coordinate patches that include T-dualities, making the background globally non-geometric. The viability of such backgrounds as potential string backgrounds then rests on the fact that T-duality is a symmetry of string theory and therefore the string theory views the background as smooth. Such arguments require a detailed knowledge of the internal background and a rigorous understanding of the appropriate duality symmetries of the string theory, which are currently absent for most six-dimensional $SU(3)$- and $SU(3)\times SU(3)$-structure backgrounds.

Ultimately, one would like to have not only an understanding of which gauged supergravities lift to string theory and how, but also a description of the vacuum solutions of these gauged supergravities from the worldsheet perspective. Whilst the reduced structure techniques pioneered in \cite{Gauntlett:2002sc,GKMW} have found very general application to the construction of ${\cal N}=1$ and ${\cal N}=2$ gauged supergravities and their vacuum solutions, the construction of a worldsheet description remains a challenge in all but the simplest cases. By contrast, compactifications which give rise to four-dimensional maximally supersymmetric supergravities have been usefully described at the worldsheet level using the doubled formalism of \cite{Hull ``A geometry for non-geometric string backgrounds''}. These maximal gauged supergravities arise from compactifications of Type II supergravity on six-dimensional $\bid$-structure backgrounds. Such backgrounds are parallelizable and do not break any of the 32 supersymmetries explicitly, resulting in ${\cal N}=8$ gauged supergravities in four dimensions. The general form for the Lagrangian of such maximal supergravities and the conditions that the gauging does not explicitly break supersymmetry\footnote{Of course, for a given solution, some degree of supersymmetry may be spontaneously broken.} were given in \cite{d=4}.

In contrast to the well-studied ${\cal N}=1$, $2$, and $8$ cases discussed above, our interest in this paper is to study the half-maximal case of ${\cal N}=4$ gauged supergravity in four dimensions. The half maximal case is of particular interest as it represents a situation in which the techniques used in compactifications on six-dimensional manifolds of $SU(2)$-structure and $\bid$-structure make contact. We will show that many of these massive four-dimensional $\mathcal{N}=4$ supergravities can be realized either as a compactification of Type IIA string theory on a manifold of $SU(2)$-structure or as a maximally supersymmetric compactification of Heterotic string theory on a manifold of $\bid$-structure and present explicit constructions of these backgrounds.

\noindent\textbf{Overview and Results}

\noindent There are, in principle, at least three routes to constructing ${\cal N}=4$ gauged supergravities by dimensional reduction from ten dimensions; we may reduce the Type I, $Spin(32)/\Z_2$ or $E_8\times E_8$ Heterotic string theories on manifolds with $\bid$-structure group, reduce Type IIA string theory (in which we include flux compactifications of M-Theory \cite{Hull:2006tp}) on $SU(2)$-structure manifolds, or reduce Type IIB (in which we include compactifications of F-Theory \cite{ReidEdwards:2006vu}) on $SU(2)$-structure manifolds. We will only consider reductions of IIA and Heterotic string theory on $SU(2)$- and $\bid$-structure backgrounds respectively here and we shall be particularly interested in realizing these reductions as compactifications on internal spaces which we shall identify.

The first step of the reduction of the Heterotic supergravity consists of a conventional Kaluza-Klein compactification on $T^4$ (with coordinates $z^m$, $m,n=1,2,3,4$), which gives a six-dimensional ${\cal N}=2$ theory with rigid $O(4,20)$ symmetry, which lifts to a $O(4,20;\Z)$ T-duality symmetry of the string theory. We then compactify on a further $T^2$, twisting by elements of this T-duality group over the cycles of the $T^2$ (with coordinates $y^i$, $i=5,6$) to give a four-dimensional ${\cal N}=4$ gauged supergravity. The six-dimensional internal space can then be thought of as a $T^4$ fibration over $T^2$ with monodromies taking values in $O(4,20;\Z)$. For example, twisting by a geometric $SL(4;\Z)\subset O(4,20;\Z)$ results in a six-dimensional internal manifold which can be thought of as a smooth $T^4$ bundle over $T^2$. For a general duality-twist reduction of this kind, the resulting ${\cal N}=4$ theory has non-Abelian gauge symmetry generated by the algebra
$$
[Z_i,Z_m]=f_{im}{}^nZ_n+M_{im}{}^aY_a+K_{imn}X^n   \qquad [Z_m,Z_n]=K_{imn}X^i   \qquad  [X^m,Z_n]=f_{im}{}^nX^i
$$
$$
[X^m,X^n]=Q_i{}^{mn}X^i \qquad  [Z_i,X^m]=f_{in}{}^mX^n+W_i{}^{ma}Y_a +Q_i{}^{mn}Z_n
$$
$$
[Z_i,Y_a]=-\delta_{ab}W_i{}^{mb}Z_m+M_{ima}X^m+S_{ia}{}^bY_b    \qquad    [Z_m,Y_a]=M_{ima}X^i \qquad [Y_a,Y_b]=S_{iab}X^i \nonumber
$$
where the generators $(Z_m,Z_i)$ are related to diffeomorphisms of the six-dimensional internal space, $(X^m, X^i)$ are related to $B$-field antisymmetric tensor transformations and the $Y_a$ ($a,b=1,2,..16$) generate the $U(1)^{16}$ internal gauge symmetry.

The gauge algebra contains a lot of information about the ten-dimensional lift of the four-dimensional supergravity. A particular challenge to realizing lower-dimensional gauged supergravities as compactifications of string theory is to give a string theoretic interpretation to the parameters, such as the structure constants of this gauge algebra. In particular, the structure constants $f_{im}{}^n$ encode information about the local geometry of the compactification manifold and the structure constants $K_{mnp}$ and $M_{mn}{}^a$ contain information about the fluxes in the internal geometry. It has become commonplace to refer to the higher-dimensional description of all such structure constants, regardless of their physical interpretation, as fluxes \cite{Shelton:2005cf,Shelton:2006fd} and we shall adopt this nomenclature here. In addition to the geometric, $K$- and $M$-fluxes, which are related to those described in \cite{KM}, we also find new $Q$-, $W$- and $S$-fluxes. It will be argued that the $Q$-flux is akin to that found for the T-folds studied in \cite{Hull ``A geometry for non-geometric string backgrounds'',Hull ``Doubled geometry and T-folds'',Dabholkar:2002sy,Dabholkar:2005ve} where the monodromy around the $T^2$ cycles include a strict T-duality and mixes metric and $B$-field degrees of freedom. The $W$-flux will be shown to be indicative of a T-fold background, but of a kind specific to Heterotic string theory and not previously studied. The $S$-flux will be understood as a non-trivial fibering of the internal $U(1)^{16}$ gauge bundle over the $T^2$. Remarkably, we shall see that all of these backgrounds can be understood in terms of a worldsheet sigma model of the form pioneered in \cite{Hull ``A geometry for non-geometric string backgrounds''} and \cite{Hull ``Doubled geometry and T-folds''}.

The IIA theory can be similarly compactified on a $K3$ manifold, giving a six-dimensional theory with $O(4,20)$ rigid symmetry, dual to the Heterotic theory compactified on $T^4$ discussed above. Performing a duality-twist reduction, as above, over the cycles of a further $T^2$ gives an ${\cal N}=4$ gauged supergravity with non-Abelian gauge symmetry generated by the Lie algebra
$$
[Z_i,J]=\mathcal{K}_i{}^AT_A  \qquad    [Z_i,\tilde{J}]=\mathcal{Q}_{i}{}^AT_A  \qquad
[Z_i,T_A]=\mathcal{D}_{iA}{}^BT_B-\mathcal{K}_{iA}\tilde{J}-\mathcal{Q}_{iA}J\nonumber
$$
$$
 \qquad [T_A,T_B]=\mathcal{D}_{ABi}X^i    \qquad  [J,T_A]=\mathcal{K}_{iA}X^i \qquad  [\tilde{J},T_A]=\mathcal{Q}_{iA}X^i\nonumber
$$
where the generators $Z_i$ and $X^i$ are as in the Heterotic case above. $J$, $\tilde{J}$ and $T_A$ generate gauge transformations of the Ramond-Ramond fields where the indices $A,B=1,2,..22$ run over the twenty-two harmonic two-cycles of the $K3$ manifold. When the monodromy takes values in the $O(3,19;\Z)$ mapping class group\footnote{The group of large diffeomorphisms.} of the $K3$, the internal background is a smooth $K3$ fibration of the kind considered in \cite{Cvetic:2007ju}. Such compactifications are characterised by the structure constants ${\cal D}_{iA}{}^B$ - a geometric flux for $SU(2)$-structure reductions. It will be shown that the ${\cal K}$-fluxes correspond to $H$-flux compactifications and the ${\cal Q}$-flux corresponds to a new non-geometric flux. We will argue that this non-geometric background can be thought of as a $K3$-fibration over $T^2$ in which the monodromy includes a Mirror Symmetry in the $K3$ fibres, giving what might be called a \emph{Mirror-fold} \cite{Kawai and Sugawara ``Mirrorfolds with K3 Fibrations''}.

The content of this paper is organi\textbf{s}ed as follows. In the next section we review the salient features of ${\cal N}=4$ gauged supergravity in four dimensions. The ungauged, massless, theory has a rigid $SU(2)\times O(6,22)$ symmetry. The general form of the Lagrangian for the gauged supergravity was found in \cite{SW} where the gauge symmetry includes a non-Abelian subgroup of the rigid $SU(2)\times O(6,22)$ and we shall only consider electric gaugings of the $O(6,22)\subset SL(2)\times O(6,22)$. In section three we discuss the key features of six-dimensional $SU(2)$-structure manifolds. In section four we briefly review evidence, at the level of the massless supergravity, for the conjectured duality between the compactification of IIA supergravity on $K3$ and Heterotic supergravity on $T^4$. In section five we consider a further duality-twist reduction, down to four dimensions, on $T^2$. In section six, the interpretation of the duality twist reductions as compactifications of ten-dimensional theories on explicitly constructed $\bid$- and $SU(2)$-structure backgrounds is discussed. Section seven considers a worldsheet construction of the Heterotic backgrounds considered in section six along the lines of the doubled formalisms introduced in \cite{Hull ``A geometry for non-geometric string backgrounds''} and \cite{Tseytlin:1990va} and section eight concludes and discusses possible avenues for future research. To streamline the arguments, the details of most calculations are included in the Appendices.

\section{${\cal N}=4$ Gauged Supergravity in Four Dimensions}

In \cite{SW} the general form of the Lagrangian for ${\cal N}=4$ gauged supergravity in four dimensions was presented. The massless Abelian gauge theory has a rigid
$SL(2)\times O(6,n)$ symmetry \cite{Bergshoeff:1985ms}, certain subgroups of which can be promoted to a non-Abelian local symmetry in a manner consistent with ${\cal N}=4$ supersymmetry. The
non-Abelian gauging breaks the rigid symmetry to a subgroup, however there is still a natural action of $SL(2)\times O(6,n)$ on the fields of the
gauged theory. In \cite{SW}, following the work of \cite{d=4,d=5,d=6,d=7} (see also \cite{Samtleben Review} for an excellent review), this fact was exploited in order to write all possible ${\cal N}=4$, four
dimensional, gauged supergravities in terms of a single, universal, Lagrangian. The action of $SL(2)\times O(6,n)$ does not preserve the
gauging, but generally maps one gauged supergravity into another. The gauging introduces the constant deformation parameters
\begin{equation}\label{constants}
f_{\alpha MNP},  \qquad  \xi_{\alpha M},
\end{equation}
where $\alpha=\pm$ are indices for a vector representation of $SL(2)$ and $M,N=1,2,..6+n$ are indices for a vector representation of
$O(6,n)$. These deformation parameters determine the gauging and must satisfy various constraints in order to be consistent with ${\cal
N}=4$ supersymmetry. The task of classifying all ${\cal N}=4$ gauged supergravities in four dimensions reduces to one of finding the constants
(\ref{constants}) compatible with these constraints. We focus on the case $n=22$ as this is of most relevance to the string theory compactifications we shall be considering. The parameters
(\ref{constants}) transform as tensors under the rigid $SL(2)\times O(6,22)$. The bosonic degrees of freedom transform under the rigid
$G=SL(2)\times O(6,22)$, whilst the fermionic degrees of freedom transform in representations of the maximal compact subgroup
$H=SO(2)\times O(6)\times O(22)\subset G$.

The field content of the theory is made up of an ${\cal N}=4$ gravity multiplet which consists of a vielbein $e_{\mu}{}^a$, four
gravitini $\psi_{\mu}{}^i$, six spin-one gravi-photons $A_{\mu}$, four spin-half fermions $\chi^i$ and a complex scalar $\tau$ respectively. To this are
coupled twenty-two ${\cal N}=4$ vector multiplets, each consisting of a vector $A_{\mu}$, four spin-half gauginos $\lambda^{ai}$ and six real
scalars. It is useful to combine the vectors from the vector and gravity multiplets and write them as ${\cal A}_{(1)}^{M}$ and the 132 scalars of the 22 vector multiplets into an array ${\cal M}_{MN}$. The
scalars $\tau$ and ${\cal M}_{MN}$ take values in the product of cosets
$$
\frac{SL(2)}{SO(2)}\times \frac{O(6,22)}{O(6)\times O(22)}.
$$
The $SL(2)$ has a fractional-linear action on $\tau$ as $\tau\rightarrow (a\tau+b)/(c\tau+d)$. This can be realized as a linear action on the
array $M_{\alpha\beta}(\tau)$ given by
$$
M_{\alpha\beta}=\frac{1}{\Im(\tau)}\left(%
\begin{array}{cc}
  |\tau^2| & \Re(\tau) \\
  \Re(\tau) & 1 \\
\end{array}%
\right),
$$
where $\Re(\tau)$ and $\Im(\tau)$ denote the real and imaginary parts of $\tau$.

Here, we shall consider electric gaugings of the $O(6,22)$, where the only non-zero deformation parameter is
$f_{+MNP}\equiv t_{MNP}$ which plays the role of structure constants in the gauge algebra. The constants $t_{MNP}$ are antisymmetric in all indices. The four-dimensional gauged Lagrangian for the
bosonic sector is then
\begin{eqnarray}\label{general sugra}
\mathscr{L}_4&=&R*1+\frac{1}{4}D{\cal M}_{MN}\wedge *D{\cal M}^{MN}+\frac{1}{2}dM_{\alpha\beta}\wedge
*dM^{\alpha\beta}-\frac{1}{2}\Im(\tau){\cal M}_{MN}{\cal F}_{(2)}^{M}\wedge *{\cal F}_{(2)}^{M}\nonumber\\
&&-\frac{1}{2}\Re(\tau)L_{MN}{\cal F}_{(2)}^{M}\wedge {\cal F}_{(2)}^{M}-g^2V*1,\nonumber
\end{eqnarray}
where the scalar potential is
\begin{equation}\label{potential}
V=\frac{1}{48 \Im(\tau)}t_{MNP}t_{QRS}\left({\cal M}^{MQ}{\cal M}^{NR}{\cal M}^{PS}-3{\cal M}^{MQ}L^{NR}L^{PS}\right)
+\frac{1}{24\Im(\tau)}t_{MNP}t^{MNP}
\end{equation}
and ${\cal F}^{M}_{(2)}$ is the gauge covariant field strength for ${\cal A}^{M}_{(1)}$. The only constraint on this gauging is the Jacobi identity $t_{
[MN|R}t_{|PQ]}{}^R=0$ so that, a priori, any Lie subgroup of $O(6,22)$ should give a consistent gauging. The invariant of $O(6,22)$ is
$L_{MN}$, which can be used to raise and lower indices on the constants\footnote{In \cite{SW} a basis was chosen in
which $L_{MN}=$diag$(-\bid_6,\bid_{22})$, the basis choice here is given by (\ref{L}).} $t_{MNP}$. In particular
$$
t_{MNP}=L_{MQ}t_{NP}{}^Q
$$
Our goal will be to understand how some of these gauged supergravities lift to compactifications of string theory.

\section{Six-dimensional manifolds with $SU(2)$-structure}
One way in which a four-dimensional, half-maximal, gauged supergravity can be realized as a string theory reduction is as IIA supergravity, reduced on a manifold with $SU(2)$-structure. In this section, we will recall some of the salient features of manifolds with reduced structure, and then discuss the specific case of a six-dimensional manifold with $SU(2)$-structure in detail. Much of what is discussed here is published elsewhere (in particular, see \cite{GMW,GMPT3,BLT} and references therein), but the $SO(3)$-symmetry of $j$ and $\omega$, and the discussion about the forms we expand the ten-dimensional fields in, are original.

One of the  main questions in the reduction of a supersymmetric theory is how much supersymmetry of the higher-dimensional theory is explicitly broken by the dimensional reduction. In the case of a ten-dimensional supergravity theory the reduction ansatz of the ten-dimensional spacetime $M^{1,9}$ is
$$
M^{1,9} = M^{1,3} \times Y,
$$
where $Y$ is a compact six-dimensional manifold. The ten-dimensional Lorentz group decomposes as
$$
SO(1,9) \to SO(1,3) \times SO(6),
$$
and the 16-dimensional Majorana-Weyl spinor representation therefore decomposes as
$$
\mathbf{16} \to (\mathbf{2},\mathbf{4})\oplus (\mathbf{\bar
2},\mathbf{\bar 4}).
$$
A ten-dimensional supersymmetry parameter $\varepsilon$ can be written as
\begin{equation}\label{spinorexpansion}
\varepsilon  = \xi_{+\iota}\otimes\eta^{\iota}_+ + \xi_{-\iota}\otimes\eta^{\iota}_- \ ,
\end{equation}
with $\xi_{\iota}$ four-dimensional spinors and $\eta^{\iota}$ the available six-dimensional spinors where the index $\iota$ runs over 1 and 2 and the subscripts $\pm$ denote the chirality of the spinor. Clearly, the amount of supersymmetry of the four-dimensional theory depends on the number of supersymmetry parameters $\varepsilon$ of the original theory and on the number of six-dimensional spinors $\eta^{\iota}$.

The spinors $\eta^{\iota}$ in equation (\ref{spinorexpansion}) need to be globally well-defined on $Y$. In other words, they need to be singlets under the structure group\footnote{The structure group is the group the transition functions of the tangent bundle $TY$ take values in.} $G\subseteq SO(6)$ of $Y$. For example, on a manifold with $SU(3)$-structure, the spinor representation decomposes as
\begin{equation}\label{spinorreduc}
SO(6)\cong SU(4)\to SU(3)\, :\qquad
\mathbf{4} \to \mathbf{3}\oplus\mathbf{1},
\end{equation}
so there is one spinor that transforms as a singlet under the structure group. This is the one spinor we can use in (\ref{spinorexpansion}).

The spinors $\eta^{\iota}$ need not be covariantly constant, but there does exist a unique metric-compatible connection $\nabla^T$ such that
$$
\nabla^T \eta =0.
$$
If $\nabla^T$ is not the Levi-Civita connection, it must be torsional. This torsion can be described by five torsion classes $\mathcal{W}_1, \dots , \mathcal{W}_5$ \cite{GMW} and so manifolds with reduced structures can be characterized by their torsion classes.

Equivalently, a manifold of $SU(n)$-structure in $n$ complex dimensions can be defined in terms of a two-form $J$ and an $n$-form $\Omega$. These forms obey the relations
$$
J\wedge\Omega= 0,	\qquad	\Omega\wedge\bar{\Omega} = i^{n(n+2)}\frac{2^n}{n}J^n.
$$
Furthermore, $J$ with one index raised is an almost complex structure $I$,
$$
I_{v}^{\phantom{w}w}\equiv J_{vx} g^{xw} \qquad
I^2 = -\bid,
$$
and with respect to this almost complex structure, $J$ is a $(1,1)$-form and $\Omega$ is an $(n,0)$-form (with $v,w=1,\dots 2n$ the indices on the real coordinates). The exterior derivatives of $J$ and $\Omega$ are determined by the torsion classes as follows:
$$
dJ	 \in \mathcal{W}_1\oplus\mathcal{W}_3\oplus\mathcal{W}_4,	\qquad	d\Omega	 \in \mathcal{W}_1\oplus\mathcal{W}_2\oplus\mathcal{W}_5.
$$
In three complex dimensions, the equivalence between the definition of $SU(n)$-structure in terms of the spinors $\eta_{\pm}$ and that given in terms of the forms $(J,\Omega)$ is given by formulas that express the forms in terms of the spinors. A manifold with $SU(3)$-structure has one globally well-defined spinor $\eta$, and in terms of this spinor the forms $J$ and $\Omega$ can be written as
$$
J_{vw}=i\eta^{\dagger}_{-}\gamma_{vw}\eta_{-},
\qquad
\Omega_{vwx} =i\eta^{\dagger}_{-}\gamma_{vwx}\eta_{+},
\qquad
v,w=1,\ldots,6\ .
$$
Using Fierz identities, it can be shown that the forms, so defined, indeed satisfy the constraints of an $SU(3)$-structure.

Let us now specialise to the case of a six real-dimensional manifold with $SU(2)$-structure. By the reasoning presented before, such a manifold has two globally well-defined spinors $\eta^{\iota}$ ($\iota=1,2$). From (\ref{spinorexpansion}) we see that every ten-dimensional supersymmetry parameter then generates two four-dimensional supersymmetry parameters; for example, dimensional reduction of type IIA supergravity on a six-dimensional manifold with $SU(2)$-structure would give a four-dimensional supergravity with 16 supercharges, or $\mathcal{N}=4$.

For manifolds with  $SU(2)$-structure one can define a pair of $SU(3)$-structures: a pair of $2$-forms $J^{\iota}$ and a pair of $3$-forms $\Omega^{\iota}$ via
$$
J^{\iota}_{vw}=i\eta^{\iota\dagger}_{-}\gamma_{vw}\eta^{\iota}_{-},
\qquad
\Omega^{\iota}_{vwx} =i\eta^{\iota\dagger}_{-}\gamma_{vwx}\eta^{\iota}_{+},
\qquad
v,w=1,\ldots,6\ .
$$
By raising an index with the metric one obtains two almost complex structures
$$
I_{\phantom{\iota}v}^{\iota\phantom{v}w}\equiv J^{\iota}_{vx} g^{xw}, \qquad
(I^{\iota})^2 = -\bid,
$$
which generically are not integrable since the Nijenhuis-tensor is not necessarily vanishing. With respect to $I^{\iota}$ the two-forms $J^{\iota}$ are $(1,1)$-forms while the
$\Omega^{\iota}$ are $(3,0)$-forms.

If the manifold has an $SU(2)$-structure, the two almost complex structures commute, $[I^1,I^2]=0$, and define an almost product structure $\pi$ via
$$
\pi_{v}^{\phantom{v}w}\equiv I_{\phantom{2}v}^{1\phantom{v}x}I_{\phantom{2}x}^{2\phantom{x}w}
\quad \textrm{with}\quad
\pi^2 = \bid.
$$
One can check that $\pi$ has four negative and two positive eigenvalues which in turn implies that the tangent space splits into a four-dimensional and a two-dimensional component. It follows, then, that this split also holds for all tensor products of tangent and cotangent spaces. A form $\chi$ on the six-dimensional manifold $Y$ can therefore always be written as a wedge product
$$
\chi  = \chi_{(2)}\wedge\chi_{(4)}
$$
of a form $\chi_{(2)}$ with its legs in the two-dimensional directions and a form $\chi_{(4)}$ with its legs in the four-dimensional directions. Furthermore, it can be shown that the almost product structure $\pi$ is integrable, which means that it is possible, on a chart $U$, to choose coordinates $\{y^i, z^m\}$ for $i=1,2$ and $m=1,\dots,4$ such that
$$
\left\{\frac{\p}{\p y^i},\frac{\p}{\p z^m}\right\}
$$
spans the tangent space to $U$. This means that on $U$, the metric can be written in the block-diagonal form
$$
ds^2 = g_{mn}(y,z)dz^mdz^n + g_{ij}(y,z)dy^idy^j.
$$
In other words, locally the six-manifold $Y$ is a product of the form $Y\simeq Y^{(2)} \times Y^{(4)}$ where $Y^{(4)}$ is a real four-dimensional manifold while $Y^{(2)}$ is a real two-dimensional manifold. The almost product structure becomes more apparent when we look at the forms that can be defined using the spinors. Since the spinors are never parallel, a globally defined complex one-form
$$
\sigma_v \equiv \sigma_v^1-i\sigma_v^2  \equiv \eta^{2\dagger}_+\gamma_v\eta^1_-
$$
exists. With this information, the four tensors $J^{\iota}$ and $\Omega^{\iota}$ can be expressed in terms of the one forms $\s^{i}$, a $(1,1)$-form $j$ and a $(2,0)$-form $\omega$ via \cite{GMW,GMPT3}
$$
J^{1,2} = j \pm  \sigma^1\wedge \sigma^2\ , \qquad
\Omega^{1,2}  = \omega\wedge (\sigma^1\pm i\sigma^2 )\ ,
$$
or equivalently \cite{BLT}
$$
j=\frac{1}{2}(J^1-J^2)\ ,\quad
\omega_{vw}=i\eta^{1\dagger}_{-}\gamma_{vw}\eta^2_{-}\ .
$$
As shown in \cite{GMW,GMPT3}, the $\s^{i}$ can be viewed as one-forms on the two-dimensional component $Y^{(2)}$ while $j$ and $\omega$ define an $SU(2)$-structure on the four-dimensional component $Y^{(4)}$.

The simplest example of a six-dimensional manifold with $SU(2)$-structure is truly a product structure: it is $T^2\times K3$. The forms $\sigma^i$ are then simply $dy^i$, and the forms $j$ and $\omega$ that determine the structure of the $K3$ are closed.

It is well-known that every $K3$ is hyperk\"ahler, i.e. it has a two-sphere of complex structures with respect to which the metric is K\"ahler; in the terminology used above this means that any $SO(3)$-rotation of the vector
$$
\left(
\begin{array}{c}
j \\
\mathrm{Re}\, \omega \\
\mathrm{Im}\, \omega
\end{array}\right)
$$
defines a complex structure on that $K3$. In Appendix \ref{SO3SpinorCalc} we show that a similar symmetry exists for a general manifold with $SU(2)$-structure: the $SU(2)$-structure is defined up to an $SU(2)$-rotation of the spinors $\eta^i$. This rotation leaves $\sigma$ invariant, but translates into an $SO(3)$-rotation on $(j, \mathrm{Re}\, \omega, \mathrm{Im}\,\omega)$.

In order to perform a dimensional reduction of a supergravity, we need to know what forms to expand the ten-dimensional fields in. Since six-dimensional manifolds with $SU(2)$-structure seem to be, in many aspects, a generalization of $T^2\times K3$, it is reasonable to base the available forms on $T^2\times K3$ as well. We therefore expand in a basis of forms spanned by two one-forms $\s^{i}$ and an arbitrary number $n$ of two-forms $\widetilde{\Omega}^A$. These two-forms contain the $(2,0) \oplus (0,2)$-form $\omega$, the $(1,1)$-form $j$ and $n-3$ further $(1,1)$-forms. We can define an intersection metric $\eta^{AB}$ as
$$
\eta^{AB}\eps^{ij}\equiv\int_{Y}\s^{i}\wedge \s^{j}\wedge\widetilde{\Omega}^A\wedge\widetilde{\Omega}^B.
$$

In contrast to the forms $(dy^i,\Omega^A)$ on $T^2\times K3$, however, the forms $(\sigma^i,\widetilde{\Omega}^A)$ of the $SU(2)$-structure manifold need not be closed. This means the most general consistent formulas for the exterior derivatives of the forms are
\begin{equation}\label{dforms2}
d\s^{i}=\frac{1}{2}\mathcal{D}_{jk}{}^i\s^{j}\wedge \s^{k}+\mathcal{D}^i{}_A\widetilde{\Omega}^A,	\qquad	 d\widetilde{\Omega}^A=\mathcal{D}_{iB}{}^A\s^{i}\wedge\widetilde{\Omega}^B.
\end{equation}
and we shall assume that $\mathcal{D}_{ij}{}^k$, ${\cal D}_A{}^i$ and ${\cal D}_{iA}{}^B$ are all constant. Furthermore, there are some additional constraints on the matrices $\mathcal{D}$. These come from requiring $d^2=0$ (integrability of $\sigma^i$ and $\widetilde{\Omega}^A$) and Stokes' theorem to hold. Let us look at what that implies for $\mathcal{D}^i{}_A=0$. Requiring $d^2=0$ then means
\begin{equation}\label{Jacobi21}
\mathcal{D}_{iB}{}^C\mathcal{D}_{jA}{}^B-\mathcal{D}_{jB}{}^C\mathcal{D}_{iA}{}^B=\mathcal{D}_{ij}{}^k\mathcal{D}_{kA}{}^C,
\end{equation}
but it does not give a similar constraint on the $\mathcal{D}_{ij}^k$ since a triple wedge product of $\s^i$'s is zero regardless of the coefficient. Instead, we can obtain the constraint
\begin{equation}\label{Jacobi22}
\mathcal{D}_{il}{}^k\mathcal{D}_{jm}{}^l - \mathcal{D}_{jl}{}^k\mathcal{D}_{im}{}^l=\mathcal{D}_{ij}{}^k\mathcal{D}_{lm}{}^l
\end{equation}
by explicitly writing out the indices $i,j$. Finally, Stokes' theorem yields the constraint
\begin{equation}\label{Stokes2}
-\eta^{AB}\mathcal{D}_{ik}{}^k=\eta^{AC}\mathcal{D}_{iC}{}^B+\eta^{BC}\mathcal{D}_{iC}{}^A.
\end{equation}

We shall be particularly interested in a class of simple examples, where $\mathcal{D}_{ij}{}^k=\mathcal{D}^i{}_A=0$ and $A=1,\dots 22$, so that
$$
d\s^{i}	 = 0,	\qquad	d\widetilde{\Omega}^A=\mathcal{D}_{iB}{}^A\s^{i}\wedge\widetilde{\Omega}^B.
$$
These Bianchi identities may be solved by
$$
\s^{i}	 = dy^i,	\qquad	\widetilde{\Omega}^A=\mathrm{exp}\,(\mathcal{D}_{iB}{}^Ay^i)\Omega^B
$$
where we have introduced the harmonic two-form $\Omega^A(z) \in H^2(Y^{(4)})$, so that $d\Omega^A=0$. We may then identify $Y^{(4)}\simeq K3$ and it is clear that $Y$ is a deformation of the $SU(2)$-holonomy manifold $T^2\times K3$ to a bundle
\begin{eqnarray}
K3\hookrightarrow &Y&\nonumber\\
&\downarrow&\nonumber\\
&T^2&
\end{eqnarray}
where $Y^{(4)}\simeq K3$, $Y^{(2)}\simeq T^2$ and $Y$, therefore, is a non-trivial $K3$-fibration over the cycles of the base $T^2$. The monodromy of the fibrations over the cycle with coordinate $y^i \sim y^i + \xi^i$ is given by $e^{\mathcal{D}_{iA}{}^B\xi^i}$ and we require that this monodromy takes values in the mapping class group of the $K3$-fibre in order for $Y$ to be a smooth bundle.

We shall see, in the sections that follow, that many compactifications on manifolds of this kind may be realized by the duality-twist reductions of \cite{Scherk How To Get Masses From Extra Dimensions} which we review in section five.

\section{${\cal N}=2$ Supergravity in Six Dimensions}
Our goal is to realize a certain class of four-dimensional, half-maximal, supergravities in which a non-Abelian subgroup of $O(6,22)$ is gauged, as
compactifications of ten-dimensional string and supergravity theories. As stated in the introduction, we shall first consider standard
Kaluza-Klein compactifications of ten-dimensional IIA and Heterotic supergravities to six dimensions. The compactifications of the Heterotic and IIA theories on
$T^4$ and $K3$ respectively, give rise to massless, Abelian, ${\cal N}=2$ theories in six dimensions. The conjectured duality between Heterotic
and IIA string theories in six dimensions is then used to identify these theories as a precursor to further dimensional reduction to four dimensions, which will be
performed in the next section. The details of the compactification to six dimensions will be of importance in understanding the lift of the four
dimensional gauged supergravities to string theory as will be discussed in section six and seven.

\subsection{Compactification of Heterotic Supergravity on $T^4$}

The bosonic sector of the Heterotic supergravity consists of a scalar dilaton $\Phi$, a two-form potential $\mathscr{B}_{(2)}$ with
associated three form field strength $\mathscr{H}_{(3)}=d\mathscr{B}_{(2)}+...$, and gauge bosons $\mathscr{A}_{(1)}^a$ with field strength $\mathscr{F}_{(2)}^a$,
taking values in the adjoint representation of either $E_8\times E_8$ or $Spin(32)/\Z_2$ \cite{Gross:1985fr,Gross:1985rr,Gross:1984dd}. It will be assumed that the
internal $E_8\times E_8$ or $Spin(32)/\Z_2$ gauge group\footnote{
The $Spin(32)/\Z_2$ Heterotic string theory is more usually referred to as the $SO(32)$ Heterotic string. The worldsheet theory does indeed have a rigid $SO(32)$ symmetry; however, the worldsheet fermions $\lambda^a$, taking values in $Spin(32)$ - the cover of $SO(32)$ - are subject to a GSO-type projection, which means that only two of the four conjugacy classes of $Spin(32)$ play a role in the spectrum of the theory \cite{Gross:1985fr}. This distinction will not play an important role here but, in keeping with our aim to describe the supergravities in terms of string theory, we shall adopt the more accurate description of the ten-dimensional gauge group, as $Spin(32)/\Z_2$ throughout.} is broken to the Cartan subgroup $U(1)^{16}$ by some mechanism, such as Wilson lines in the toroidal compactification. The internal gauge fields $\mathscr{A}^a$ then take values in the Lie algebra of $U(1)^{16}$
where $a,b=1,2,...16$. In fact, sixteen of the 496 generators of $E_8\times E_8$ or $Spin(32)/\Z_2$ can be identified as isometries of the Cartan torus $T^{16}$, with the remaining 480 generators related to massless solitons \cite{Gross:1985fr}. We may therefore relate this $U(1)^{16}$ directly to the geometric symmetries of the Cartan torus. This perspective will be important in section seven where we consider the worldsheet description of the backgrounds discussed in the following sections.

The supergravity may be thought of as a low energy, weak coupling, effective field theory for the Heterotic string. The low energy description is truncated to first order in $\alpha'$, where the internal gauge fields first appear. A one-loop calculation in the
string coupling gives further corrections, involving the spin connection, at first order in $\alpha'$. Such corrections are important for issues
such as anomaly cancelation in the theory and any full treatment should take such corrections into account \cite{Green:1984sg}\footnote{See also Chapter 13 of \cite{Green:1987mn} for further discussion.}. Here, we shall
neglect such string loop effects and consider only the tree level contributions to the effective action, leaving a more complete analysis to be
considered at a later date. The bosonic sector of the effective ten-dimensional theory we shall consider is given, in string frame, by the Lagrangian
\begin{eqnarray}\label{heterotic}
\mathscr{L}^{Het}_{10}=e^{-\Phi}\left(\mathscr{R}*1+*d\Phi\wedge
d\Phi-\frac{1}{2}\mathscr{H}_{(3)}\wedge*\mathscr{H}_{(3)}-\frac{1}{2}\delta_{ab}\mathscr{F}_{(2)}^a\wedge*\mathscr{F}_{(2)}^b\right),
\end{eqnarray}
where we have set $\alpha'=1$ and the field strengths are given by
$$
\mathscr{F}_{(2)}^a=d\mathscr{A}_{(1)}^a,  \qquad \mathscr{H}_{(3)}=d\mathscr{B}_{(2)}-\frac{1}{2}\delta_{ab}\mathscr{A}_{(1)}^a\wedge\mathscr{F}_{(2)}^a.
$$
The particular case of interest is that in which four of the space coordinates are compactified into a $T^4$ with internal coordinates $z^m$, where
$m,n=6,7,8,9$. The standard Kaluza-Klein reduction ansatz is used, for which the reduction ansatz for the fields are the zero modes of harmonic
expansions on the internal space, and as such the ansatz does not depend on the coordinates $z^m$. The details of this
reduction are given in the Appendix \ref{6DN2Theory}. Inserting the reduction
ansatz given in (\ref{B3}) into the Lagrangian (\ref{heterotic}) and integrating over the $T^4$ gives the effective \textbf{${\cal N}=2$} theory in six dimensions. This six-dimensional theory has a rigid $O(4,20)$ symmetry \cite{Nishino:1986dc} and the fields can be
combined into multiplets of this rigid symmetry so that the Lagrangian can be written in a manifestly $O(4,20)$ invariant way \cite{Nishino:1986dc}
\begin{equation}\label{HetSO420}
\mathscr{L}^{Het}_{6}=e^{-\widehat{\phi}}\left(\widehat{R}*1+*d\widehat{\phi}\wedge d\widehat{\phi}+\frac{1}{4}d\widehat{{\cal M}}_{IJ}\wedge
*d\widehat{{\cal M}}^{IJ}-\frac{1}{2}\widehat{\mathcal{H}}_{(3)}\wedge *\widehat{\mathcal{H}}_{(3)}-\frac{1}{2}\widehat{{\cal
M}}_{IJ}\widehat{\mathcal{F}}^I_{(2)}\wedge*\widehat{\mathcal{F}}^J_{(2)}\right),\nonumber
\end{equation}
where the indices $I,J$ run from $1$ to $24$. The two- and three-form $O(4,20)$-covariant field strengths are given by
\begin{equation}\label{H and F}
\widehat{\mathcal{H}}_{(3)}=d\widehat{C}_{(2)}-\frac{1}{2}L_{IJ}\widehat{{\cal A}}_{(1)}^I\wedge d\widehat{{\cal A}}^J_{(1)}  \qquad
\widehat{{\cal F}}^I_{(2)}=d\widehat{\cal A}^I_{(1)}
\end{equation}
where the potential $\widehat{C}_{(2)}$ is related to the $B$-field $\mathscr{B}$.

The vector fields coming from the reduction of the gauge fields $\mathscr{A}^a$, the off-diagonal parts of the metric, and the $B$-field components with one leg on the $T^4$ and the other in the six-dimensional spacetime, combine into the $O(4,20)$ vector $\widehat{\cal A}^I_{(1)}$. The array of scalars $\widehat{\cal M}_{IJ}$ is given by
\begin{eqnarray}\label{Heterotic M}
\widehat{{\cal M}}_{IJ}=\left(%
\begin{array}{ccc}
  \widehat{g}_{mn}+\widehat{A}_{ma}\widehat{A}_n{}^a-\widehat{C}_{mp}\widehat{g}^{pq}\widehat{C}_{qn} & \widehat{C}_{mn}\widehat{g}^{np} & \widehat{A}_m{}^a-\widehat{C}_{mn}\widehat{g}^{np}\widehat{A}_p{}^a \\
  -\widehat{g}^{mp}\widehat{C}_{pn} & \widehat{g}^{mn} & -\widehat{g}^{mn}\widehat{A}_n{}^a \\
  \widehat{A}_m{}^a+\widehat{A}_p{}^a\widehat{g}^{pn}\widehat{C}_{nm} & -\widehat{A}_n{}^a\widehat{g}^{nm} & \delta_{ab}+\widehat{A}_{ma}\widehat{g}^{mn}\widehat{A}_{nb} \\
\end{array}%
\right),
\end{eqnarray}
where $\widehat{g}_{mn}$  are the metric moduli for the $T^4$, $\widehat{C}_{mn}$ is related to the $B$-field with both legs on the $T^4$ and
$\widehat{A}_m{}^a$ are the components of the gauge field $\mathscr{A}^a$ along the $T^4$. The scalars $\widehat{{\cal M}}_{IJ}$ take values in
the Grasmannian $O(4,20)/(O(4)\times O(20))$ and the inverse of $\widehat{\cal M}_{IJ}$ is given by $\widehat{\cal M}^{IJ}=L^{IK}\widehat{\cal M}_{KL}L^{LJ}$ where
\begin{equation}\label{L het}
L_{IJ}=\left(\begin{array}{ccc} 0 & \bid_{4} & 0 \\ \bid_{4} & 0 & 0 \\ 0 & 0 & \bid_{16}
\end{array}\right)\nonumber \end{equation}
is the invariant of $O(4,20)$.

The six-dimensional theory has an Abelian $U(1)^{24}$ gauge symmetry, with gauge bosons $\widehat{\cal A}^I_{(1)}$,  where $U(1)^{16}\subset U(1)^{24}$ arises from the internal gauge symmetry inherited directly from the ten-dimensional theory. In addition, a $U(1)^4$ comes from diffeomorphisms $z^m\rightarrow z^m+\omega^m$ of the internal $T^4$ and a further $U(1)^4$ arises from the antisymmetric tensor transformations of the ten-dimensional $B$-field where one leg of the $B$-field lies
along a cycle in the $T^4$ and the other along the six-dimensional non-compact spacetime. Let us denote the sixteen generators of the ten-dimensional (internal) gauge transformations by $Y_a$, the four generators of diffeomorphisms
along the cycles of the torus as $Z_m$ and the generators of the antisymmetric tensor transformations as $X^m$. The generators may be arranged
into an $O(4,20)$  vector
\begin{eqnarray}\label{generator decomposition}
T_I=\left(%
\begin{array}{c}
  Z_m \\
  X^m \\
  Y_a \\
\end{array}%
\right),
\end{eqnarray}
which generates the full $U(1)^{24}$ with Abelian gauge algebra $[T_I,T_J]=0$.

\subsection{Compactification of IIA Supergravity on $K3$}

\noindent The Lagrangian of the bosonic sector of Type IIA supergravity in ten dimensions is
\begin{eqnarray}
 \mathscr{L}^{IIA}_{10} &=&   e^{-\Phi}\left( \mathscr{R}*1 + d\Phi\wedge *d\Phi - \frac{1}{2}d\mathscr{B}_{(2)}\wedge\ast d\mathscr{B}_{(2)}
-\frac{1}{2}d\mathscr{A}_{(1)}\wedge\ast d\mathscr{A}_{(1)} \right. \nonumber\\
&&- \left.\frac{1}{2}(d\mathscr{C}_{(3)} - \mathscr{A}_{(1)}\wedge d\mathscr{B}_{(2)})\wedge\ast (d\mathscr{C}_{(3)}- \mathscr{A}_{(1)}\wedge
d\mathscr{B}_{(2)}) -\frac{1}{2} \mathscr{B}_{(2)}\wedge d\mathscr{C}_{(3)}\wedge d\mathscr{C}_{(3)}\right),\nonumber
\end{eqnarray}
where $\Phi$ is the dilaton, $\mathscr{B}_{(2)}$ is the Kalb-Ramond field and $\mathscr{A}_{(1)}$ and $\mathscr{C}_{(3)}$ are Ramond-Ramond fields.
The $O(4,20)$-invariant, six-dimensional, \textbf{${\cal N}=2$} theory above can also be obtained from a Kaluza-Klein reduction of IIA supergravity on
$K3$.

The $K3$ manifold is characterised by twenty-two harmonic two cycles, nineteen of which are self-dual, the remaining three are anti-self-dual. We denote these two cycles by
$\Omega^A$, where $A,B=1,2,...22$. The Hodge dual of $\Omega^A$ is also a two-form and can be expanded in the basis $\{ \Omega^A \}$ as $*\Omega^A=H^A{}_B\Omega^B$. The array of coefficients $H^A{}_B$ take values in the Grassmannian
$$
\fdh{A}{B} \in\frac{SO(3,19)}{SO(3)\times SO(19)}
$$
and encode fifty-seven of the fifty-eight metric moduli of the $K3$ appearing as scalar fields in the six-dimensional effective theory. It is not hard to show that $\fdh{A}{B}$ satisfies
$H^A{}_CH^C{}_B=\delta^A{}_B$ and $\eta_{[A|C}H^C{}_{|B]}=0$, where $\eta^{AB}$ is the intersection matrix for $K3$
$$
\eta^{AB}\equiv\int_{K3}\Omega^A\wedge \Omega^B,
$$
and $\eta_{AB}$ is defined as its inverse. A final metric modulus, $e^{-\rho}$, controls the overall volume of the $K3$. When compactifying on a $K3$, the moduli $H^A{}_B$ and $\rho$, may depend on the six non-compact directions. We will later give the explicit dependence of these moduli fields on the two torus directions, and the fields with this specific form of dependence will be denoted by $\widetilde{H}^A{}_B$ and $\widetilde{\rho}$ and we shall henceforth use $H^A{}_B$ and $\rho$ to denote moduli that do not depend on the $T^2$ coordinates $y^i$.

The six-dimensional supergravity Lagrangian is given by
\begin{eqnarray}\label{IIA in 6D}
\mathscr{L}_{6}^{IIA}&=&e^{-\widetilde{\phi}}\left(\widetilde{R}*1+*d\widetilde{\phi}\wedge d\widetilde{\phi}+\frac{1}{4}d\widetilde{{\cal M}}_{IJ}\wedge
*d\widetilde{{\cal M}}^{IJ}-\frac{1}{2}\widetilde{\mathcal{H}}_{(3)}\wedge *\widetilde{\mathcal{H}}_{(3)}-\frac{1}{2}\widetilde{{\cal
M}}_{IJ}\widetilde{\mathcal{F}}^I_{(2)}\wedge*\widetilde{\mathcal{F}}^J_{(2)}\right)\nonumber\\
&&-\frac{1}{2}L_{IJ}\widetilde{\mathcal{B}}_{(2)}\wedge \widetilde{\mathcal{F}}_{(2)}^I\wedge \widetilde{\mathcal{F}}_{(2)}^J,
\end{eqnarray}
where the field strengths are
$$
\widetilde{\mathcal{H}}_{(3)}=d\widetilde{\mathcal{B}}_{(2)},  \qquad  \widetilde{\mathcal{F}}_{(2)}^I=d\widetilde{\mathcal{A}}^I_{(1)}.
$$
The relationship between $\widetilde{\mathcal{B}}$ and the ten-dimensional $B$-field is given in Appendix D, as are other details of this reduction. The Ramond potential $\mathscr{A}_{(1)}$ and the part of the Ramond field
$\mathscr{C}_{(3)}$ which wraps the twenty-two harmonic cycles $\Omega^A$ combine with the six-dimensional Hodge-dual of the three-form Ramond field to
give the $O(4,20)$ vector $\widetilde{\mathcal{A}}^I_{(1)}$.

The matrix $\widetilde{{\cal M}}_{IJ}$ takes values in the Grassmannian $O(4,20)/(O(4)\times O(20))$ and is given by
\begin{equation}\label{IIAK3Scalars}
\widetilde{{\cal M}}_{IJ}=\left(
    \begin{array}{ccc}
 e^{-\widetilde{\rho}} + \widetilde{H}^{AB}\widetilde{b}_A\widetilde{b}_B + e^{\widetilde{\rho}} \widetilde{C}^2
        & e^{\widetilde{\rho}} \widetilde{C}
        & -\widetilde{H}^{C}{}_{B}\widetilde{b}_C - e^{\widetilde{\rho}} \widetilde{b}_B\widetilde{C}  \\
 e^{\widetilde{\rho}} \widetilde{C}
        & e^{\widetilde{\rho}}
        & - e^{\widetilde{\rho}} \widetilde{b}_B  \\
 -\widetilde{H}^{B}{}_{A}\widetilde{b}_B - e^{\widetilde{\rho}} \widetilde{b}_A\widetilde{C}
        & - e^{\widetilde{\rho}} \widetilde{b}_A
        & \eta_{AC}\widetilde{H}^{C}{}_{B} + e^{\widetilde{\rho}} \widetilde{b}_A\widetilde{b}_B
    \end{array}\right),
\end{equation}
where $\widetilde{b}_A$ are the twenty-two components of the $B$-field which wrap the harmonic cycles $\Omega^A$,
$\widetilde{C}=\frac{1}{2}\eta^{AB}\widetilde{b}_A\widetilde{b}_B$ and the indices on $\widetilde{H}^A{}_B$ are raised and lowered using $\eta^{AB}$ and its inverse $\eta_{AB}$ respectively. The symmetric matrix of scalars $\widetilde{\cal M}_{IJ}$ satisfy $\widetilde{{\cal
M}}_{IK}L^{KL}\widetilde{{\cal M}}_{LJ}=L_{IJ}$ with $L_{IJ}$, the invariant of $O(4,20)$, given by
\begin{equation}
L_{IJ}=\left(\begin{array}{ccc}
0   & -1    & 0 \\
-1  & 0 & 0 \\
0   & 0 & \eta_{AB} \end{array}\right).\nonumber
\end{equation}

The gauge algebra of this theory is $U(1)^{24}$, where $U(1)^{22}\subset U(1)^{24}$ is generated by antisymmetric tensor transformation of the Ramond-Ramond fields with parameters
$\Lambda^A$ associated to each of the harmonic two cycles of the $K3$. A further $U(1)$ is inherited directly from ten dimensions as the Abelian
gauge transformation of the $\mathscr{A}_{(1)}$ Ramond field $\delta\mathscr{A}_{(1)}=d\Lambda$. We denote the generator of this transformation
by $J$. In six dimensions the three form part of the Ramond field $\mathscr{C}_{(3)}$ is dual to a one form $\tilde{C}_{(1)}$ and so a final $U(1)$ comes
from the Abelian gauge transformations of this field $\delta\widetilde{C}_{(1)}=d\widetilde{\lambda}_{(0)}$, generated by $\tilde{J}$. These
generators can be written as a $O(4,20)$ vector $T_I$
\begin{eqnarray}
T_I=\left(%
\begin{array}{c}
  J \\
  \tilde{J} \\
  T_A \\
\end{array}%
\right).
\end{eqnarray}
which generate the Abelian gauge algebra $[T_I,T_J]=0$.

\subsection{Heterotic/IIA Duality}

At the level of the supergravity, one can show that the Heterotic and IIA compactifications considered above give equivalent six-dimensional Lagrangians. It has been conjectured \cite{Hull and Townsend``Enhanced gauge symmetries in superstring theories,Seiberg}
that the full quantum string theories are in fact dual, a conjecture which we shall assume to be true in what follows. Evidence for the duality can, for example, be found in the six-dimensional supergravities and a study of the BPS sectors as discussed in \cite{Sen1995,Sen:1998kr,Harvey:1995rn,Duff:1995wd} and references therein. We review the basic supergravity argument below. An excellent introduction to this conjectured duality and its consequences may be found in \cite{Kiritsis:1998hy}.

The IIA Lagrangian $\mathscr{L}_{6}^{IIA}$ can be brought to the form of the Heterotic Lagrangian $\mathscr{L}_{6}^{Het}$ by first identifying
\begin{equation*}
\widetilde{{\cal M}}_{IJ} \to \widehat{{\cal M}}_{IJ}, \qquad \widetilde{\mathcal{A}}^I_{(1)} \to \widehat{\mathcal{A}}^I_{(1)}.
\end{equation*}
The next step is to dualise the three form $\widetilde{\mathcal{H}}_{(3)}$. We introduce the term
$$
d\widehat{C}_{(2)}\wedge\widetilde{\cal H}_{(3)}
$$
into the Lagrangian (\ref{IIA in 6D}). The two-form $\widehat{C}_{(2)}$ is a Lagrange multiplier which imposes the Bianchi
identity $d\widetilde{\cal H}_{(3)}=0$, the variation of the Lagrangian with respect to the field strength $\widetilde{\mathcal{H}}_{(3)}$
gives\footnote{Up to total derivative terms which vanish in the action.}
$$
\widehat{\mathcal{H}}_{(3)}=d\widehat{C}_{(2)}-\frac{1}{2}L_{IJ}\widehat{{\cal A}}_{(1)}^I\wedge d\widehat{{\cal
A}}^J_{(1)}=e^{-\widehat{\phi}}\widetilde{\mathcal{H}}_{(3)},
$$
which we recognize as the three-form field strength in the six-dimensional Heterotic theory (\ref{H and F}). Substituting this back into the six-dimensional IIA Lagrangian, and changing the sign of the dilaton $\widetilde{\phi}\rightarrow -\widehat{\phi}$, gives the Lagrangian of the six-dimensional Heterotic theory. We see then that, at the level of classical supergravities, the theories are indeed equivalent descriptions of the same physics.

\section{Double Duality Twist Reduction Over $T^2$}
We have seen that the Heterotic theory compactified on $T^4$ and the IIA theory compactified on $K3$ give the same supergravity with effective
Lagrangian in six dimensions given by
\begin{eqnarray}\label{D=6 lagrangian}
\mathscr{L}_{6}=e^{-\widehat{\phi}}\left(\widehat{R}*1+*d\widehat{\phi}\wedge d\widehat{\phi}+\frac{1}{4}d\widehat{{\cal M}}_{IJ}\wedge
*d\widehat{{\cal M}}^{IJ}-\frac{1}{2}\widehat{\mathcal{H}}_{(3)}\wedge *\widehat{\mathcal{H}}_{(3)}-\frac{1}{2}\widehat{{\cal
M}}_{IJ}\widehat{F}^I_{(2)}\wedge*\widehat{F}^J_{(2)}\right).
\end{eqnarray}
As noted in \cite{Nishino:1986dc}, the theory has $O(4,20)$ rigid symmetry, a discrete subgroup of which lifts to a U-duality symmetry of the full
string theory \cite{Hull and Townsend Unity of superstring dualities}. In this section we consider a further reduction on $T^2$, twisting by
two commuting elements of the discrete $U$-duality subgroup $O(4,20;\mathbb{Z})\subset O(4,20)$ over the two cycles of the $T^2$, to give an
effective theory in four dimensions. Let $y^i$, $i=1,2$ be the $T^2$ coordinates. The reduction ansatz for the duality twist compactifications
introduces a $y^i$-dependence in the fields according to their transformation properties under $O(4,20)$ \cite{Dabholkar:2002sy,Scherk How To Get Masses From Extra Dimensions} so that, for the scalar and
vector fields, $\widehat{\cal M}_{IJ}$ and $\widehat{\cal A}^I_{(1)}$, the reduction ansatz is given by
\begin{equation}\label{ansatz}
\widehat{{\cal M}}_{IJ}(x,y)=\mathcal{O}_I{}^K(y)\mathcal{M}_{KL}(x)\mathcal{O}^L{}_J(y),  \qquad  \widehat{\mathcal{A}}^I(x,y)=\mathcal{O}^I{}_J(y)\mathcal{A}^J(x),
\end{equation}
where $\mathcal{O}^I{}_J(y)=\exp(N_{iJ}{}^Iy^i)$, with $N_I{}^J$ taking values in the Lie algebra of $SO(4,20)$. Following established convention, we shall refer to ${\cal O}^I{}_J$ as the \emph{twist} matrix and $N_{iI}{}^J$ as the \emph{mass} matrix. The notation we adopt is that the fields with hats or tildes are $y$-dependent.

The structure constants $N_{iJ}{}^I$ encode the monodromy around the $i=1,2$ directions\footnote{One can consider these directions as
describing the canonical $\alpha$ and $\beta$ cycles of the torus.}
\begin{eqnarray}
N_{iJ}{}^I=\left(%
\begin{array}{cc}
 \alpha_J{}^I, &
  \beta_J{}^I
\end{array},%
\right)
\end{eqnarray}
where $e^{\alpha}$ is the $SO(4,20)$ monodromy around the $y^1\sim y^1+1$ direction and $e^{\beta}$ is that around the $y^2\sim y^2+1$
direction where $[\alpha,\beta]=0$. The condition that the two twists commute is
$[\alpha,\beta]_I{}^J=\alpha_I{}^K\beta_K{}^J-\beta_I{}^K\alpha_K{}^J=2N_{I[i|}{}^KN_{|j]K}{}^J=0$ which is equivalent to a Bianchi identity for the twist matrix
\begin{equation}
d^2(e^{N\cdot y})_I{}^J=(e^{N\cdot y})_I{}^LN_{Li}{}^KN_{jK}{}^Jdy^i\wedge dy^j=0.
\end{equation}

Putting the reduction ansatz (\ref{ansatz}) into the six-dimensional Lagrangian (\ref{D=6 lagrangian}) and integrating over the $y^i$-coordinates gives the Lagrangian for an ${\cal N}=4$, four-dimensional gauged supergravity. This Lagrangian may be written in an $O(6,22)$ covariant way
\begin{eqnarray}
\mathscr{L}_4&=&e^{-\phi}\left(R*1+*d\phi\wedge d\phi+\frac{1}{2}*\mathcal{H}_{(3)}\wedge \mathcal{H}_{(3)}+\frac{1}{4}*D{\cal M}_{MN}\wedge
D{\cal M}^{MN}\right. \nonumber\\ &&- \left.\frac{1}{2}{\cal M}_{MN}*\mathcal{F}_{(2)}^M\wedge\mathcal{F}_{(2)}^N\right)+V*1.
\end{eqnarray}
The scalar potential is given by
\begin{eqnarray}
V=e^{-\phi}\left(- \frac{1}{12}{\cal M}^{MQ}{\cal M}^{NT}{\cal M}^{PS}t_{MNP}t_{QTS}+ \frac{1}{4}{\cal M}^{MQ}L^{NT}L^{PS}t_{MNP}t_{QTS}\right),
\end{eqnarray}
where $L_{MQ}t_{NP}{}^Q=t_{MNP}$ are the structure constants for the gauged supergravity. At first glance, there seems to be a discrepancy between this scalar potential and the general one given by (\ref{potential}). The expressions are
reconciled if we note that $t_{MNP}t_{QTS}L^{MQ}L^{NT}L^{PS}=N_{iIJ}N_{jKL}L^{ij}L^{IK}L^{JL}=0$ since $L^{ij}=0$. The scalar fields ${\cal
M}_{MN}$ take values in the coset $O(6,22)/(O(6)\times O(22))$ where
\begin{equation} {\cal M}_{MN}= \left(\begin{array}{ccc}
g_{ij}+{\cal M}_{IJ}{\cal A}^I_{i}{\cal A}^J_{j}+g^{kl}C_{ik}C_{jl} & g^{ik}C_{jk} & g^{jk}C_{ij}L_{IK}{\cal A}^K_{k}+{\cal M}_{IK}{\cal A}^K_{i} \\
g^{ik}C_{jk} & g^{ij} & g^{ij}L_{IK}{\cal A}^K_{j}  \\
 g^{jk}C_{ij}L_{JK}{\cal A}^K_{k}+{\cal M}_{JK}{\cal A}^K_{i}   &  g^{ij}L_{IK}{\cal A}^K_{j} & {\cal
 M}_{IJ}+g^{ij}L_{IK}L_{JL}{\cal A}^K_{i}{\cal A}^L_{j}
\end{array}\right).\nonumber
\end{equation}
The $O(6,22)$ invariant is
\begin{equation}\label{L}
L_{MN}=\left(\begin{array}{ccccc} 0 & \bid_2 & 0 \\ \bid_2 & 0 & 0 \\ 0 & 0 & L_{IJ}
\end{array}\right). \end{equation}

In four dimensions, we may write this in the Einstein frame using the four-dimensional Weyl rescaling
\begin{equation}
g_{\mu\nu}(x)\rightarrow e^{\phi(x)}g_{\mu\nu}(x)
\end{equation}
and then dualizing the $\mathcal{H}_{(3)}$ field to a scalar $\chi$ as described in Appendix E.3. Introducing the axio-dilaton $\tau=\chi_{}+ie^{-\phi}$ and defining
\begin{equation}
M_{\alpha\beta}=e^{\phi}\left(%
\begin{array}{cc}
  \chi_{}^2+e^{-2\phi} & \chi_{} \\
  \chi_{} & 1 \\
\end{array}%
\right),
\end{equation}
the Lagrangian may then be written as
\begin{eqnarray}\label{O(d,d) Lagrangian}
\mathscr{L}_4&=&R*1+\frac{1}{4}dM_{\alpha\beta}*dM^{\alpha\beta}+\frac{1}{4}D{\cal M}_{MN}\wedge *D{\cal M}^{MN}\nonumber\\
&&-\frac{1}{2}\Im(\tau){\cal M}_{MN}*\mathcal{F}_{(2)}^M\wedge\mathcal{F}_{(2)}^N+\frac{1}{2}\Re(\tau)L_{MN}\mathcal{F}_{(2)}^M\wedge\mathcal{F}_{(2)}^N+V*1.
\end{eqnarray}
Written in this form, it is easier to see that the scalars
$M_{\alpha\beta}$ and $\mathcal{M}_{IJ}$ parameterize the space
\begin{equation}
\frac{SL(2)}{U(1)}\times \frac{O(6,22)}{O(6)\times O(22)}.
\end{equation}

\subsection{Gauge Algebra}
Introducing generators for the diffeomorphisms $y^i\rightarrow y^i+\omega^i$ of the $T^2$ as $Z_i$ and the antisymmetric tensor transformations of the $B$-field,
with one leg along the $T^2$ and the other in the non-compact four-dimensional spacetime, as $X^i$, the gauge algebra of this four-dimensional
gauged supergravity is
\begin{eqnarray}\label{O(d,d+16) Lie algebra}
\left[Z_i,T_I\right]&=&N_{Ii}{}^JT_J,   \qquad\qquad  \left[T_I,T_J\right]=N_{IJi}X^i,
\end{eqnarray}
with all other commutators vanishing. We have defined
\begin{equation}
N_{IJi}=-N_{JIi}=L_{IK}N_{Ji}{}^K.
\end{equation}
The gauge generators can be combined into a $O(6,22)$ vector $T_M$ as
\begin{eqnarray}\label{generator decomposition 2}
T_M=\left(%
\begin{array}{c}
  Z_i \\
  X^i \\
  T_I \\
\end{array}%
\right),
\end{eqnarray}
the algebra (\ref{O(d,d+16) Lie algebra}) may be written as $[T_M,T_N]=t_{MN}{}^PT_P$ where the structure constants of the gauge group are
\begin{equation}
t_{iI}{}^J=N_{iI}{}^J,   \qquad  N_{IJi}=L_{JK}f_{iI}{}^K,
\end{equation}
and all other structure constants are zero. The derivation of this gauge algebra is given in Appendix E.2.

\section{Lifting to String Theory}
In this section we elucidate the structure of the internal space given by the duality-twist construction described above. Our main goal will be
to understand the lift of the four-dimensional gauged supergravity to a ten-dimensional string theory background. We shall see that, whilst many
of the duality-twists give supergravities which can be lifted to a compactification of string theory on a conventional manifold, the generic
situation cannot be understood in this way. Instead, we can understand the lift of the supergravity as a string theory on a non-geometric
background. Such backgrounds do not have a realization as a compactification of ten-dimensional supergravity, but nonetheless are good candidates for consistent
string theory backgrounds.

The string theory sigma model describes the embedding of the two-dimensional worldsheet into the ten-dimensional target spaces described above where the $O(4,20;\Z)\subset O(4,20)$ acts as a perturbative duality symmetry of the string theory. Therefore, in order for the four-dimensional supergravity to have a physical ten-dimensional interpretation, in which the string theory is globally defined, the monodromies around the $T^2$ cycles must take values in the discrete $O(4,20;\Z)$ subgroup of $O(4,20)$. As we shall see, this restricts the possible duality-twist backgrounds considerably.

\subsection{Duality-Twist Compactifications of Heterotic String Theory}

The duality-twist reduction leads to a natural interpretation of the four-dimensional supergravities as arising from a compactification of
Heterotic theory on a six-dimensional background. In particular, we can think of the six-dimensional space as a $T^4$ fibration with
monodromies along the two cycles of the $T^2$ base. For there to be a well-defined string theory background, the monodromies $e^{\alpha}$ and $e^{\beta}$ are
required to commute and take values in the discrete T-duality group $O(4,20;\mathbb{Z})$.

Not all of these backgrounds will be geometries in the conventional sense and there are many examples of duality-twist reductions lifting to non-geometric compactifications of string theory \cite{Hull ``A geometry for non-geometric string backgrounds'',Dabholkar:2002sy,Dabholkar:2005ve}. As a warm-up, let us recall the example of a compactification of bosonic string theory on $T^d$.  The resulting theory will have an $O(d,d)$ rigid symmetry which can be used to perform a duality-twist reduction, over a circle $S^1_x$ with coordinate $x \sim x+1$, of the kind described in the previous section and in \cite{Dabholkar:2002sy,Dabholkar:2005ve,Hull:2007jy}. The monodromy $e^N$ takes values in $O(d,d;\mathbb{Z})\subset O(d,d)$, the discrete T-duality group of the string theory in the fibres. It is useful to decompose the $2d\times 2d$ twist matrix $N_I{}^J$ into $SL(d;\Z)$ blocks  \cite{Hull:2007jy}
\begin{eqnarray}
N_I{}^J=\left(%
\begin{array}{cc}
  f_m{}^n & K_{mn}  \\
  Q^{mn} & -f_n{}^m \\
  \end{array},%
\right)
\end{eqnarray}
where $\exp(f_m{}^n)\in SL(d;\Z)$. The resulting gauge algebra is \cite{Hull:2007jy}
\begin{eqnarray}
[Z_x,Z_m]=f_m{}^nZ_n+K_{mn}X^n,   \qquad [Z_m,Z_n]=K_{mn}X^x,   \qquad  [X^m,Z_n]=f_m{}^nX^x, \nonumber
\end{eqnarray}
\begin{eqnarray}
[X^m,X^n]=Q^{mn}X^x, \qquad  [Z_x,X^m]=f_n{}^mX^n.
 \nonumber
\end{eqnarray}
where $Z_x$ and $Z_m$ generate diffeomorphisms of the $S^1_x$ and $T^d$ cycles respectively and $X^x$ and $X^m$ generate antisymmetric tensor transformations of the $B$-field. As discussed in \cite{Hull:2007jy}, a compactification in which $Q^{mn}=0$, can be understood as a reduction on a conventional torus bundle, in which the fibres are patched together by a large diffeomorphism, and $K_{mn}$ gives a non-trivial flux $H=\frac{1}{2}K_{mn}dx\wedge dz^m\wedge dz^n$ for the $H$-field strength. Compactifications in which $Q^{mn}\neq 0$ have monodromies for which the theory in the $T^4$ fibres must be patched together by a more general element of $O(d,d;\mathbb{Z})$ involving T-dualities. Such backgrounds, called T-folds, are conventional geometries in a contractible patch (or, for example, on the cover $S^1_x\rightarrow \R_x$), but will not be a conventional geometry globally. There is evidence that such backgrounds, with only one type of structure constant turned on, are related by T-duality \cite{Kachru:2002sk,Shelton:2005cf,Hull:2007jy,Hull:2006qs}, a relationship which may be summarized diagrammatically as
\begin{equation}\label{sequence1}
K_{mn}\rightarrow f_m{}^n \rightarrow Q^{mn},
\end{equation}
where the arrows denote a T-duality along a $z^m$-direction of the torus fibre.

We now return to the Heterotic theory and consider an open contractible patch $U_{\alpha}$ on the base $T^2$. The sigma model, describing the embedding of the Heterotic worldsheet into the region $U_{\alpha} \times T^4$, has the discrete T-duality symmetry $O(4,20;\mathbb{Z})$. This sigma model determines the local physics of the string theory, a global description requires one to specify the transition functions between coordinates in different patches, $U_{\alpha}$ and $U_{\beta}$ say, on the overlap $U_{\alpha} \cap U_{\beta}$. From the duality twist construction we know that the transition functions ${\cal O}_{\alpha\beta}$ will be elements
of $O(4,20;\mathbb{Z})$ and so in order to understand the background we need to understand the action of $O(4,20;\mathbb{Z})$ on the conformal field theory in the fibres. The monodromy around a cycle with coordinate $y^i\sim y^i+\xi^i$ is given by $\exp (\xi^iN_{iI}{}^J)$. Following the description of the Bosonic theory above, we decompose the twist matrix as
\begin{eqnarray}\label{twist}
N_{iI}{}^J=\left(%
\begin{array}{ccc}
  f_{im}{}^n & K_{imn} & M_{im}{}^b \\
  Q_i{}^{mn} & -f_{in}{}^m & W_i{}^{bm} \\
  -W_{ia}{}^n & -M_{ina} & S_{ia}{}^b \\
\end{array}%
\right),
\end{eqnarray}
where $M_{ina}=\delta_{ab}M_{in}^b$, $K_{mni}=-K_{nmi}$, $Q_i{}^{mn}=-Q_i{}^{nm}$ and $W_{ia}{}^{m}=\delta_{ab}W_i{}^{bm}$.
This is the most general form for which $N_{iIJ}=N_{iI}{}^KL_{KJ}=-N_{iJI}$, where
\begin{eqnarray}
N_{iIJ}=\left(%
\begin{array}{ccc}
  K_{imn} & f_{im}{}^n & M_{im}{}^a \\
  -f_{in}{}^m & Q_i{}^{mn} & W_i{}^{am} \\
  -M_{inb} & -W_{ib}{}^{n} & S_i{}_{ab} \\
\end{array}%
\right),
\end{eqnarray}
where $S_{iab}=\delta_{ac}S_{ib}{}^c=-S_{iba}$. There are further quadratic constraints on the elements of (\ref{twist}) coming from the Jacobi identity $N_{[i|I}{}^KN_{j]K}{}^J=0$. Decomposing the generators $T_M$ using (\ref{generator decomposition 2}) and (\ref{generator
decomposition}), the four-dimensional gauge algebra (\ref{O(d,d+16) Lie algebra}) can then be written as
\begin{eqnarray}
[Z_i,Z_m]=f_{im}{}^nZ_n+M_{im}{}^aY_a+K_{imn}X^n,   \qquad [Z_m,Z_n]=K_{imn}X^i,   \qquad  [X^m,Z_n]=f_{im}{}^nX^i, \nonumber
\end{eqnarray}
\begin{eqnarray}
[X^m,X^n]=Q_i{}^{mn}X^i, \qquad  [Z_i,X^m]=f_{in}{}^mX^n+W_i{}^{ma}Y_a+Q_i{}^{mn}Z_n,
 \nonumber
\end{eqnarray}
\begin{eqnarray}
[Z_i,Y_a]=-\delta_{ab}W_i{}^{mb}Z_m+M_{ima}X^m+S_{ia}{}^bY_b,    \qquad    [Z_m,Y_a]=M_{ima}X^i, \qquad [Y_a,Y_b]=S_{iab}X^i. \nonumber
\end{eqnarray}
As mentioned in the previous section, the generators $Z_i$ and $Z_m$ are related to the diffeomorphisms $y^i\rightarrow y^i+\omega^i$ and $z^m\rightarrow z^m+\omega^m$. The $X^i$ and $X^m$ are related to antisymmetric tensor transformations of the $B$-field with one leg in the $y^i$ and $z^m$ directions respectively. The $Y_a$ generate the $U(1)^{16}$ internal gauge symmetry. Our task now is to understand what it means physically to `switch on' the structure constants (or `fluxes') in the above gauge algebra.

Upon circumnavigating the cycles of the base $T^2$ the effect of the twist matrix ${\cal O}_I{}^J=\exp (N_{iI}{}^Jy^i)$ on the fields on the $T^4$ is $
\widehat{{\cal M}}= {\cal O}^T{\cal M}{\cal O} $ where $\widehat{\cal M}_{IJ}$ is given by (\ref{Heterotic M}). To understand the effect of the
duality twist on the $T^4$ fields, it is considerably easier to work in terms of the vielbein ${\cal V}:O(4,20)\rightarrow O(4)\times
O(20)$ given by $\widehat{{\cal M}}=\widehat{{\cal V}}^T\widehat{{\cal V}}$ where $\widehat{{\cal M}}$ is given by (\ref{Heterotic M}) and
\begin{eqnarray}\label{vielbein}
\widehat{{\cal V}}^{\Lambda}{}_I=\left(%
\begin{array}{ccc}
  \widehat{e}^m{}_{\alpha} & -\widehat{e}^m{}_{\alpha}\widehat{C}_{mn} & -\widehat{e}^m{}_{\alpha}\widehat{A}^a_m \\
  0 & \widehat{e}_m{}^{\alpha} & 0 \\
  0 & \widehat{A}_m^a & \delta_a{}^b \\
\end{array}%
\right)
\end{eqnarray}
where $e:GL(4)\rightarrow O(4)$ is the vielbein for the $T^4$ with metric
$\widehat{g}_{mn}=\delta_{\alpha\beta}\widehat{e}^{\alpha}{}_m\widehat{e}^{\beta}{}_n$ and $\widehat{C}_{mn}$ denotes
\begin{equation}
\widehat{C}_{mn}=\widehat{B}_{(0)mn}+\frac{1}{2}\delta_{ab}\widehat{A}^a_{(0)m}\widehat{A}^b_{(0)n}.
\end{equation}
The duality twist reduction ansatz is then simply $\widehat{{\cal V}}(y)={\cal V}\cdot{\cal O}(y)$.

\subsubsection{Geometric Flux $f_{im}{}^n$}

Consider first the case where the only non-zero structure constant in \ref{twist} is $N_{im}{}^n=f_{im}{}^n$. The duality-twist matrix is
\begin{eqnarray}
{\cal O}_I{}^J(y,f)=\left(%
\begin{array}{ccc}
  \exp(y^if_{im}{}^n) & 0 & 0 \\
  0 & \exp(-y^if_{in}{}^m) & 0 \\
  0 & 0 & \delta_{ab} \\
\end{array}%
\right),
\end{eqnarray}
and it is equivalent to multiplying, or twisting, every $z^m$ fibre index  in the reduction ansatz by $\exp(y^if_{im}{}^n)$. For example, the
metric of the $T^4$ becomes
$$
\widehat{g}_{mn}(y)=(e^{y^if_i})_{m}{}^pg_{pq}(e^{y^jf_j})^{q}{}_n,
$$
with line element
$$
ds^2=\widehat{g}_{mn}(y)dz^m\otimes dz^n
$$
on the $T^4$ fibres. Equivalently, this twist may be implemented by twisting the harmonic forms on the $T^4$ fibres
$$
dz^m\rightarrow \sigma^m=(e^{-y^if_i})^{m}{}_ndz^n,  \qquad  dy^i\rightarrow \sigma^i=dy^i,
$$
so that they are no longer harmonic but satisfy the Maurer-Cartan equations
$$
d\sigma^m+f_{in}{}^m \sigma^i\wedge\sigma^n=0,  \qquad  d\sigma^i=0.
$$
The reduction ansatz for the fields is then given by replacing $(dz^m,dy^i)$ in the standard Kaluza-Klein reduction with $(\sigma^m,\sigma^i)$. For example the reduction ansatz for the metric becomes
$$
ds^2=g_{mn}\sigma^m\otimes\sigma^n.
$$
The one-forms $(\sigma^i,\sigma^m)$ are the natural left-invariant one forms on the group manifold $G$ with algebra generated by the vector fields $K_m=(e^{y^if_i})_{m}{}^n\partial_n$ and
$K_i=\partial_i$, dual to the forms $(\sigma^i,\sigma^m)$, which satisfy the commutator relations
$$
[K_i,K_m]=f_{im}{}^nK_n, \qquad  [K_m,K_n]=0.
$$

The reduction can then be thought of as a standard compactification on the six-dimensional group manifold $G$. In fact, one need only require
that the manifold is locally of the form $G$. Globally the internal manifold can be the twisted torus $\G\backslash G$ where $\G\subset G$ is a discrete
subgroup of $G$, acting from the left, such that $\G\backslash G$ is compact \cite{Hull:2005hk}. Such discrete groups, which ensure that $\G\backslash G$ is compact, are called
cocompact groups. Examples of flux such compactifications on twisted tori were studied at length in \cite{Hull:2006tp,ReidEdwards:2006vu,KM,Hull:2005hk}. Generally the group $G$ will not be compact and so a
duality-twist reduction with such a monodromy can only be realized as a compactification when such a cocompact subgroup can be found.

We see then that such a ${\cal N}=4$ gauged supergravity lifts to a compactification of Heterotic string theory on a six-dimensional twisted torus $\G\backslash G$ which may be thought of as a topologically twisted $T^4$ bundle over $T^2$.

\subsubsection{Compactifications with $H$-flux}

We now consider the case where the only non-zero structure constant in (\ref{twist}) is $N_{imn}=K_{imn}$. For this choice of structure constant, the calculation of the twist matrix is simplified
greatly by the fact that $N_{iI}{}^JN_{jJ}{}^K=0$, so that ${\cal O}_I{}^J(y,K)=\delta_I{}^J+y^iN_{iI}{}^J$, or
$$
{\cal O}_{I}{}^J(y,K)=\left(%
\begin{array}{ccc}
  \delta^m{}_n & y^iK_{imn} & 0 \\
  0 & \delta_m{}^n & 0 \\
  0 & 0 & \delta_{ab} \\
\end{array}%
\right).
$$
The duality twist, with this parameter, reproduces the result of the following reduction ansatz for the vielbein $\widehat{\cal V}(y)={\cal V}{\cal O}(y)$:
\begin{eqnarray}
\widehat{{\cal V}}^{\Lambda}{}_I(y)=\left(%
\begin{array}{ccc}
  e^m{}_{\alpha} & -e^m{}_{\alpha}\left(C_{mn}-K_{mni}y^i\right) & -e^m{}_{\alpha}A^a_m \\
  0 & e_m{}^{\alpha} & 0 \\
  0 & A_m^a & \delta_a{}^b \\
\end{array}%
\right),
\end{eqnarray}
which may be summarized as the reduction ansatz
$$
\widehat{g}_{mn}(x,y)= g_{mn}(x),   \qquad  \widehat{B}_{(0)mn}(x,y)= B_{(0)mn}(x)-K_{mni}y^i, \qquad  \widehat{A}^a_m(x,y)= \widehat{A}^a_m(x).
$$
thus, the effect of this duality-twist reduction is equivalent to introducing a constant $H$-flux
$$
\widehat{H}=H+\frac{1}{2}K_{mni}dz^m\wedge dz^n\wedge dy^i.
$$
and the gauged supergravity lifts to a Kaluza-Klein compactification on $T^6$ with constant $H$-flux on three cycles of the six-torus.

\subsubsection{Compactification with $F$-Flux}

A slightly less trivial example is the monodromy corresponding to $N_{im}{}^a=M_{im}{}^a$. It is not hard to show that the twist matrix in this case is given by ${\cal
O}_I{}^J(y,M)=\delta_I{}^J+y^iN_{iI}{}^J+\frac{1}{2}y^iy^jN_{iI}{}^K N_{jK}{}^J$ so that
$$
{\cal O}(y,M)=\left(%
\begin{array}{ccc}
  \delta^m{}_n & -\frac{1}{2}y^iy^jM_{im}{}^aM_{jna} & y^iM_{im}{}^a \\
  0 & \delta_m{}^n & 0 \\
  0 & -y^iM_{imb} & \delta^a{}_b \\
\end{array}%
\right).
$$
The twist matrix is of the same form as the vielbein ${\cal V}$, so that the monodromy will preserve the form of ${\cal V}$ and not mix the
components. This is a clue that the monodromy has a geometric action on the internal fields. The vielbein reduction
ansatz
\begin{eqnarray}
\widehat{{\cal V}}^{\Lambda}{}_I(y)=\left(%
\begin{array}{ccc}
  e^m{}_{\alpha} & -e^m{}_{\alpha}\left(B_{mn}+\frac{1}{2}\delta_{ab}(A_{m}{}^a-M_{mi}{}^ay^i)(A_{n}{}^b-M_{ni}{}^by^i)\right) & -e^m{}_{\alpha}\left(A_{m}{}^a-M_{mi}{}^ay^i\right) \\
  0 & e_m{}^{\alpha} & 0 \\
  0 & A_{m}{}^a-M_{mi}{}^ay^i & \delta_a{}^b \\
\end{array}%
\right)\nonumber
\end{eqnarray}
reproduces the results of this monodromy, and this duality-twist ansatz is equivalent to the reduction ansatz
$$
\widehat{g}_{mn}(x,y)= g_{mn}(x),  \qquad  \widehat{B}_{(0)mn}(x,y)= B_{(0)mn}(x), \qquad  \widehat{A}^a_m(x,y)= A^a_m(x)-M_{mni}y^i.
$$
This duality twist reduction therefore can be thought of as adding a constant flux to the $U(1)^{16}$ gauge bosons
$$
\widehat{F}^a_{(2)}=F^a_{(2)}-M_{im}{}^ady^i\wedge dz^m.
$$
so that the half-maximal gauged supergravity with structure constants $M_{im}{}^a$ can be lifted to a compactification of Heterotic string theory on $T^6$ with constant flux on the $F$-field strengths wrapping $T^2$ cycles inside the $T^6$. Compactifications on twisted tori with such fluxes on the $F$-field and also on the $H$-field as discussed above were studied in detail in
\cite{KM} for more general geometric backgrounds than the torus fibrations considered here.

\subsubsection{Non-geometric Flux $Q_i{}^{mn}$}

Let us now consider the duality-twist arising from setting all structure constants in (\ref{twist}) to zero, with the exception of
$N_i{}^{mn}=Q_i{}^{mn}=(\alpha^{mn},\beta^{mn})$. The only requirement is that the monodromies around the cycles of the $T^2$ take values in the T-duality group $O(4,4;\mathbb{Z})\subset O(4,20;\Z)$. Such a duality-twist reduction of the supergravity lifts to string theory compactified on a T-fold background of the kind discussed in \cite{Hull ``A geometry for non-geometric string backgrounds'',Hull ``Doubled geometry and
T-folds'',Hull:2007jy}. Let us consider this in more detail. The twist matrix is
\begin{eqnarray}
{\cal O}_{I}{}^J(y,Q)=\left(%
\begin{array}{ccc}
  \delta^m{}_n & 0 & 0 \\
  y^iQ_i{}^{mn} & \delta_m{}^n & 0 \\
  0 & 0 & \delta^a{}_b \\
\end{array}%
\right).
\end{eqnarray}
The twist matrix is not of the same form as the vielbein and so will not generally preserve the upper triangular form of the vielbein
(\ref{vielbein}). This is an indication that the background corresponding to this gauging is non-geometric.

$L_{IJ}$ is the invariant of $O(4,20)$ given in (\ref{L het}) and acts as a T-duality along
all directions in the $T^4$ \cite{Giveon:1994fu}. The easiest way to see that this
reduction is equivalent to a $T^4$ fibration over $T^2$, the monodromies of which includes a T-dualities, is to note that this twist matrix can be
realized as the $H$-flux twist matrix ${\cal O}(K)$, conjugated by the action of T-duality along all of the $T^4$ fibre coordinates
\begin{equation}\label{Q}
{\cal O}^I{}_J(Q)=L^{I}{}_{K}\,{\cal O}^K{}_L(K)\,L^{L}{}_{J},
\end{equation}
where $Q_i{}^{mn}=\delta^{mp}\delta^{nq}K_{ipq}$. Let the periodicities of the $\alpha$ and $\beta$ cycle coordinates on the $T^2$ be $y^i\sim y^i+\xi^i$, then the
monodromy in $\mathcal{M}_{IJ}$, as we circumnavigate the cycles of the $T^2$, is given by $\mathcal{M}_{IJ}\sim {\cal
O}_{I}{}^K(Q)\mathcal{M}_{KL}{\cal O}^{L}{}_J(Q)|_{y^i=\xi^i}$, where the twist matrices are evaluated at $y^i=\xi^i$. Locally, over a simply
connected region of the base $T^2$, the theory in the fibres is described by a free Heterotic CFT on $T^4$. As we circumnavigate the base
the theory is identified by this monodromy, which is an action of $O(4,20;\mathbb{Z})$ which includes a T-duality. Since the action of T-duality is a
symmetry of the $T^4$ CFT the background is smooth from the point of view of the string theory, even though it is not a conventional geometry and does not admit a ten-dimensional supergravity description.

\subsubsection{Non-geometric Flux $W_i{}^{ma}$}

The flux $W_i{}^{ma}$ plays a similar role to the flux $Q_i{}^{mn}$. It is not hard to show that
\begin{eqnarray}
{\cal O}(y,W)=\left(%
\begin{array}{ccc}
  \delta^m{}_n & 0 & 0 \\
  -\frac{1}{2}y^iy^jW_i{}^{ma}W_{ja}{}^n & \delta_m{}^n & y^iW_i{}^{ma} \\
  -y^iW_{ia}{}^n & 0 & \delta^a{}_b \\
\end{array}%
\right),
\end{eqnarray}
which is equivalent to the $F$-flux twist matrix ${\cal O}(M)$ considered above, conjugated
by a T-duality
\begin{equation}
{\cal O}(y,W)_I{}^J=L^{JK}{\cal O}_K{}^L(M)L_{IL}.
\end{equation}
The background is a T-fold in the spirit of the previous example, in that the monodromy includes a T-duality and can be thought of as a
conventional flux compactification, seen through the distorting lens of a series of T-dualities along all fibre coordinates. However, this Heterotic T-fold background has no analogue in the Type II or bosonic string theories due to the
important role the internal gauge fields $A^a$ play which are only present in the Heterotic and Type I string theories. It is
interesting to note that the quadratic part of the twist, $-\frac{1}{2}y^iy^jW_i{}^{ma}W_{ja}{}^n$, plays a similar role to
the non-geometric flux $y^iQ_i{}^{mn}$ in a conventional T-fold, except there is a symmetry as opposed to an antisymmetry of the un-contracted
indices.

\subsubsection{Topological Twisting of the Cartan Torus}
If we treat the Cartan torus of the $E_8\times E_8$ or $Spin(32)/\Z_2$ as a bona-fide geometry, then $S_{ia}{}^b$ can be thought of as a geometric flux which describes a
non-trivial fibration of the Cartan torus over the base $T^2$
\begin{eqnarray}
\mathcal{O}_{I}{}^J(y,S)=\left(%
\begin{array}{ccc}
  \delta_m{}^n & 0 & 0 \\
  0 & \delta^m{}_n & 0 \\
  0 & 0 & \exp(S_{ia}{}^by^i) \\
\end{array}%
\right).
\end{eqnarray}
We can then think of this gauged supergravity, where all structure constants except $S_{ia}{}^b$ vanish, as arising from a standard geometric compactification on $T^6$ in which the $U(1)^{16}$ fibration over the internal $T^2$ is not trivial but has
a topological twist with monodromy $\exp(S_{ia}{}^b\xi^i)$ as we circumnavigate the cycles of the $T^2$. The monodromy must take values in
$O(16;\mathbb{Z})$ in order for the gauge bundle to be smooth.

It is worth noting that this interpretation of the different components has only limited viability. While we have given here an interpretation of every single component, this cannot be used to interpret every combination of components. Some combinations, such as a combination of a geometric flux $f$- and an $H$-flux, can be interpreted in this way (as an $H$-flux compactification on a twisted torus), but others, for example a combination of $H$- and $Q$-flux may not have so straightforward an interpretation.

\subsection{Duality-Twist Compactifications of Type-IIA String Theory}

We now consider lifting the gauged supergravities (\ref{O(d,d) Lagrangian}) to IIA string theory. Although the procedure we shall follow, of analyzing the effect of specific classes of monodromies on the fields in the fibre, is similar, there are important qualitative differences to the Heterotic case discussed above. For the Heterotic case, the metric moduli of the $T^4$ fibres are in one-to-one correspondence with the components of the $T^4$ metric. We may therefore consider the monodromy to act directly on the metric of the $T^4$. Indeed, in the previous section, we saw that it was possible to show how the $SL(4;\Z)$ monodromy acts on the coordinates of the $T^4$ fibre. The fact that we can describe how the coordinates $z^m$ transform under the monodromy will allow us to construct a worldsheet description of these backgrounds in section seven. By contrast, the moduli of the $K3$ are not in one-to-one correspondence with the components of the $K3$ metric and deducing the direct action of the monodromy on the coordinates of the $K3$ fibre goes beyond our present considerations. Instead, the monodromy acts directly on the two-forms $\Omega^A$ and $b_A$ and the volume modulus $\rho$. This is enough to give an accurate description of how the $K3$ is fibred over the $T^2$ base.

As before, the possible mass matrices $N_{iI}{}^J$ take values in the Lie algebra of $SO(4,20)$ such that the monodromies around the $T^2$ cycles are in the U-duality group $O(4,20;\mathbb{Z})\subset O(4,20)$. The $K3$
fibres have mapping class group (the group of large diffeomorphisms) $O(3,19;\mathbb{Z})$. The generators of $O(4,20;\mathbb{Z})$ can be decomposed in terms of the generators of $O(3,19;\mathbb{Z})$. Similar to the Heterotic decomposition of $N_{iI}{}^J$ according to $SL(4;\Z)$ discussed above, it can be shown that a general twist element taking values in the Lie algebra of the continuous group $SO(4,20)$, can be written as
\begin{eqnarray}
N_{iI}{}^J=\left(%
\begin{array}{ccc}
  -\Lambda_i & 0 & \mathcal{K}_{i}{}^B \\
  0 & \Lambda_i & \mathcal{Q}_i{}^{B} \\
  -\mathcal{Q}_{iA} & -\mathcal{K}_{iA} & \mathcal{D}_{iA}{}^B \\
\end{array}%
\right), \qquad  N_{iIJ}=\left(%
\begin{array}{ccc}
  0 & -\Lambda_i & \mathcal{K}_{iA} \\
  \Lambda_i & 0 & \mathcal{Q}_{iA} \\
  -\mathcal{K}_{iA} & -\mathcal{Q}_{iA} & \mathcal{D}_{iAB} \\
\end{array}%
\right).
\end{eqnarray}
As we will now show, the requirement that the monodromy is in $O(4,20;\mathbb{Z})$ restricts the form, or \emph{texture}\footnote{The term \emph{texture} is often used to describe the qualitative features of quark and neutrino mass matrices. The term seems well-suited to also describe the qualitative features of the mass and twist matrices here.}, of the twist matrix further. Let us consider first the monodromy with mass matrix $N_{i}=\Lambda_i$. The twist matrix is
\begin{eqnarray}
{\cal O}_{I}{}^J(y,\Lambda)=\left(%
\begin{array}{ccc}
  e^{-\Lambda_iy^i} & 0 & 0 \\
  0 & e^{\Lambda_iy^i} & 0 \\
  0 & 0 & \delta^A{}_B \\
\end{array}%
\right)
\end{eqnarray}
which, upon circumnavigating the cycles of the $T^2$, simply has the effect of re-scaling the volume of the $K3$ fibre and corresponds to the
reduction ansatz
\begin{equation}
\widetilde{H}^A{}_B(x,y)= H^A{}_B (x), \qquad  \widetilde{\rho}(x,y)= \rho(x)+2\Lambda_iy^i,  \qquad  \widetilde{b}_A(x,y)= b_A(x).
\end{equation}
A priori, one might expect this to provide a reasonable reduction ansatz, however the constraint that the monodromy is in $O(4,20;\mathbb{Z})$ means
that $e^{\Lambda}$ and $e^{-\Lambda}$ must be integral. This is only the case for $\Lambda=0$, and so, only in the trivial case, can such a reduction ansatz be realized as a compactification of string theory.

Thus, the most general mass matrix $N_{iI}{}^J$, such that $\exp(N_{iI}{}^J\xi^i) \in O(4,20;\mathbb{Z})$ is then
\begin{eqnarray}
N_{iI}{}^J=\left(%
\begin{array}{ccc}
  0 & 0 & \mathcal{K}_{i}{}^B \\
  0 & 0 & \mathcal{Q}_i{}^{B} \\
  -\mathcal{Q}_{iA} & -\mathcal{K}_{iA} & \mathcal{D}_{iA}{}^B \\
\end{array}%
\right),
\end{eqnarray}
where $\mathcal{K}_{iA}\xi^i=\eta_{AB}\mathcal{K}_i{}^B\xi^i\in\Z_{3,19}$, $\mathcal{Q}_{iA}\xi^i=\eta_{AB}\mathcal{Q}_i{}^B\xi^i\in\Z_{3,19}$ and
$e^{\mathcal{D}_{iA}{}^B\xi^i}\in O(3,19;\mathbb{Z})$ and we have taken the identifications of the $T^2$ coordinates to be $y^i\sim y^i+\xi^i$. The gauge algebra of the general gauged supergravity, arising from a reduction of this kind, is then
\begin{eqnarray}
[Z_i,J]=\mathcal{K}_i{}^AT_A,  \qquad    [Z_i,\tilde{J}]=\mathcal{Q}_{i}{}^AT_A,  \qquad
[Z_i,T_A]=\mathcal{D}_{iA}{}^BT_B-\mathcal{K}_{iA}\tilde{J}-\mathcal{Q}_{iA}J,\nonumber
\end{eqnarray}
\begin{eqnarray}
 \qquad [T_A,T_B]=\mathcal{D}_{ABi}X^i,    \qquad  [J,T_A]=\mathcal{K}_{iA}X^i, \qquad  [\tilde{J},T_A]=\mathcal{Q}_{iA}X^i,\nonumber
\end{eqnarray}
where the physical significance of the generators was discussed in sections two and three. Note that the requirement that the monodromy be an element of $O(4,20;\mathbb{Z})$
means that the commutator $[J,\tilde{J}]=\Lambda_iX^i=0$, a result which would not have been possible from a direct consideration of the four-dimensional Lagrangian.

As for the Heterotic case, it is useful to define a vielbein ${\cal V}:O(4,20)\rightarrow O(4)\times O(20)$ such that ${\cal M}={\cal V}\cdot {\cal V}^T$ where ${\cal M}$ is given by (\ref{IIAK3Scalars}). The $y$-dependent array $\widetilde{\cal M}$ is given in terms of the twisted vielbein $\widetilde{\cal V}(y)={\cal O}(y)\cdot{\cal V}$. The vielbein can be given by
\begin{eqnarray}\label{IIA vielbein}
{\cal V}^{\Lambda}{}_I=\left(%
\begin{array}{ccc}
  e^{-\frac{1}{2}\rho} & e^{\frac{1}{2}\rho}C & -b_A\nu^A{}_a \\
 0 & e^{\frac{1}{2}\rho} & 0 \\
  0 & -e^{-\frac{1}{2}\rho}b_A & \nu_{Aa} \\
\end{array}%
\right),
\end{eqnarray}
where $\nu:SO(3,19)\rightarrow SO(3)\times SO(19)$ is a vielbein such that $H^{AB}=\nu^A{}_a(\nu^T)^{aB}$ where $\nu_{Aa}=\eta_{AB}\nu^B{}_a$.

\subsubsection{Geometric $O(3,19;\mathbb{Z})$ Twisted Reduction}

The gauged supergravity with structure constants $N_{iA}{}^B=\mathcal{D}_{iA}{}^B$ corresponds to duality twist compactifications with twist matrix
\begin{eqnarray}
{\cal O}_{I}{}^J(y,{\cal D})=\left(%
\begin{array}{ccc}
  1 & 0 & 0 \\
  0 & 1 & 0 \\
  0 & 0 & e^{\mathcal{D}_{iA}{}^By^i} \\
\end{array}%
\right)
\end{eqnarray}
and can be realized as a smooth, geometric, $K3$ fibration over $T^2$ for $\exp(\mathcal{D}_{iA}{}^B)\in O(3,19;\mathbb{Z})$, for $i=1,2$. The action of this monodromy on the $K3$ moduli gives the reduction ansatz
\begin{equation}
\widetilde{H}^A{}_B(x,y)= (e^{\mathcal{D}_iy^i})^A{}_CH^C{}_D(x)(e^{\mathcal{D}_iy^i})^D{}_B,    \qquad  \widetilde{\rho}(x,y)=\rho(x),	 \qquad	 \widetilde{b}_A(x,y)=(e^{\mathcal{D}^T_iy^i})_A{}^Bb_B(x).
\end{equation}
To see what this means, recall the definition of $H^A{}_B$ in terms of the holomorphic two-forms defined on the $K3$ fibres as
$*\Omega^A=H^A{}_B\Omega^B$.  In order for this relation on the four-dimensional $K3$ to hold globally on the six-dimensional $SU(2)$-structure internal space, this equation must be replaced by $*\widetilde{\Omega}^A=\widetilde{H}^A{}_B\widetilde{\Omega}^B$ where the $\widetilde{\Omega}^A=
(e^{\mathcal{D}_iy^i})^A{}_B\Omega^B$. Thus the duality-twist reduction with mass matrix $\mathcal{D}_{iA}{}^B$ is equivalent to an $SU(2)$-structure compactification on a
six-dimensional manifold where the harmonic two-forms $\Omega^A$ of the the $K3$ manifold are replaced by the two-forms $\widetilde{\Omega}^A(y)$ which are
not closed but satisfy
\begin{equation}
d\widetilde{\Omega}^A-\mathcal{D}_{iB}{}^A\widetilde{\Omega}^B\wedge dy^i=0.
\end{equation}

We encountered this $SU(2)$-structure manifold earlier in section three. To see why the structure constants $\mathcal{D}_{iB}{}^A=(\alpha^A{}_B,\beta^A{}_B)$ must take values in the generators of the discrete subgroup $SO(3,19;\mathbb{Z})$, recall that although $K3$, like a Calabi-Yau three-fold, does not have any continuous isometries, it does have discrete isometric symmetries. These large diffeomorphisms preserve the lattice $\mathbb{Z}_{3,19}$ and generate the discrete group $SO(3,19;\mathbb{Z})$. Let the $T^2$ coordinates be given by $y^1\sim y^1+1$ and $y^2\sim y^2+1$. Upon circumnavigating the cycles of the $T^2$ base, the $K3$ comes back to itself up to a the action of $\exp(\alpha^A{}_B)$ along the $y^1$ cycle and $\exp(\beta^A{}_B)$ around the $y^2$ cycle. The background will only be smooth if $\alpha^A{}_B$ and $\beta^A{}_B$ are elements of the mapping class group $SO(3,19;\mathbb{Z})$ of the $K3$ fibres\footnote{This is analogous to the simpler three-dimensional nilmanifold construction in a $T^2$ is fibred over a circle with monodromy in $SL(2;\Z)$ - the mapping class group of the $T^2$ fibre \cite{Hull:2007jy}.}.

\subsubsection{Compactifications with $H$-flux}

\noindent Let us now consider the gauged supergravity with non-zero structure constants $\mathcal{K}_{iA}$ for $i=1,2$. The
twist matrix in this case is
\begin{eqnarray}
{\cal O}_{I}{}^J(\mathcal{K},y)=\left(%
\begin{array}{ccc}
  1 & -\frac{1}{2}\mathcal{K}_{iA}\mathcal{K}_j{}^Ay^iy^j & \mathcal{K}_i{}^Ay^i \\
  0 & 1 & 0 \\
  0 & -\mathcal{K}_{iB}y^i & \delta^A{}_B \\
\end{array}%
\right)
\end{eqnarray}
which is equivalent to the reduction ansatz
\begin{equation}
\widetilde{H}^A{}_B(x,y)= H^A{}_B(x), \qquad \widetilde{\rho}(x,y) = \rho(x), \qquad \widetilde{b}_A(x,y)= b_A(x)+\mathcal{K}_{Ai}y^i,
\end{equation}
and so it is clear that this gauged supergravity arises from a compactification on the trivial bundle $K3\times T^2$ with constant $H$-flux
\begin{equation}
\widetilde{H}= H + \mathcal{K}_{iA}dy^i \wedge\Omega^A.
\end{equation}
The geometric reductions discussed above where ${\cal D}_{iA}{}^B, {\cal K}_{iA}\neq 0$ only make use of the subgroup $O(3,19;\mathbb{Z})\ltimes \mathbb{Z}_{3,19}\subset O(4,20;\mathbb{Z})$ as topological twists in the $K3$ fibration over $T^2$. We shall refer to this as the geometric monodromy subgroup. This is in contrast with the other monodromies in $O(4,20;\mathbb{Z})$ which produce backgrounds which cannot be thought of geometric compactifications in the standard sense. It is to these non-geometric compactifications to which we turn next.

\subsubsection{Mirror-folds and Non-Geometric Backgrounds with $\mathcal{Q}_i{}^A$ Flux}

\noindent The description of these backgrounds as $K3$ fibrations over $T^2$ suggests that there should be a family of $IIA$ reductions constructed as $K3$ fibrations over $T^2$ with
monodromy taking values, not just in the  geometric subgroup $O(3,19;\mathbb{Z})\ltimes\mathbb{Z}_{3,19}\subset O(4,20;\mathbb{Z})$, but in the full duality group $O(4,20;\mathbb{Z})$.
In particular, we are interested in duality-twist reductions where the twist matrix takes the form
\begin{eqnarray}
{\cal O}^I{}_J(y,\mathcal{Q})=\left(%
\begin{array}{ccc}
  1 & 0 & 0 \\
  -\frac{1}{2}y^iy^j\mathcal{Q}_{iA}\mathcal{Q}_j{}^A & 1 & y^i\mathcal{Q}_i{}^A \\
  -y^i\mathcal{Q}_{iB} & 0 & \delta^A{}_B \\
\end{array}%
\right),
\end{eqnarray}
which produces a gauged supergravity with structure constants $Q_i{}^A$. The first thing we notice is that this twist matrix will not preserve the form of the vielbein (\ref{IIA vielbein}), i.e. the zero entries in the matrix ${\cal V}$ will generally not be preserved in the twisted vielbein $\widetilde{\cal V}$. An $O(4)\times O(20)$ transformation can be used to restore $\widetilde{\cal V}$ to the form (\ref{IIA vielbein}) but, as for the Heterotic T-fold example considered in the last section, this mixing of $\rho$, $b_A$ and $H^A{}_B$ in the reduction ansatz, is indicative of a twist giving rise to a non-geometric background. Another similarity between the T-fold monodromy and the case we are considering here is that the twist matrix may be given in terms of the $H$-flux twist matrix ${\cal O}(y,{\cal K})$, conjugated by $L_{IJ}$
$$
{\cal
O}^I{}_J(y,\mathcal{Q})=L_{JK}\,{\cal O}^K{}_L(y,\mathcal{K})\,L^{LI},	\qquad	{\cal Q}_i{}^A={\cal K}_{iB}\eta^{BA},
$$
which is reminiscent of the relationship between ${\cal O}(y,K)$ and ${\cal O}(y,Q)$ in (\ref{Q}). In (\ref{Q}), the action of the $O(4,20)$ invariant $L_{IJ}$ was identified as a T-duality along all cycles of the $T^4$ fibre. Here, although the interpretation is not as clear, we will argue that gauged supergravities with structure constants $N_i{}^A={\cal Q}_i{}^A$ lift to a compactification on a non-geometric background which exhibits many characteristics similar to that of the T-fold.

It was shown in \cite{Seiberg} that the moduli space of ${\cal N}=(4,4)$ conformal field theories, describing embeddings of the worldsheet
into $K3$ manifolds with a $B$-field, is given by
$$
O(4,20;\mathbb{Z})\backslash O(4,20)/(O(4)\times O(20))
$$
where the discrete U-duality group $O(4,20;\mathbb{Z})$ is
generated by the following symmetry transformations \cite{Aspinwall:1996mn,Aspinwall:1994rg}:

\begin{list}{}{}
\item\emph{Mapping Class Group}:
$O(3,19;\mathbb{Z})$ is the mapping class group of a $K3$ fibres. The background given by the duality-twist with mass matrix ${\cal D}_{iA}{}^B$ has been shown above to generate all such fibrations with monodromies in $O(3,19;\Z)$.
\item\emph{Integral $B$-Field Shifts}:
The $H$-flux background, given by the duality-twist reduction with mass matrix ${\cal K}_{iA}$ gives a monodromy to the $B$-field $b_A\sim b_A+{\cal K}_{iA}\xi^i$, where ${\cal K}_{iA}\xi^i$ takes values in the discrete lattice $\mathbb{Z}_{3,19}$. All elements of $O(4,20;\mathbb{Z})$ corresponding to integral $B$-field shifts are therefore accounted for by the mass matrix ${\cal K}_{iA}$.
\item\emph{Mirror Symmetry}:
The mass matrices ${\cal D}_{iA}{}^B$ and ${\cal K}_{iA}$ together account for the geometric subgroup $O(3,19;\mathbb{Z})\ltimes\mathbb{Z}_{3,19}\subset O(4,20;\mathbb{Z})$. In addition, there is a version of Mirror Symmetry for $K3$ manifolds, which gives a $\Z_2$ contribution to the duality group $O(4,20;\mathbb{Z})$. In fact, it has been shown\footnote{See \cite{Aspinwall:1994rg} and references contained therein.} that the geometric symmetries $O(3,19;\mathbb{Z})\ltimes\mathbb{Z}_{3,19}$ and this $\Z_2$ Mirror map generate the full $O(4,20;\mathbb{Z})$ duality group.
\end{list}

The familiar Mirror symmetry relation between Calabi-Yau three-folds involves the exchange of complex and complexified K\"{a}hler structures on mirror pairs of manifolds\footnote{The K\"{a}hler form $J$ is complexified by including the $B$-field to give the complexified K\"{a}hler form $J+ib$.}. For $K3$, one may choose any one of the real two-forms $j$, $Re(\omega)$ and $Im(\omega)$ to be paired with the $B$-field to give a complexified K\"{a}hler structure, the other two give a complex structure. One may think of Mirror Symmetry as exchanging these complex and complexified K\"{a}hler structures, as is the case with Mirror Symmetry for Calabi-Yau three-folds. However, this split of the hyperk\"{a}hler structure into complex and K\"{a}hler structures is not unambiguous and the $SO(3)$ relates different choices of complex structure to each other. A clear definition of Mirror symmetry for ${\cal N}=(4,4)$ string theory on $K3$ manifolds is therefore not as straightforward as that for Calabi-Yau three-folds. In spite of these complications, it is reasonable to expect that the action of the invariant element $L_{IJ}$ is related to this $\Z_2$ mirror symmetry. Inspired by the T-fold construction, we conjecture that this background, with a duality-twist given by the mass matrix $\mathcal{Q}_i{}^A$, can be interpreted as a Mirror-fold.

The idea of a Mirror-fold is not new (see, for example, \cite{Dabholkar:2002sy}) and in \cite{Kawai and Sugawara ``Mirrorfolds with K3 Fibrations''} a CFT for the interpolating orbifold corresponding to a Mirror-fold was explicitly constructed. The Mirror-fold proposed here is a smooth, non-geometric, string theory background, given by a $K3$ fibration over $T^2$ in which the $K3$ fibres are patched together by a transition function which includes a Mirror Symmetry.

There is possibly a much closer relationship between the T-fold constructions of \cite{Hull ``A geometry for non-geometric
string backgrounds''} and the Mirror-fold proposed here in the case where the $K3$ is elliptically fibred. In this case the $K3$ is a $T^2$ fibration over $\C$P$^1\simeq S^2$, where the complex structure of the fibres is a holomorphic function of the base coordinates. Mirror symmetry for the $K3$ can be thought of as T-duality along both cycles of the elliptic $T^2$ fibres \cite{SYZ}. We may consider performing the T-duality fibre-wise, despite the absence of continuous isometries in the $K3$. In \cite{Hull ``A geometry for non-geometric
string backgrounds''}, a three-dimensional T-fold was constructed as a $T^2$ fibration over $S^1$, where the theory in the $T^2$ fibres is patched, upon circumnavigating the base, by a double T-duality along the fibres of the $T^2$. If we now think of this $T^2$ as corresponding to the $T^2$ fibres of an elliptically fibred $K3$, then we can think of this five-dimensional Mirror-fold as a particular example of a three-dimensional T-fold, fibred over $\C$P$^1$ base. The generalization to a six-dimensional Mirror-fold, by including an extra $S^1$ is straightforward.

Similar to the heterotic case, the interpretation of individual components of the mass matrix does not have a straightforward generalization to multiple non-zero components. It is possible to reproduce a dimensional reduction with both ${\cal D}_{iA}{}^B$ and ${\cal K}_{iA}$ non-zero as a reduction of IIA supergravity on a $K3$ bundle with an $H$-flux, but an interpretation of a dimensional reduction with, for example, ${\cal K}_{iA}$ and ${\cal Q}_{iA}$ both switched on is not so readily given.

\subsection{Duality of the Different Backgrounds}

In the previous two subsections we considered the lifting different gauged supergravities to compactifications, possibly with fluxes, of string theory. Of the reductions considered, it was useful to split the mass matrices into different classes and we shall be concerned here with understanding the way in which these different classes of compactifications may be related to each other. In particular, we shall discuss to what extent the action of the group $O(4,20;\Z)$ on the space of all ${\cal N}=4$ gauged supergravities can be thought of as a duality symmetry of the string theory.

The massless ${\cal N}=4$ gauged supergravity has a rigid $SL(2)\times O(6,22)$ symmetry which acts on the bosonic degrees of freedom (the fermions transform under the maximal compact subgroup) where the discrete subgroup $SL(2;\Z)\times O(4,22;\Z)$ is conjectured to lift to a U-duality symmetry of the ten dimensional string theory \cite{Hull and Townsend Unity of superstring dualities}. As shown in \cite{SW}, the gauged theory can be written in an $SL(2)\times O(6,22)$-covariant way.

The action of $SL(2)\times O(6,22)$ does not preserve the gauging and the structure constants of the gauge group transform covariantly under $SL(2)\times O(6,22)$ as
$$
t_{\alpha MN}{}^P\rightarrow U_{\alpha}{}^{\beta} U_M{}^Q U_N{}^T t_{\beta QT}{}^S(U^{-1})_S{}^P	\qquad	\xi_{\alpha M}\rightarrow U_{\alpha}{}^{\beta}U_M{}^N\xi_{\beta N}
$$
where $U_{\alpha}{}^{\beta}\in SL(2)$ and $U_M{}^N\in O(6,22)$. The structure constants play the role of masses and charges in the four dimensional supergravity, data which we usually think of as fixed for a given effective theory. Thus, the action of $SL(2)\times O(6,22)$ maps one gauged supergravity into another, inequivalent gauged supergravity; however, one might conjecture that the action of the discrete subgroup $SL(2;\Z)\times O(6,22;\Z)$ still lifts to a symmetry of the string theory. Another possibility is that only a subgroup of $SL(2;\Z)\times O(6,22;\Z)$ survives as a duality symmetry of the string theory.

The gaugings arising from the $O(4,20;\Z)$ duality-twist reductions considered in this paper give rise to ${\cal N}=4$ gauged supergravities, the gauge groups for which are characterized by the structure constants
$$
t_{MN}{}^P=\delta_M{}^i\delta_N{}^I\delta^P{}_JN_{iI}{}^J
$$
where $M=1,2,..28$, $i=1,2$ and $I,J=5,6,...28$. Here, we shall only consider the action of $O(4,20;\Z)\subset O(6,22;\Z)$ on these structure constants and leave a discussion of the action of the full $O(6,22;\Z)$ for section eight. Under this action of $O(4,20;\Z)$ the structure constants transform as
$$
t_{iI}{}^J\rightarrow  U_I{}^K t_{iK}{}^L(U^{-1})_L{}^J
$$
where $U^I{}_J\in O(4,20;\Z)$ (recall that for the gaugings considered here $\xi_{\alpha M}=0$). This is equivalent to the action on the mass matrix $N_{iI}{}^J\rightarrow  U_I{}^K N_{iK}{}^L(U^{-1})_L{}^J$.

Within $O(4,20;\Z)$ are sets of discrete $\Z_2$ duality symmetries which one might call \emph{strict dualities}. These include the $\Z_2$ T-duality symmetries discovered by Buscher in \cite{Buscher ``A Symmetry of the String Background Field Equations''} or the $K3$ Mirror Symmetries discussed in the previous sub-section. For example, the $O(d,d;\Z)$ discrete symmetry of the Bosonic string theory compactified on a $d$ dimensional torus background may be decomposed into the action of $SL(d;\Z)$ (the mapping class group of $T^d$), discrete (constant) $B$-field shifts\footnote{which preserve the partition function} and a set of $d$ $\Z_2$ strict T-duality symmetries which exchange momentum and winding modes. We consider now how the action of such strict $\Z_2$ dualities in the $O(4,20;\Z)$-action of $U_I{}^J$ are expected to relate the backgrounds considered in the previous sub-sections.

\subsubsection{Heterotic T-Duality}

For the Heterotic string
compactified on $T^4$, the $O(4,20;\Z)$ is a T-duality symmetry of the theory and may be decomposed into the $SL(4;\Z)$ mapping class group of the $T^4$, discrete shifts in the $B$-field and gauge potential $A^a$ and a set of strict $\Z_2$ T-duality symmetries. The action of the strict T-dualities relates different gaugings and is summarized in the following
diagram
\begin{eqnarray}\label{sequence2}
&K_{imn}&\rightarrow f_{im}{}^n\rightarrow Q_i{}^{mn}\nonumber\\
&\downarrow &\qquad \downarrow\nonumber\\
&M_{im}{}^a&\rightarrow W_i{}^{ma}\nonumber\\
&\downarrow &\nonumber\\
&S_{ia}{}^b&
\end{eqnarray}
where the horizontal arrows denote the action of a $\Z_2\subset O(4,4;\Z)\subset O(4,20;\Z)$ strict duality which is common to the Heterotic and Bosonic string theories. The vertical arrows denote the action of those strict dualities that do not have a counterpart in the Bosonic theory, i.e. those T-dualities which directly involve the sixteen gauge fields $A^a$. One may think of this as a Heterotic generalization of the bosonic T-duality sequence (\ref{sequence1}). There is some evidence \cite{Hull ``Doubled geometry and T-folds''} that, for certain choices of structure constants, some of these
backgrounds are indeed T-dual to each other. In particular, it was shown in \cite{Hull:2006qs} that certain $H$-flux backgrounds, characterized by the structure constants $K_{imn}$, are physically equivalent as string backgrounds, to the twisted tori, characterized by the structure constants $f_{im}{}^n$.

\subsubsection{IIA Mirror Symmetry}

The IIA case is much simpler as there is only one strict duality - the $K3$ Mirror Symmetry - which, we conjecture, acts as
$$
{\cal K}_{iA}\rightarrow {\cal Q}_i{}^A
$$
If we treat the Mirror Symmetry as two T-dualities along the fibres of an elliptically fibred $K3$ \cite{SYZ}, then we can see that there is some similarity with the bosonic T-duality sequence (\ref{sequence1}).

\section{Heterotic Worldsheet Theory}

A principal concern of this investigation has been to realize a class of half-maximal gauged supergravities within the framework of Kaluza-Klein
theory and thus give a higher-dimensional interpretation to the these four-dimensional effective theories. The ultimate goal of this programme is to obtain such gauged supergravities, not just from higher-dimensional supergravities, but also as full string theory scenarios. We have seen that many of the half-maximal gauged supergravities do not lift to compactifications of field theory on a smooth manifold, but are possible candidates for non-geometric string theory backgrounds.

As a first step in the study of such backgrounds from the string theory perspective, we present in this section a sigma model description in
which the backgrounds studied in previous sections can be described. Due to the difficulties of an explicit construction of sigma models on $K3$
the discussion will be confined to the Heterotic theory, although we believe a corresponding study of the IIA model in the orbifold limit of the $K3$ fibres is
tractable. It would also be interesting to see if the IIA description can be studied from the perspective of the gauged linear sigma model along the
lines of \cite{Witten ``Phases of N = 2 theories in two dimensions'',Hori ``Mirror symmetry''}. The challenge here would be to correctly apply
the duality-twist in a way that did not affect the proper renormalization group flow of the linear sigma model. We hope to return to these
issues at a later date.

The approach that we will take to study the Heterotic theory is based on the doubled formalism of \cite{Hull ``A geometry for non-geometric string
backgrounds'',Hull ``Doubled geometry and T-folds''}. We double the internal $T^4$ coordinates $z^m$ as in \cite{Hull ``A geometry for non-geometric string backgrounds''} by introducing dual coordinates $\tilde{z}_m$ for a dual torus $\widetilde{T}^4$, conjugate to the winding modes wrapping the cycles of $T^4$. A self-duality
constraint is then imposed to ensure the doubled theory describes the correct numbers of degrees of freedom.

We shall see that the $O(p,q;\mathbb{Z})$-covariance of the heterotic theory, where $p\neq q$, introduces new
issues not found in the $O(p,p;\Z)$-covariant models studied in \cite{Hull ``A geometry for non-geometric string backgrounds'',Hull ``Doubled
geometry and T-folds''}. In particular, we have to consider the doubling of the left-moving scalars taking values in $U(1)^{16}$. These scalars $\chi_L{}^a$ can be thought of as describing embeddings of the left-moving string modes into the Cartan torus, $T^{16}_c$, of $E_8\times E_8$ or $Spin(32)/\Z_2$. A priori, it is
not clear how to treat these chiral fields in a doubled formalism. The solution we propose here is to double these degrees of freedom by introducing right-moving modes $\chi_R{}^a$, also describing embedding into the Cartan torus. The self duality constraint becomes a chirality constraint on the fields $\chi^a$
which ensures that the theory has the correct number of chiral degrees of freedom. The coordinates on the doubled fibres are
$\mathbb{X}^I=\left(z^m,\tilde{z}_m,\chi^a\right)$ where the `internal' bosonic coordinates $\chi^a$ take values in the
Cartan torus of $E_8\times E_8$ or $Spin(32)/\Z_2$. Initially we impose no chirality constraint on the $\chi^a$ so we may think of the
internal coordinates as having been `doubled'. In this way, we may now treat the Cartan torus $T_c^{16}$ of $E_8\times E_8$ or $Spin(32)/\Z_2$ as we would a conventional target space geometry and consider embeddings of the worldsheet $\Sigma$ into this $T^{24}\simeq T^4\times \widetilde{T}^4\times T^{16}_c$.

\subsection{Heterotic Doubled Geometry}

Consider an $O(4,20;\mathbb{Z})$-twisted $T^4$ fibration over $T^2$. We are interested in constructing a doubled formalism which encodes the data of this background, including the gauge fields, geometrically. We denote the $26$-dimensional $T^{24}$ fibration over $T^2$ by ${\cal T}$, where
\begin{eqnarray}
T^{24}\simeq T^4\times \widetilde{T}^4\times T^{16}_c\hookrightarrow &{\cal T}&\nonumber\\
&\downarrow&\nonumber\\
&T^2&
\end{eqnarray}

The coordinates on ${\cal T}$ are given by $(y^i,\mathbb{X}^I)$, where $y^1\sim y^1+1$ and $y^2\sim y^2+1$ are coordinates on the $T^2$ base and $\mathbb{X}^I$ (the index $I$ runs from $1$ to $24$) are coordinates on the doubled fibres. For a bundle with a monodromy taking values in $O(4,20;\mathbb{Z})$, the doubled coordinates are subject to the identifications
$$
\left(y^i\,,\,\mathbb{X}^I\right)\sim \left(y^i+\xi^i\,,\,(e^{-N\cdot\xi})^I{}_J\mathbb{X}^J\right),	\qquad	 \left(y^i\,,\,\mathbb{X}^I\right)\sim \left(y^i\,,\,\mathbb{X}^I+\alpha^I\right),
$$
where $\xi^i=(1,1)$ and the cycles of the doubled torus fibre can be normalized such that all entries in $\alpha^I$ are also unity.

In order to recover a conventional description of the background a polarization $\Pi$ must be chosen - a projection that selects which of the twenty-four $\mathbb{X}^I$ are to be identified as the four spacetime coordinates. This may be written as
\begin{equation*}
z^m=\Pi^m{}_I\mathbb{X}^I.
\end{equation*}
It is generally not possible to define the polarization globally. There are two cases in particular in which a global description of the spacetime cannot be given. The first, where we consider the only non-zero entry in the twist matrix to be $N_i{}^{mn}=Q_i{}^{mn}$, then the coordinates on ${\cal T}$ are subject to the identifications
$$
\left(y^i\,,\,z^m\,,\,\tilde{z}_m\,,\,\chi^a\right)\sim \left(y^i+\xi^i\,,\,z^m+\x^iQ_i{}^{mn}\tilde{z}_n\,,\,\tilde{z}_m\,,\,\chi^a\right),
$$
and we see that the monodromy mixes the $z^m$ and $\tilde{z}_m$ coordinates together and so the polarization is not well-defined under this monodromy. This phenomenon has been discussed for the bosonic string in \cite{Dabholkar:2002sy,Dall'Agata:2007sr}. A second example, which does not arise for the bosonic string, occurs when the only non-zero entry in the twist matrix is $N_i{}^{am}=W_i{}^{am}$ so that the coordinates on ${\cal T}$ are then subject to the identifications
$$
\left(y^i\,,\,z^m\,,\,\tilde{z}_m\,,\,\chi^a\right)\sim \left(y^i+\xi^i\,,\,z^m-\frac{1}{2}\xi^i\x^jW_i{}^{ma}W_{ja}{}^n\tilde{z}_n-\xi^iW_{ia}{}^m\chi^a\,,\,\tilde{z}_m\,,\,\chi^a+\xi^iW_i{}^{ma}\tilde{z}_m\right),
$$
and we see explicitly that the monodromy mixes the $z^m$, $\tilde{z}_m$ and $\chi^a$ coordinates and therefore does not preserve the polarization.

Once a polarization is chosen, the metric $\widehat{g}_{mn}(y)$, $B$-field $\widehat{B}_{mn}(y)$,  and gauge fields $\widehat{A}_m{}^a(y)$ on the $T^4$ fibres define a $y$-dependent metric $\widehat{\cal M}_{IJ}(y)=(e^{-N\cdot y})_I{}^K{\cal M}_{KL}(e^{-N\cdot y})^L{}_J$ where the $y$-independent ${\cal M}$ can be written, in this polarization, as
\begin{equation}\label{SigmaHetM}
{\cal M}_{IJ}= \left(\begin{array}{ccc}
g_{mn}-C_{mp}g^{pq}C_{qn}+A_{ma}A_n{}^a & C_{mn}g^{np} & A_m{}^a-C_{mn}g^{np}A_p{}^a \\
-g^{mp}C_{pn} & g^{mn} & -g^{mn}A_n{}^a \\
A_m{}^a+A_p{}^ag^{pn}C_{nm} & -A_n{}^ag^{nm} &
\delta_{ab}+A_{ma}g^{mn}A_{nb}
\end{array}\right),
\end{equation}
where $C_{mn}=B_{mn}+\frac{1}{2}A_{ma}A_n{}^a$. The $T^4$ metric, $B$-field and gauge fields, each with one leg on the $T^4$, define a connection of the $T^{24}$ bundle $\widehat{\cal A}^I(y)=(e^{-N\cdot y})^I{}_J{\cal A}^J{}_idy^i$, where
$$
{\cal A}^I{}_i= \left(\begin{array}{ccc}
A^m{}_i \\B_{mi}+B_{mn}A^n{}_i \\ A^a{}_i
\end{array}\right).
$$

\subsection{Sigma Model for the Doubled Torus Fibration}

Following the
strategy pioneered in \cite{Hull ``A geometry for non-geometric string backgrounds''} (see also \cite{Duff:1989tf,Duff:1990hn}), a sigma model, describing the embedding of a worldsheet $\Sigma$ into the doubled target space ${\cal T}$, is given by
\begin{eqnarray}\label{doubled lagrangian}
S[y^i,\mathbb{X}^I]&=&\frac{1}{4}\oint_{\Sigma}\widehat{{\cal M}}(y)_{IJ}d\mathbb{X}^I\wedge *d\mathbb{X}^J+\frac{1}{4}\oint_{\Sigma}\Theta_{IJ}d\mathbb{X}^I\wedge
d\mathbb{X}^J+\frac{1}{2}\oint_{\Sigma}d\mathbb{X}^I\wedge*\widehat{J}_I(y)\nonumber\\
&&+\frac{1}{2}\oint_{\Sigma}G_{ij}dy^i\wedge*dy^j+\frac{1}{2}\oint_{\Sigma}B_{ij}dy^i\wedge dy^j,
\end{eqnarray}
where
$$
\widehat{J}_I(y)=\widehat{{\cal M}}(y)_{IJ}\widehat{\cal A}^J-*L_{IJ}\widehat{\cal A}^J(y),	\qquad	 G_{ij}=g_{ij}+\frac{1}{2}{\cal M}_{IJ}{\cal A}^I{}_i{\cal A}^J{}_j,
$$
and $g_{ij}$ is the metric on the $T^2$ base. $\widehat{\cal A}^I$ now denotes the pull-back of the connection form to the worldsheet, so that $\widehat{\cal A}^I=\widehat{\cal A}^I{}_i\partial_{\alpha}y^id\sigma^{\alpha}$ where $\sigma^{\alpha}=(\tau,\sigma)$ are coordinates on $\Sigma$ and $d=d\sigma^{\alpha}\partial_{\alpha}$ is the worldsheet exterior derivative. We choose to omit contributions to the sigma model coming from embedding $\Sigma$ into the four-dimensional noncompact space. These contributions are important in finding exact solutions based on these fibrations that can be included without difficulty. Their omission here is simply for clarity of exposition. We also define
\begin{equation}\label{theta}
\Theta_{IJ}=\left(%
\begin{array}{ccc}
  0 & \bid_4 & 0 \\
  -\bid_4 & 0 & 0 \\
  0 & 0 & 0 \\
\end{array}%
\right), \qquad  L_{IJ}=\left(%
\begin{array}{ccc}
  0 & \bid_4 & 0 \\
  \bid_4 & 0 & 0 \\
  0 & 0 & \bid_{16} \\
\end{array}%
\right).
\end{equation}

It is useful to define the one forms on ${\cal T}$
$$
P^i=dy^i,	\qquad	{\cal P}^I=\left(e^{N_iy^i}\right)^I{}_Jd\mathbb{X}^J.
$$
These forms are globally defined on ${\cal T}$ and satisfy the worldsheet Bianchi identities
\begin{equation}\label{Bianchi2}
d P^i=0,	\qquad	d{\cal P}^I-N_{iJ}{}^IP^i\wedge {\cal P}^J=0.
\end{equation}
The sigma model may be conveniently written in terms of these one-forms as
\begin{eqnarray}\label{hello}
S[y^i,\mathbb{X}^I]&=&\frac{1}{4}\oint_{\Sigma}{\cal M}_{IJ}{\cal P}^I\wedge *{\cal P}^J+\frac{1}{4}\oint_{\Sigma}\Theta_{IJ}d\mathbb{X}^I\wedge
d\mathbb{X}^J+\frac{1}{2}\oint_{\Sigma}{\cal P}^I\wedge*J_I\nonumber\\
&&+\frac{1}{2}\oint_{\Sigma}G_{ij}dy^i\wedge*dy^j+\frac{1}{2}\oint_{\Sigma}B_{ij}dy^i\wedge dy^j,
\end{eqnarray}
where
$$
J_I={\cal M}_{IJ}{\cal A}^J-*L_{IJ}{\cal A}^J.
$$
The correct number of physical degrees of freedom are ensured by the imposition of the self-duality constraint \cite{Hull ``A geometry for non-geometric string backgrounds''}
\begin{equation}\label{constraint}
{\cal P}^I=L^{IJ}\left({\cal M}_{JK}*{\cal P}^K+*J_J\right).
\end{equation}
This constraint is compatible with the equations of motion of the action (\ref{hello}) and Bianchi identities (\ref{Bianchi2}). In the classical theory, once a polarization is chosen, the self-duality constraint may be used to eliminate the auxiliary degrees of freedom $\tilde{z}_m$ and $\chi_R{}^a$ in the equations of motion. One then finds a conventional description of the system described by a set of equations of motion written in terms of the physical fields $z^m$ and $\chi_L{}^a$. We now turn to the more involved issue of imposing this constraint in the quantum theory.

\subsubsection{Constraining the Quantum Theory}
There is a $U(1)^{24}$ isometry symmetry of the target space $\mathbb{X}^I\rightarrow\mathbb{X}^I+\epsilon^I$ which is manifest in the sigma model as a rigid symmetry of the two-dimensional field theory. For the bosonic string, it was shown in \cite{Hull ``Doubled geometry and T-folds''}, that a self-duality constraint of the form (\ref{constraint}) can be imposed in the quantum theory by gauging a maximally isotropic subgroup of the rigid $U(1)^{24}$ symmetry. The rigid symmetry we wish to gauge is a $U(1)^{12}$ which acts on the coordinates as
$$
\delta z^m=0,	\qquad	\delta\tilde{z}_m=\epsilon_m,	\qquad	\delta \chi^a=\epsilon^a.
$$
This may at first appear to produce a gauging of $U(1)^{20}$; however, we shall require that $d\epsilon^a=*d\epsilon^a$ so that $\epsilon^a$ only describes eight independent parameters and therefore only gauges a $U(1)^8$. Including also the four parameters $\epsilon_m$, there are then twelve independent parameters, corresponding to the gauge group $U(1)^{12}$. The effect of this chirality condition on the gauge parameter will be that only the right-moving part of $\chi^a$ will be gauged. Following \cite{Hull ``Doubled geometry and T-folds''}, the gauging proceeds by minimal coupling in the kinetic term of the sigma model
$$
d\mathbb{X}^I\rightarrow{\cal D}\mathbb{X}^I=d\mathbb{X}^I+C^I,
$$
where the gauge worldsheet one-forms are $C^I=(0,C_m,C^a)$ which transform as
$$
\delta C_m=-d\epsilon_m,	\qquad	\delta C^a=-d\epsilon^a,
$$
and $C^a=*C^a$ is chiral. One must also introduce the Wess-Zumino term
$$
\frac{1}{2}L_{IJ}{\cal P}^I\wedge {\cal C}^J.
$$
where we have introduced the duality-twisted one-forms ${\cal C}^I=\left(e^{N\cdot y}\right)^I{}_JC^J$. We shall choose to set the background one-forms ${\cal A}^I$ to zero. This is done to simplify the exposition and the general case, in which ${\cal A}^I\neq 0$, follows in a straight-forward manner.

The gauged sigma model is
\begin{eqnarray}\label{gauged sigma model}
S[y^i,\mathbb{X}^I,C_m,C^a]&=&\frac{1}{4}\oint_{\Sigma}{\cal M}_{IJ}{\cal P}^I\wedge *{\cal P}^J+\frac{1}{4}\oint_{\Sigma}\Theta_{IJ}d\mathbb{X}^I\wedge
d\mathbb{X}^J+\frac{1}{2}\oint_{\Sigma}{\cal P}^I\wedge*{\cal J}_I+\frac{1}{2}L_{IJ}{\cal P}^I\wedge {\cal C}^J\nonumber\\
&&+\frac{1}{4}{\cal M}_{IJ}{\cal C}^I\wedge*{\cal C}^J+\frac{1}{2}\oint_{\Sigma}G_{ij}dy^i\wedge*dy^j+\frac{1}{2}\oint_{\Sigma}B_{ij}dy^i\wedge dy^j,
\end{eqnarray}
where
$$
{\cal J}_I={\cal M}_{IJ}{\cal P}^J-L_{IJ}*{\cal P}^J.
$$
Following \cite{Hull ``Doubled geometry and T-folds''}, one can show that, by completing the square in $C_m$ that the action splits into two parts
$$
S[y^i,\mathbb{X}^I,C_m,C^a]=S[y^i,z^m,C^a]+S[\Lambda_m],
$$
where $\Lambda_m=\left(e^{N\cdot y}\right)_m{}^nd{\tilde z}_n+{\cal C}_m+...$ appear quadratically in the action. In the path integral one performs a change of variables in the integration measure from a functional integration over $C_m$ to one over $\Lambda_m$. We may then integrate out the $\Lambda_m$. The determinant coming from integrating out these fields, in addition to the Jacobian factor arising from the change of variables in the integration measure, will contribute to a shift in the dilaton as described in \cite{Hull ``Doubled geometry and T-folds''}. One then finds that the remaining chiral gauge field $C^a$ acts as a Lagrange multiplier in the path integral, constraining the right-moving part of $\chi^a$ to vanish and giving a sigma model for the physical fields $(y^i, z^m,\chi_L^a)$, where $\chi_L^a$ is the left-moving part of $\chi^a$. Since $\chi^a$ is chiral, it is often easier to write these degrees of freedom in terms of a chiral fermion $\lambda^a$.

\subsubsection{Example: Heterotic Doubled Formalism in a Flat Background}

Let us consider explicitly the procedure of recovering the standard formulation from the doubled geometry in the following example of a trivial fibration in which the mass matrix  $N_{iI}{}^J$ is zero, i.e. the internal space is simply $T^6\simeq T^4\times T^2$. Imposing the constraint by gauging is subtle in the case where $N_{iI}{}^J\neq 0$ and the $U(1)^{12}$ isometry group is not always well-defined on the doubled geometry ${\cal T}$. In particular, the target space vector fields which generate the $U(1)^{12}$ isometries are defined in each $T^{24}$ fibre, but may not be well defined on the full bundle ${\cal T}$. Such issues where not addressed in \cite{Hull ``Doubled geometry and T-folds''} and we shall not elaborate on them further here, except to point out that a detailed discussion of these issues will be presented in \cite{Hull and Reid-Edwards coming soon}. For the simple example presented here, where $N_{iI}{}^J=0$, no such global issues arise and we can follow the procedure outlined in \cite{Hull ``Doubled geometry and T-folds''}. We first consider the case where the background is given by the spacetime fields ${\cal M}_{IJ}$ and ${\cal A}^I{}_i$, all $(y^i,\mathbb{X}^I)$-independent. It will aid the clarity of the exposition to assume also that ${\cal A}^I{}_i=0$, although this condition may be relaxed. The constraint (\ref{constraint}) is imposed by gauging a
null subgroup of the algebra $\left[T_I,T_J\right]=0$. The action for the gauged theory is $S[y^i,\mathbb{X}^I,C^I]=S[y^i]+S[\mathbb{X}^I,C^I]$ where the theory on the base is given by the action
$$
S[y^i]=\frac{1}{2}\oint_{\Sigma}g_{ij}dy^i\wedge *dy^j+\frac{1}{2}\oint_{\Sigma}B_{ij}dy^i\wedge dy^j,
$$
and the theory in the $T^4\times\widetilde{T}^4\times T^{16}_c$ fibres is given by
\begin{eqnarray}
S[\mathbb{X}^I,{\cal C}^I]&=&\frac{1}{4}\oint_{\Sigma}{\cal M}_{IJ}{\cal D}\mathbb{X}^I\wedge *{\cal D}\mathbb{X}^J+\frac{1}{4}\oint_{\Sigma}\Theta_{IJ}d\mathbb{X}^I\wedge
d\mathbb{X}^J+\frac{1}{2}\oint_{\Sigma}L_{IJ}d\mathbb{X}^I\wedge {\cal C}^J,\nonumber
\end{eqnarray}
where the invariant derivative ${\cal D}\mathbb{X}^I=d\mathbb{X}^I+C^I$ can be written as
\begin{equation}
{\cal D}\mathbb{X}^I=\left(%
\begin{array}{ccc}
  dz^m & d\tilde{z}_m+C_m & d\chi^a+C^a \\
\end{array}%
\right).\nonumber
\end{equation}

For the trivial $T^4$ fibration over $T^2$, which we consider in this example, the theory on the base will not play an important role and we shall focus on the theory in the fibres given by the Lagrangian $\mathscr{L}(\mathbb{X}^I,{\cal C}^I)$. Expanding this Lagrangian out using (\ref{SigmaHetM}) and (\ref{theta}) gives
\begin{eqnarray}
\mathscr{L}(\mathbb{X}^I,{\cal C}^I)&=&\frac{1}{4}\left(g_{mn}+A_m{}^aA_{na}-C_{mp}g^{pq}C_{qn}\right)dz^m\wedge*dz^n
+\frac{1}{4}g^{mn}{\cal D}\tilde{z}_m\wedge*{\cal D}\tilde{z}_n\nonumber\\
&&+ \frac{1}{4}\left(\delta_{ab}+A_{ma}g^{mn}A_{nb}\right){\cal D}\chi^a\wedge*{\cal D}\chi^b -\frac{1}{2}g^{mp}C_{pm}{\cal D}\tilde{z}_m\wedge*dz^n
\frac{1}{2}A_{na}g^{nm}{\cal D}\chi^a\wedge*{\cal D}\tilde{z}_m\nonumber\\
&&+\frac{1}{2}\left(A_{ma}+A_{pa}g^{pn}C_{nm}\right){\cal D}\chi^a\wedge*dz^m +\frac{1}{2}dz^m\wedge d\tilde{z}_m +\frac{1}{2}dz^m\wedge
C_m+\frac{1}{2}d\chi^a\wedge C_a.\nonumber
\end{eqnarray}
where ${\cal D}\tilde{z}_m=d\tilde{z}_m+C_m$ and ${\cal D}\chi^a=d\chi^a+C^a$. Completing the square in $C_m$ we have
\begin{eqnarray}
\mathscr{L}&=&\frac{1}{2}G_{mn}dz^m\wedge*dz^n +\frac{1}{2}B_{mn}dz^m\wedge
dz^n+\frac{1}{4}{\cal D}\chi^a\wedge*{\cal D}\chi_a\nonumber\\
&&+\frac{1}{2}A_{ma}{\cal D}\chi^a\wedge \left(*dz^m-dz^m\right) +\frac{1}{2}d\chi^a\wedge C_a+\frac{1}{4}g^{mn}\Lambda_m\wedge *\Lambda_n,\nonumber
\end{eqnarray}
where
\begin{equation}
\Lambda_m=C_m+d\tilde{z}_m-g_{mn}*dz^n -C_{mn}dz^n-A_{ma}{\cal D}\chi^a\nonumber
\end{equation}
gives a $\frac{1}{2}\ln(\det(g^{mn}))$ contribution to the dilaton when we integrate over the gauge fields $C_m$ in the path integral. The metric $G_{mn}$ of the sigma model is given by
$$
G_{mn}=g_{mn}+\frac{1}{2}A_m{}^aA_{na}
$$
and is the form of the effective metric required in order for Green-Schwarz anomaly cancellation to be consistent with supersymmetry, as found in
\cite{Hull and Townsend``WORLD SHEET SUPERSYMMETRY AND ANOMALY CANCELLATION IN THE HETEROTIC STRING''}. It is helpful to define the light-cone
worldsheet coordinates
\begin{equation}
\zeta^{\pm}=\frac{1}{2}\left(\tau\pm\sigma\right),  \qquad  \partial_{\pm}=\frac{\partial}{\partial \zeta^{\pm}}=\partial_{\tau}\pm\partial_{\sigma}.\nonumber
\end{equation}
Recalling that $C^a=*C^a$, we see that $C_-{}^a=0$ and the Lagrangian then becomes
\begin{eqnarray}
\mathscr{L}&=&-G_{mn}\partial_+z^m\partial_-z^n -B_{mn}\partial_+z^m\partial_-z^n -\frac{1}{2}\partial_+\chi^a\partial_-\chi_a
-A_{ma}\partial_+\chi^a\partial_-z^m
\nonumber\\
&&-\left(\partial_-\chi_a+A_{ma}\partial_-z^m\right)C_+{}^a.\nonumber
\end{eqnarray}
We see that, as expected, the chiral one-form $C_+{}^a$ acts as a Lagrange multiplier to impose the constraint
\begin{equation}\label{covariant chirality}
\partial_-\chi^a+A_m{}^a\partial_-z^m=0.
\end{equation}
In the $A_m{}^a=0$ background this is just the familiar constraint that $\chi^a$ is chiral, i.e. a left-mover. The condition (\ref{covariant chirality}) is a covariantization of this chirality constraint. Imposing the constraint gives the
Lagrangian
\begin{eqnarray}
\mathscr{L}&=&-G_{mn}\partial_+z^m\partial_-z^n -B_{mn}\partial_+z^m\partial_-z^n +\frac{1}{2}A_{ma}\partial_+\chi^a\partial_-z^m.\nonumber
\end{eqnarray}
We encounter the familiar problem that a chiral scalar does not have a Lorentz-covariant kinetic term. In the next sub-section we consider a
solution to this problem following Tseytlin \cite{Tseytlin:1990va} and propose a Lagrangian in which manifest Lorentz invariance is lost. Alternatively, one may write
the scalars $\chi^a$ as chiral fermions $\lambda^a$ where
$$
\partial_+\chi^a=2\bar{\lambda}\rho_+T^a\lambda.
$$
 $T^a$ is a generator in the Cartan subalgebra of $E_8\times E_8$ or $Spin(32)/\Z_2$ and $\rho_{\pm}=\rho_0\pm\rho_+$ are worldsheet gamma
matrices in light-cone coordinates. Substituting for $\partial_+\chi^a$ in the Lagrangian above and introducing a kinetic term for the chiral
fermions, the Lagrangian becomes
$$
\mathscr{L}=-(g_{ij}+B_{ij})\partial_+y^i\partial_-y^j-(G_{mn}+B_{mn})\partial_+z^m\partial_-z^n  +i\bar{\lambda}\rho^+\partial_+\lambda
+\bar{\lambda}\rho_+T_a\lambda A_m{}^a\partial_-z^m,
$$
where the contribution from the theory on the $T^2$ base, with coordinates $y^i$, has been included. This is the conventional form of the sigma model for a Heterotic string on a flat background with constant gauge and $B$-field \cite{Sen:1985eb}.

\subsubsection{Non-Covariant Formalism}

In the (worldsheet) Lorentz-convariant sigma model considered above, the chirality constraint on the bosons $\chi^a$ must be imposed at the level of the equations of motion. There is a manifestly
duality-symmetric formalism proposed by Schwarz and Sen \cite{Schwarz:1993mg,Schwarz:1993vs}, building on the earlier work of Tseytlin
\cite{Tseytlin:1990va} in which the chirality constraint is imposed at the level of the Lagrangian. The drawback of this formalism is that
general covariance of the worldsheet is not manifest. In \cite{Schwarz:1993mg,Schwarz:1993vs} the duality-covariant Lagrangian for a reduction of Heterotic String theory on $T^d$ was proposed
\begin{eqnarray}
\mathscr{L}&=&\frac{1}{2}g_{ij}\eta^{\alpha\beta}\partial_{\alpha}i^i\partial_{\beta}y^j
-\frac{1}{2}L_{IJ}D_{\tau}\mathbb{X}^ID_{\sigma}\mathbb{X}^J -\frac{1}{2}{\cal M}_{IJ}D_{\sigma}\mathbb{X}^ID_{\sigma}\mathbb{X}^J
\nonumber\\
&&+\frac{1}{2}\epsilon^{\alpha\beta}\left(B_{ij}\partial_{\alpha}y^i\partial_{\beta}y^j- L_{IJ}{\cal
A}^I{}_i\partial_{\alpha}y^iD_{\beta}\mathbb{X}^J\right)
\end{eqnarray}
where
\begin{equation}
D_{\alpha}\mathbb{X}^I=\partial_{\alpha}\mathbb{X}^I+{\cal A}^I{}_i\partial_{\alpha}y^i
\end{equation}
In order to describe the $O(4,20;\mathbb{Z})$-twisted reduction of the kind we have been examining, it is a simple matter to twist the embedding fields $\partial_{\alpha}\mathbb{X}^I$ according to the duality-twist reduction ansatz. The non-Lorentz-covariant version of the action (\ref{gauged sigma model}) is then given by
\begin{eqnarray}
\mathscr{L}&=&\frac{1}{2}g_{ij}\eta^{\alpha\beta}\partial_{\alpha}y^i\partial_{\beta}y^j -\frac{1}{2}L_{IJ}\widehat{{\cal
P}}_{\tau}{}^I\widehat{{\cal P}}_{\sigma}{}^J -\frac{1}{2}{\cal M}_{IJ}\widehat{{\cal P}}_{\sigma}{}^I\widehat{{\cal P}}_{\sigma}{}^J
+\frac{1}{2}\epsilon^{\alpha\beta}\left(B_{ij}\partial_{\alpha}y^i\partial_{\beta}y^j- L_{IJ}{\cal
A}^I{}_i\partial_{\alpha}y^i\widehat{{\cal P}}_{\beta}{}^J\right)\nonumber
\end{eqnarray}
where we have defined
$$
\widehat{\cal P}_{\alpha}{}^I=\left(e^{N\cdot y}\right)^I{}_JD_{\alpha}\mathbb{X}^J.
$$
Duality-twist models of this kind, for the Bosonic String, have recently been studied in \cite{Dall'Agata:2008qz}.

\section{Conclusion, Discussion and Future Directions}

In this article, we have shown that certain four-dimensional gauged $\mathcal{N}=4$ theories can be obtained from dimensional reduction of either IIA or Heterotic supergravity. Our starting point has been the ungauged $\mathcal{N}=2$ supergravity in six dimensions obtained by either compactifying IIA supergravity on a $K3$ or by compactifying Heterotic supergravity on a $T^4$. We have used the $O(4,20)$-symmetry of this theory to perform a duality twist reduction on a further $T^2$, leading to a gauged $\mathcal{N}=4$ supergravity in four dimensions.

We have analyzed the possible gaugings of the four-dimensional theory and interpreted the components of the structure constants in terms of both IIA and Heterotic theories. A twist matrix can be divided into different classes for a IIA origin and a Heterotic origin. For IIA reductions, we were able to identify the three different classes of mass matrix: ${\cal D}_{iA}{}^B$, ${\cal K}_{iA}$ and ${\cal Q}_i{}^A$, as defining a manifold of $SU(2)$-structure, an $H$-flux and what might be called a Mirror-flux respectively. For Heterotic reductions, we found a geometric flux, $H$ and $F$-fluxes, two new classes of T-fold and a flux corresponding to a topological twisting of the Cartan torus of the the $U(1)^{16}$ internal gauge group over spacetime.

It should be stressed that the interpretation we gave to each class of twist holds individually for each case where the only non-zero elements in the mass matrix $N_{iI}{}^J$ take values in any one of the classes ${\cal D}_{iA}{}^B$, ${\cal K}_{iA}$ or ${\cal Q}_i{}^A$. However, for more general cases, where the mass matrix $N_{iI}{}^J$ is composed of non-zero elements from more than one of the classes, the interpretation may be more subtle. For example, if a mass matrix has components of the geometric class ${\cal D}_{iA}{}^B$ and components of the flux ${\cal K}_{iA}$ switched on, then the interpretation of the duality-twist reduction as a compactification is simple - we can think of it as a compactification on a $K3$ bundle with non-trivial $H$-flux. More generally, we might consider a reduction in which elements of the class ${\cal D}_{iA}{}^B$ and, say, ${\cal Q}_i{}^A$ are switched on. In this case the interpretation of the reduction is not straightforward. It is not immediately clear how to make sense of such a $K3$ fibration with Mirror flux and these, more general, situations must be interpreted on a case by case basis. Similar considerations also apply to the Heterotic reductions.

Finally, we have written down a worldsheet sigma model for the Heterotic theory. The model we presented uses a generalization of the doubled formalism pioneered in \cite{Hull ``A geometry for non-geometric string backgrounds'',Tseytlin:1990va,Duff:1989tf,Duff:1990hn} for both the space-time coordinates and the chiral Heterotic fields. In this way, it is possible to write down a Heterotic worldsheet theory for the backgrounds discussed here.

IIA string theory compactified on $K3$ and Heterotic string theory compactified on $T^4$ are conjectured to be dual descriptions of the same physics. Whether or not this implies that IIA string theory on the six-dimensional $SU(2)$-structure manifolds and Heterotic string theory on the six-dimensional $\bid$-structure manifolds we described should also be conjectured to be dual remains unclear. One might argue that since the duality is clearly true for the theory in a single $T^4$ or $K3$ fibre, then the proposed equivalence of the descriptions in terms of the IIA and Heterotic theories should follow as a simple application of fibre-wise duality. An illustrative example, from fibre-wise T-duality, where much more is understood, demonstrates that such reasoning may be too naive. Consider a $T^d$ bundle with base $S^1$ of the kind constructed in \cite{Hull:2007jy}. There, one can demonstrate that two descriptions are T-dual by gauging an isometry of the $T^d$ background as described by \cite{Buscher ``A Symmetry of the String Background Field Equations''}. The \emph{proof} of T-duality requires the existence of a well-defined isometry which generates an invariance of the string background. One can show \cite{Hull and Reid-Edwards coming soon} that if one gauges an isometry that is well-defined in the $T^d$ fibres, yet is not well-defined on the $(d+1)$-dimensional bundle, then the result is a globally non-geometric background, such as a T-fold. One point of view is that the T-duality is still possible and indeed, there is evidence from Mirror symmetry \cite{SYZ} that T-duality does not require the existence of globally well-defined isometries. Furthermore, there is evidence that the existence of an isometry, even locally, is not a requirement for T-duality to be possible \cite{Dabholkar:2005ve}. Another, more conservative, perspective is that, in some cases, the monodromy acts as an obstruction to the existence of the structure - a well-defined isometry - that allows T-duality to be possible.

Given this note of caution from T-duality, one might hesitate to proclaim that the IIA and Heterotic compactifications considered in this paper are dual descriptions of the same Physics. At the very least, one may worry that a certain class of monodromies may obstruct the duality. The added difficulty here is that the strong/weak duality cannot be derived from any known worldsheet construction, such as the gauging of isometries in T-duality, and so it is unclear what structure (playing the role of the global isometry in T-duality) one might wish to preserve in order to ensure the duality holds. For the T-duality case discussed above, we know that if the monodromy does not obstruct the existence of a global isometry, then T-duality still holds. If the monodromy does provide an obstruction, then we are on uncertain ground. By contrast, it is difficult to say anything about the validity of the IIA/Heterotic duality in any of the cases we have been considering here. One way in which the proposal for such a duality might be checked would to be to directly calculate and compare non-perturbative effects such as BPS-states for the generalized case, akin to the tests that were done for the original IIA/heterotic duality.

If the Heterotic/IIA duality holds in the generalized case treated in this article, it would have some interesting implications for string theory compactifications on generalized manifolds. For example, some compactifications on the IIA side that are geometric would be dual to non-geometric compactifications of the Heterotic theory and vice versa. This would provide strong evidence that non-geometric backgrounds play a central role in string and M-theory.

In section four we reviewed some of the arguments, coming from supergravity, for a strong/weak coupling duality between the Heterotic string on $T^4$ and the Type IIA string on $K3$. By contrast, the IIB theory, compactified on $K3$, leads to a six-dimensional chiral theory which cannot be directly related to the the Heterotic and IIA string. However, there is evidence that a further Kaluza-Klein compactification on $T^2$, down to four dimensions, allows for a link to the IIB theory to be established and the duality between the Heterotic and IIA theories in four-dimensions is extended to a triality between the IIA, IIB and Heterotic theories\footnote{The two Heterotic strings are T-dual to each other.}. This triality has many remarkable features and was studied at length in \cite{DLR}. In particular, there is a $\left(SL(2;\Z)\right)^3\subset SL(2;\Z)\times SO(6,22;\Z)$ symmetry for which the roles of the three $SL(2;\Z)$'s are permuted under the triality. For example, in the Heterotic compactification, there is a $SL(2;\Z)_{\tau}\times SL(2;\Z)_{\omega}\subset SO(6,22;\Z)$ which acts on the complex and complexified K\"{a}hler structures ($\tau$ and $\omega$ respectively) of the $T^2$ base and arises from the
$$
\left(SL(2;\Z)_{\tau}\times SL(2;\Z)_{\omega}\right)/\Z_2\simeq O(2,2;\Z)
$$
T-duality symmetry of the metric and $B$-field on the $T^2$. The third $SL(2;\Z)$ is the manifest $SL(2;\Z)$ symmetry in the reduced action which acts linearly on the Heterotic axio-dilaton $M_{\alpha\beta}$.

A natural question is whether or not this triality occurs in the compactifications we have been considering here, where the fibration of $K3$ or $T^4$ over the $T^2$ base is not trivial, but has monodromy in $O(4,20;\Z)$. This question may be phrased in a slightly different way: can we lift the gauged supergravities considered here to compactifications of IIB string theory? One could consider performing a T-duality along one of
the circles of the $T^2$ to obtain a description in terms of IIB string theory from the IIA theory. However, a monodromy around a cycle of the $T^2$ will be an obstruction to the existence of an Abelian isometry along that direction. In the absence of
an isometric direction in which to perform the duality, it is not clear that a T-dual description even exists\footnote{A potential difficulty, even for isometric duality, arises from the suggestion in \cite{Aspinwall and
Plesser   ``T-duality can fail''} that the existence of an isometry is not enough to allow for the application of the Buscher rules in certain
circumstances. In particular, some evidence was presented which suggested the expected application of T-duality along the $T^2$ of even a $T^2\times
K3$ background fails. We shall not comment further on this here, but refer the interested reader to \cite{Aspinwall and Plesser ``T-duality can fail''} for
details.}. If we twist around only one of the
$T^2$ cycles, i.e. we set $\beta_I{}^J=0$, then there is a non-trivial gauging with structure constant $t_{1I}{}^J=\alpha_I{}^J$. Then, the
circle with coordinate $y^1\sim y^1+1$ has monodromy $\exp(\alpha_I{}^J)$ and the circle with coordinate $y^2$ has an Abelian isometry. We may then perform a (possibly fibre-wise) T-duality along $y^2$ to give a dual IIB description.

More generally, one could consider dualizing along a cycle with non-trivial monodromy. There is some evidence that such a
non-isometric duality can be performed \cite{Dabholkar:2005ve} and, in this case, would give rise to Heterotic compactifications with what has come to be known as `$R$-flux'. For example, if we consider a generic double duality-twist compactification of the Heterotic string and dualize along both directions of the $T^2$ base we expect to find a theory with gauge algebra
\begin{eqnarray}
[X^i,Z_m]=Q_{m}{}^{in}Z_n+W_{m}{}^{ia}Y_a+f_{mn}{}^iX^n   \qquad [Z_m,Z_n]=f_{mn}{}^iZ_i   \qquad  [X^m,Z_n]=Q_{m}{}^{ni}Z_i \nonumber
\end{eqnarray}
\begin{eqnarray}
[X^m,X^n]=R^{mni}Z_i \qquad  [X^i,X^m]=Q_{n}{}^{mi}X^n+V^{ima}Y_a+R^{imn}Z_n
 \nonumber
\end{eqnarray}
\begin{eqnarray}\label{alg}
 [X^i,Y_a]=-\delta_{ab}V^{imb}Z_m+W_{ma}{}^iX^m+G_a{}^{ib}Y_b    \qquad    [Z_m,Y_a]=W_{ma}{}^iZ_i \qquad [Y_a,Y_b]=G_{ab}{}^iZ_i
\end{eqnarray}
The diagram (\ref{sequence2}) is then extended to
\begin{eqnarray}\label{sequence3}
&K_{imn}&\rightarrow f_{im}{}^n\rightarrow Q_i{}^{mn}\rightarrow R^{imn}\nonumber\\
&\downarrow &\qquad \downarrow\qquad\quad\downarrow\nonumber\\
&M_{im}{}^a&\rightarrow W_i{}^{ma}\rightarrow V^{ima}\nonumber\\
&\downarrow &\qquad\downarrow\nonumber\\
&S_{abi}&\rightarrow G_{ab}{}^i\nonumber\\
&\downarrow&\nonumber\\
&h_{ab}{}^c&\nonumber
\end{eqnarray}
where the horizontal arrows now allow for a dualization along the $y^i$ directions and we have included the obvious generalization to accommodate structure constants of the form $h_{ab}{}^c$.

One might wonder how such structure constants $h_{ab}{}^c$ may be realized in compactifications. One possibility is that at special points in the moduli space of the Heterotic and Type II compactifications, the symmetry group is enhanced \cite{Hull and Townsend``Enhanced gauge symmetries in superstring theories}. For example, the $U(1)^{16}$ internal symmetry of the Heterotic theories may be enhanced to non-Abelian $E_8\times E_8$ or $Spin(32)/\Z_2$. At these enhanced points, the algebra includes the commutator $[Y_a,Y_b]=h_{ab}{}^cY_c+...$ where the index $a$ now runs from $1$ to $n\leq 496$, the dimension of the enhanced symmetry groups, and $h_{ab}{}^c$ are structure constants for $E_8\times E_8$ or $Spin(32)/\Z_2$ or some sub-group. Compactifications with non-zero $f_{im}{}^n$ and $h_{ab}{}^c$ were considered in \cite{Lu:2006ah} and are discussed briefly in Appendix A.2. It would be interesting to study this symmetry enhancement, in a duality-covariant manner using the doubled sigma model introduced in section seven.

One may think of backgrounds for which all structure  constants except $R^{imn}$ vanish as a $T^4$ fibration over the coordinates $\tilde{y}_i$ of the dual torus $\widetilde{T}^2$, conjugate to the winding modes of the Heterotic string around the $T^2$ \cite{Dabholkar:2005ve}. Similarly, one can view a reduction which gives the structure constant $V^{ima}$ as a Heterotic generalization of $R$-flux backgrounds in which the gauge fields of the $U(1)^{16}$ internal symmetry play an important role. The structure constants $G_{ab}{}^i$ indicate a background in which the internal ${U(1)^{16}}$ fibration is topologically twisted over the cycles of the dual $\widetilde{T}^2$.

From the Type II perspective, we can think of such $R$-flux backgrounds as $K3$ fibrations over the coordinates $\tilde{y}_i$, conjugate to the winding modes of the IIA string around the $T^2$. We conjecture that ${\cal N}=4$ gauged supergravities with such structure constants lift to compactifications of string theory on such locally non-geometric backgrounds. The corresponding gauge algebra for the lower-dimensional supergravities is then
\begin{eqnarray}
[X^i,J]=\mathcal{W}^i{}_A\eta^{AB}T_B,  \qquad    [X^i,\tilde{J}]=\mathcal{S}^{iA}T_A,  \qquad
[X^i,T_A]=\mathcal{R}_{AB}^{i}\eta^{BC}T_C-\mathcal{W}^i{}_A\tilde{J}-\eta_{AB}\mathcal{S}^{iB}J,\nonumber
\end{eqnarray}
\begin{eqnarray}
 \qquad [T_A,T_B]=\mathcal{R}_{AB}^{i}Z_i,    \qquad  [J,T_A]=\mathcal{W}^i{}_AZ_i, \qquad  [\tilde{J},T_A]=\eta_{AB}\mathcal{S}^i{}_BZ_i,\nonumber
\end{eqnarray}
We can think of the ${\cal W}$- and ${\cal S}$-fluxes as being similar to the ${\cal K}$- and ${\cal Q}$-fluxes, but involving the dual torus $\widetilde{T}^2$. For example, the reduction ansatz for the $B$-field in the $\cal W$-flux background is $\hat{b}_A=b_A+{\cal W}_A{}^i\tilde{y}_i$, the correct interpretation of which is not clear. The ${\cal R}$-flux denotes a smooth $K3$ fibration over the cycles of the dual torus $\widetilde{T}^2$. It remains to be seen if such constructions can provide a basis for good Type II string backgrounds.

Unlike the duality-twist backgrounds constructed in section five, which can be understood as a conventional background in a contractible patch, the only picture we have of the backgrounds constructed as a fibration over $\widetilde{T}^2$ is through the doubled formalism. Following the construction of the $R$-flux doubled geometry in \cite{Hull:2007jy}, the doubled geometry for such Heterotic backgrounds would be given by $\cX=\G\backslash \cG$ where $\cG$ is the (non-compact) group manifold for the gauge algebra (\ref{alg}) and $\G\subset\cG$ is a discrete (cocompact) subgroup, acting from the left, such that $\cX$ is compact. The gauge algebra could then be written as
$$
[X^i,T_I]=N_I{}^{iJ}T_J	\qquad	[T_I,T_J]=N_{IJ}{}^iZ_i
$$
with all other commutators vanishing. The globally defined, left-invariant, one-forms on $\cX$ are
$$
Q_i=d\tilde{y}_i+\frac{1}{2}N_{iIJ}\mathbb{X}^Id\mathbb{X}^J	\qquad	P^i=dy^i	\qquad	{\cal P}^I=\left(e^{N\cdot \tilde{y}}\right)^I{}_Jd\mathbb{X}^J
$$
where the coordinates on the $28$-dimensional doubled space $\cX$ are $(y^i,\tilde{y}_i,\mathbb{X}^I)$. For example, in the Heterotic case we have seen that $\mathbb{X}^I=(z^m,\tilde{z}_m,\chi^a)$ as in the previous section. We shall return to a supersymmetric sigma model description of the doubled geometry $\cX$ elsewhere.

\begin{center}
\textbf{Acknowledgments}
\end{center}
We would like to thank Jan Louis, Dan Waldram and Stefan Groot Nibbelink for helpful discussions and Andrei Micu for helpful comments and for pointing out a number of typos in an earlier version of the manuscript. This work was supported by the Deutsche Forschungsgemeinschaft (DFG) in the SFB 676 ``Particles, Strings and the Early Universe''.

\begin{appendix}

\section{Generalized Reductions and Compactifications}\label{OtherCompactifications}
In this Appendix we briefly review various features of the most prominent generalized reduction techniques.

\subsection{$SU(3)$- and $SU(3)\times SU(3)$-Structure Reductions}

As an example of the methodology and limitations of constructing lower-dimensional supergravities by reduction algorithms, we recall here $SU(3)$- and $SU(3)\times SU(3)$-structure reductions. The Canonical example is of a compactification on a three-dimensional compact Calabi-Yau manifold. One can find a (symplectic) basis of harmonic three-forms $(\alpha^{(0)}{}_I,\beta^{(0)I})$ where $I=0,1,..h^{1,2}+1$ and a basis of two-forms $\omega^{(0)}{}_A$ and, Hodge dual, four-forms $\tilde{\omega}^{(0)A}$, where $A=0,1,..h^{1,1}+1$. The effective, massless, four-dimensional supergravity is obtained by expanding the ten-dimensional form-fields and fluctuations of the ten-dimensional metric in terms of this basis of forms. This construction has been well-studied and is equivalent to a compactification on a Calabi-Yau manifold followed by a truncation to the zero modes of the Kaluza-Klein spectrum.

Massive supergravities can be constructed by adding constant fluxes to Ramond-Ramond and $H$-field strengths, wrapping harmonic cycles of the Calabi-Yau. For example, one can add constant electric ($e_{I}$) and magnetic ($m^I$) fluxes to the $H$-field strength $H=e_I\beta^{(0)I}-m^I\alpha^{(0)}{}_I+...$. Such reductions have a clear interpretation as a flux-compactifications on compact Calabi-Yau manifolds. Following \cite{GLMW}, it is then natural to consider expanding the reduction ansatz in terms of a non-harmonic, twisted basis of forms
$$
\omega_A=e^{B}\omega^{(0)}{}_A	\qquad	\tilde{\omega}^A=e^{B}\tilde{\omega}^{(0)A}	\qquad	\alpha_I=e^{B}\alpha^{(0)}{}_I	 \qquad	 \beta^I=e^{B}\beta^{(0)I}
$$
where $dB=e_I\beta^I-m^I\alpha_I$ is the flux on the $H$-field strength. Up to terms which vanish under the Mukai pairing\footnote{The $O(6,6)$-invariant Mukai pairing $\langle\,,\,\rangle$ gives
$$
\int_{CY_3}\langle\alpha_I,\beta^J\rangle=\delta^I{}_J	\qquad	 \int_{CY_3}\langle\omega_A,\tilde\omega^B\rangle=\delta_A{}^B
$$} these twisted forms of mixed degree satisfy
\begin{eqnarray}
d\alpha_I\sim e_I\tilde{\omega}^0	&\qquad &	d\beta^I\sim m^I\tilde{\omega}^0	\qquad	d\omega_0= m^I\alpha_I-e_I\beta^I
\end{eqnarray}
where all other forms are harmonic. In particular, if $m^I=0$ so that the flux is purely electric then, under the conjectured mirror symmetry which involves the exchange of mixed-odd and mixed-even degree forms, this background is dual to a particular class of $SU(3)$-structure manifolds, called half-flat manifolds as discussed in \cite{GLMW}. On the assumption that such a generalization of Mirror Symmetry does hold in this case, the fact that we can construct a rigorous Kaluza-Klein compactification for a flux compactification on a Calabi-Yau lends credence to the idea that the half-flat reduction does indeed lift to a genuine Kaluza-Klein compactification.

More generally, one can consider reductions on $SU(3)\times SU(3)$-structure backgrounds for which the reduction ansatz is expanded in terms of the forms $\alpha_I$, $\beta^I$, $\omega_A$ and $\tilde{\omega}^A$ where
\begin{eqnarray}
d\alpha_I\sim p_I{}^A\omega_A+e_{IA}\tilde{\omega}^A	&\qquad &	d\beta^I\sim q^{IA}\omega_A+m_A{}^I\tilde{\omega}^A\nonumber\\
d\omega_A\sim m_A{}^I\alpha_I-e_{IA}\beta^I	&\qquad &	d\tilde{\omega}^A\sim -q^{IA}\alpha_I+p_I{}^A\beta^I\nonumber
\end{eqnarray}
where the notion of the exterior derivative on the internal space $d:\bigwedge^p\rightarrow\bigwedge^{p+1}$ is generalized to one in which $d:\bigwedge^p\rightarrow\bigwedge^{p\pm1}$. Although it is clear that such a reduction ansatz, in which the degrees of forms are not preserved, cannot be realized as a compactification on a conventional manifold, there is little understanding of precisely what the internal background is. There is some evidence \cite{Grana:2006hr,Waldram O(dd)} that such non-geometric backgrounds should share many of the qualitative characteristics of non-geometric ${\bid}$-structure backgrounds constructed in \cite{Kachru:2002sk,Dabholkar:2002sy,Dabholkar:2005ve} and discussed briefly below.

\subsection{$\bid$-Structure Flux Compactifications and Duality-Twist Reductions}

$\bid$-structure compactifications of Heterotic supergravity to four dimensions have been studied by many authors \cite{KM,Lu:2006ah,Porrati:1989jk}, giving rise to a plethora of new constructions of ${\cal N}=4$ gauged supergravities. Of particular relevance to the $\bid$-structure reductions studied here, are those first presented in the seminal work of Scherk and Schwarz \cite{Scherk How To Get Masses From Extra Dimensions}. There, two types of dimensional reduction were introduced; the first, which has become known as a compactification on a twisted torus, involved a twisting of the frame bundle over the internal manifold in the reduction ansatz so that the ansatz is written in terms of the forms $\sigma^m=\sigma^m{}_i(y)dy^i$, where $m,n=1,2,..d$ are frame indices, $y^i$ ($i,j=1,2,..d$) are coordinates on the $d$-dimensional internal manifold. In order for the reduction to be consistent (in the sense that solutions to the lower dimensional equations of motion lift to solutions of the full, higher-dimensional theory), the $y$-dependent vielbeins $\sigma^m{}_i$ must be such that the one forms $\sigma^m$ satisfy the structure equation
$$
d\sigma^m+\frac{1}{2}f_{np}{}^m\sigma^n\wedge \sigma^p=0
$$
where $f_{np}{}^m$ are structure constants for some, possibly non-compact, $d$-dimensional group $G$. We see then that the condition for consistency is a parallelizability condition and the twisted torus is an $\bid$-structure manifold. So, for example, the reduction ansatz for the metric takes the form
$$
g(y)=g_{mn}\sigma^m\otimes \sigma^n
$$
Because of the T-duality symmetric role $H$-fluxes and the structure constants $f_{mn}{}^p$ play in such reductions, the structure constants are often referred to a `geometric fluxes'. It was shown in \cite{Hull:2005hk} that such reductions could be understood as compactifications on a \emph{twisted torus}, $\G\backslash G$, where $\G\subset G$ is a discrete group, acting on the left, such that $\G\backslash G$ is compact. This identification of the twisted torus as the compactification manifold for Scherk-Schwarz reductions elevates the Scherk-Schwarz reduction algorithm to the status of a Kaluza-Klein compactification scenario.

In \cite{KM}, compactifications of Heterotic supergravity on $d$-dimensional twisted tori were considered with fluxes for the $H$-field and internal gauge field strength switched on so that
$$
H=\frac{1}{6}K_{mnp}\sigma^m\wedge\sigma^n\wedge\sigma^p+...	\qquad	F^a=\frac{1}{2}M_{mn}{}^a\sigma^m\wedge\sigma^n+...
$$
where $K_{mnp}$ and $M_{mn}{}^a$ are constants. The case of interest here is $d=6$. In \cite{KM}, it was assumed that Wilson lines break the Heterotic gauge groups from $E_8\times E_8$ or $Spin(32)/\Z_2$ to $U(1)^{16}$ (and we have assumed the same is true in the Heterotic reductions to be considered in this paper). It was shown that the resulting four-dimensional, ${\cal N}=4$, gauged supergravity could be written in a manifestly $O(6,22)$-covariant form. The resulting gauged supergravity has a scalar potential and non-Abelian gauge group with Lie algebra\footnote{It was shown in \cite{Hull:2006tp} that the local symmetry of the supergravity was in fact generated by a Lie-algebroid, which contains this Lie algebra.}
\begin{equation}\label{KM}
 [Z_m,Z_n]=f_{mn}{}^pZ_p+K_{mnp}X^p+M_{mn}^aY_a	\qquad	[X^m,Z_n]=f_{np}{}^mX^p	\qquad	[Y_a,Z_m]=\delta_{ab}M_{mn}{}^bX^n
\end{equation}
where the $Z_m$ ($m,n=1,2,...6$) are related to diffeomorphisms of the twisted torus, the $X^m$ are related to $B$-field antisymmetric tensor transformations and the $Y_a$ ($a,b=1,2,..16$) generate the $U(1)^{16}$ internal gauge symmetry. All other commutators vanish. One can see that the gauge algebra contains a lot of information about the ten-dimensional lift of the four-dimensional supergravity. In particular, the structure constants $f_{mn}{}^p$ tell us what the local geometry of the compactification manifold is and the structure constants $K_{mnp}$ and $M_{mn}{}^a$ contain information about the fluxes in the internal geometry.

A complementary situation, in which not all of the internal gauge group is broken (but no fluxes were turned on) was considered in \cite{Lu:2006ah}. The resulting ${\cal N}=4$ gauged supergravity has a scalar potential and a non-Abelian gauge symmetry generated by the Lie algebra
\begin{equation}\label{Pope}
[Z_m,Z_n]=f_{mn}{}^pZ_p	\qquad	[X^m,Z_n]=f_{np}{}^mX^p	\qquad	[Y_a,Y_b]=h_{ab}{}^cY_c
\end{equation}
where $h_{ab}{}^c$ are the structure constants for the non-Abelian internal gauge symmetry group $H$ where now the indices $a,b,c$ run from one to the dimension of $H$. Again we see that the structure constants $f_{mn}{}^p$ encode the local structure of the compactification manifold and $h_{ab}{}^c$ tells us the subgroup $H$ of $E_8\times E_8$ or $Spin(32)/\Z_2$ which is preserved by the compactification. From such reasoning, it is possible to deduce a string theory origin of many gauged supergravities.

Duality-twist dimensional reduction was introduced in \cite{Scherk How To Get Masses From Extra Dimensions} and was reviewed in section five. Excellent discussions of this procedure may be found in \cite{Dabholkar:2002sy,Hull:Gauged D = 9 supergravities and Scherk-Schwarz reduction,Hull:Massive string theories from M-theory and F-theory}. The general idea is to perform the compactification of the higher-dimensional supergravity on two stages: the first consists of a conventional Kaluza-Klein compactification from $D+d+1$ dimensions down to $D+1$ dimensions on some $d$-dimensional manifold, followed by a truncation to the zero modes of the Kaluza-Klein spectrum. The resulting $(D+1)$-dimensional supergravity has a rigid symmetry symmetry group $K$ which will generally lift to a discrete U-duality symmetry group $K(\Z)\subset K$ of the string theory \cite{Hull and Townsend Unity of superstring dualities}. The second step of the reduction involves compactification down to $D$ dimensions on a circle, where we introduce a topological twist such that the monodromy around the circle is an element of $K(\Z)$. Since $K(\Z)$ is a symmetry of the string theory, the string theory is well defined on the $(d+1)$-dimensional internal background. It can be shown that the twist by $K(\Z)$ does not explicitly break any supersymmetry so that, if the $d$-dimensional manifold has, for example, $SU(n)$-holonomy, the $(d+1)$-dimensional background will have $SU(n)$-structure. Moreover, if the $d$-dimensional manifold is of $\bid$-holonomy, then the $(d+1)$-dimensional background is of $\bid$-structure and admits a consistent truncation of the Kaluza-Klein spectrum. Such reductions have become known as \emph{duality-twist} reductions and, interpreting such duality-twist reductions in terms of compactifications of string theory is not always as simple as for the twisted torus compactifications reviewed above.

Generally the action of $K(\Z)$ on the $d$-dimensional manifold will not be geometric and the $(d+1)$-dimensional internal background will not be a Riemannian manifold, but may be non-geometric. If however, $K(\Z)$ does act geometrically then, it was shown in \cite{Hull:2005hk} that the $(d+1)$-dimensional background is a twisted torus of the kind described above. In \cite{Porrati:1989jk} duality-twist reductions of Heterotic supergravity, where the theory was reduced to five dimensions and then twisted over a circle with monodromy in the geometric $SL(5;\Z)$, were considered\footnote{Consistent truncations to ${\cal N}=1$ models, characterised by a positive semi-definite potential, were also studied in \cite{Porrati:1989jk}.}. Such reductions give rise to gauged ${\cal N}=4$ supergravities as described above and may be thought of as a compactification on a six-dimensional twisted torus \cite{Hull:2006tp}.

\section{$SO(3)$-symmetry of six-dimensional manifolds with $SU(2)$-structure}\label{SO3SpinorCalc}

Here, we want to show that the definition of $j$ and $\omega$ in terms of spinors has an inherent $SO(3)$-symmetry which is analogous to the hyperk\"ahler-structure of $K3$. In fact, we will show the existence of a symmetry operator $g$ acting on the spinors, whose action translates into an $SO(3)$-rotation on the vector
\begin{equation*}
\left(\begin{array}{c}
		j \\
		{\rm Re}\, \omega \\
		{\rm Im}\, \omega \end{array}\right),
\end{equation*}
and that leaves the one-form $\sigma$ invariant.

We start by noting that there is an arbitrariness in our choice of spinors. Any linear combination of the two globally well-defined spinors $\eta^1, \eta^2$, as long as it leaves the lengths invariant, would yield the two-forms $j, {\rm Re}\, \omega, {\rm Im}\, \omega$ and a one-form $\sigma$. Let us consider a complex $2\times 2$-matrix $g$ acting on $\eta_{-}\equiv (\eta^1_{-}, \eta^2_{-})$ such that
\begin{equation*}
\eta^{\dagger}_{-}\eta_{-}\to \eta^{\dagger}_{-}g^{\dagger}g\eta_{-}=\eta^{\dagger}_{-}\eta_{-}.
\end{equation*}
This means $g\in SU(2)$, so we can write it as
\begin{equation*}
g=\left(	
	\begin{array}{cc}
		a 	& b \\
		-b^{\ast} 	& a^{\ast}
	\end{array}\right),
\end{equation*}
with $a,b \in \mathbb{C}$ satisfying $|a|^2+|b|^2=1$. While the negative chirality spinors transform as
\begin{align*}
\eta_{-} 		& \to g\eta_{-},\\
\eta^{\dagger}_{-}	& \to \eta^{\dagger}_{-}g^{\dagger},
\end{align*}
the positive chirality-spinors, defined as
\begin{equation*}
\eta_+^{\dagger} = \eta_-^TC,
\end{equation*}
with $C$ the charge conjugation matrix, transform as
\begin{align*}
\eta_{+} 		& \to g^{\ast}\eta_{+},\\
\eta^{\dagger}_{+}	& \to \eta^{\dagger}_{+}g^{{\rm T}}.
\end{align*}

This $SU(2)$-rotation of the spinors corresponds to an $SO(3)$-rotation of the three two-forms $j, {\rm Re}\, \omega, {\rm Im}\, \omega$. Let $G(g)$ be the action of $g$ on any spinor bi-linear given by the $SU(2)$-rotation $g$ of the spinors, written as a matrix multiplying the vector
\begin{equation*}
\left(\begin{array}{c}
		j \\
		{\rm Re}\, \omega \\
		{\rm Im}\, \omega \end{array}\right).
\end{equation*}
We want to consider the action of $g$ on this vector. In terms of spinors, we have
\begin{align*}
j_{vw} 		&= \frac{i}{2}\left(\eta_{-}^{1\dagger}\gamma_{vw}\eta_{-}^{1} - \eta_{-}^{2\dagger}\gamma_{vw}\eta_{-}^{2}\right), \\
{\rm Re}\, \omega_{vw} 	&= \frac{i}{2}\left(\eta_{-}^{1\dagger}\gamma_{vw}\eta_{-}^{2} + \eta_{-}^{2\dagger}\gamma_{vw}\eta_{-}^{1}\right),\\
{\rm Im}\, \omega_{vw}	&= \frac{1}{2}\left(\eta_{-}^{1\dagger}\gamma_{vw}\eta_{-}^{2} - \eta_{-}^{2\dagger}\gamma_{vw}\eta_{-}^{1}\right),
\end{align*}
and writing $a=t+ix$, $b=y+iz$ we find
\begin{equation*}
G(g)	=\left(
	\begin{array}{ccc}
		t^2+x^2-y^2-z^2	& 2ty + 2xz & 2xy-2tz \\
		2xz - 2ty & t^2-x^2-y^2+z^2 & 2tx+2yz \\
		2xy + 2tz & 2yz-2tx & t^2-x^2+y^2-z^2
	\end{array}\right).
\end{equation*}
For $g\in SU(2)$ we have $t^2+x^2+y^2+z^2=1$, and using this we can calculate that $GG^T=1$. This means $G\in O(3)$, and since det $G=1$ for $t=1, x=y=z=0$, we conclude $G\in SO(3)$.

On the other hand, the one-form $\sigma$ is left invariant under this rotation. To show this we need that
\begin{equation*}
\sigma_v\equiv\eta^{2\dagger}_{-}\gamma_v\eta^1_{+}=-\eta^{1\dagger}_{-}\gamma_v\eta^2_{+}.
\end{equation*}
This can be shown as follows:
\begin{align*}
\eta^{2\dagger}_{-}\gamma_v\eta^1_{+}	& = (\eta^{2}_{+})^TC\gamma_v\eta^1_{+} = - (\eta^{2}_{+})^T\gamma^T_vC\eta^1_{+} \\
& = - (\eta^{2}_{+})^T\gamma^T_v(\eta^{1\dagger}_{-})^T = - \eta^{1\dagger}_{-}\gamma_v\eta^2_{+}.
\end{align*}
We calculate
\begin{equation*}
G(g)\sigma_v=(-b\eta^{1\dagger}_{-} + a\eta^{2\dagger}_{-})\gamma_v(a^{\ast}\eta^1_{+} + b^{\ast}\eta^2_{+})=(|a|^2+|b|^2)\sigma_v=\sigma_v.
\end{equation*}

\section{Kaluza-Klein Compactification of the Heterotic Theory on $T^4$}\label{6DN2Theory}
The Heterotic theory in ten dimensions is given, in string frame, by the Lagrangian
\begin{eqnarray}\label{ten dim L for 2a}
\mathscr{L}_{10}=e^{-\Phi}\left(\mathscr{R}*1+*d\Phi\wedge
d\Phi-\frac{1}{2}\mathscr{H}_{(3)}\wedge*\mathscr{H}_{(3)}-\frac{1}{2}\delta_{ab}\mathscr{F}_{(2)}^a\wedge*\mathscr{F}_{(2)}^b\right),
\end{eqnarray}
where
\begin{equation*}
\mathscr{F}_{(2)}^a=d\mathscr{A}_{(1)}^a,  \qquad  \mathscr{H}_{(3)}=d\mathscr{B}_{(2)}-\frac{1}{2}\delta_{ab}\mathscr{A}_{(1)}^a\wedge\mathscr{F}_{(2)}^a.
\end{equation*}
We compactify the spacetime of the theory on $T^4$ using the standard Kaluza-Klein reduction ansatz
\begin{eqnarray}\label{B3}
ds^2_{10}&=&ds^2_{6}+g_{mn}\nu^m\otimes\nu^n,\nonumber\\
\mathscr{B}_{(2)}&=&\widehat{B}_{(2)}+\widehat{B}_{(1)m}\wedge\nu^m+\frac{1}{2}B_{mn}\nu^m\wedge\nu^n,\nonumber\\
\mathscr{A}_{(1)}^a&=&\widehat{A}^a_{(1)}+\widehat{A}^a_{m}\nu^m,\nonumber\\
\Phi&=&\widehat{\phi}+\frac{1}{2}\ln (g),
\end{eqnarray}
where
\begin{eqnarray*}
\nu^m=dz^m-\widehat{A}_{(1)}^m,  \qquad  \widehat{F}^m_{(2)}=d\widehat{A}_{(1)}^m.
\end{eqnarray*}
The graviphoton of the reduction is $\widehat{A}_{(1)}^m$ and $z^m$ ($m=6,7,8,9$) are coordinates on $T^4$. The notation is that a field
$\Psi_{(p)}$ is of degree $p$ in ten dimensions and a field $\widehat{\psi}_{(p)}$ is of degree $p$ in six dimensions, where the subscript for scalars is suppressed.

Inserting the reduction ansatz (\ref{B3}) into the Lagrangian (\ref{ten dim L for 2a}) gives the effective theory in six dimensions
\begin{eqnarray*}
\label{B-field lagrangian}\mathscr{L}_{6}&=&e^{-\widehat{\phi}}\left(\widehat{R}*1+*d\widehat{\phi}\wedge
d\widehat{\phi}-\frac{1}{2}dg^{mn}\wedge *dg_{mn}-\frac{1}{2}g_{mn}*\widehat{F}^m_{(2)}\wedge \widehat{F}_{(2)}^n
-\frac{1}{2}\widehat{H}_{(3)}\wedge *\widehat{H}_{(3)} \right.
\nonumber\\
&&\left.- \frac{1}{2}g^{mn}\widehat{H}_{(2)m}\wedge *\widehat{H}_{(2)n}-\frac{1}{2}g^{mn}g^{pq}\widehat{H}_{(1)mp}\wedge
*\widehat{H}_{(1)nq}\right.\nonumber\\
&&\left.-\frac{1}{2}\delta_{ab}\widehat{F}_{(2)}^a\wedge*\widehat{F}_{(2)}^b
-\frac{1}{2}\delta_{ab}g^{mn}\widehat{F}_{(1)m}^a\wedge*\widehat{F}_{(1)n}^b\right),
\end{eqnarray*}
where
\begin{eqnarray}\label{H field strengths}
\widehat{H}_{(3)}&=&d\widehat{B}_{(2)}+\widehat{B}_{(1)m}\wedge \widehat{F}^m_{(2)}-\frac{1}{2}\delta_{ab}\widehat{A}^a_{(1)}\wedge
d\widehat{A}^b_{(1)}-\frac{1}{2}\delta_{ab}\widehat{A}^a_{m}\widehat{A}^b_{(1)}\wedge \widehat{F}^m_{(2)},
\nonumber\\
\widehat{H}_{(2)m}&=&d\widehat{B}_{(1)m}+B_{mn}\widehat{F}^n_{(2)}-\frac{1}{2}\delta_{ab}\widehat{A}^a_{(1)}\wedge d\widehat{A}^b_{m}
-\frac{1}{2}\delta_{ab}\widehat{A}^a_{m}
d\widehat{A}^b_{(1)}-\frac{1}{2}\delta_{ab}\widehat{A}^a_{m}\widehat{A}^b_{n}\widehat{F}^m_{(2)},
\nonumber\\
\widehat{H}_{(1)mn}&=&dB_{mn}-\frac{1}{2}\delta_{ab}\widehat{A}^a_{m}d\widehat{A}^b_{n},
\nonumber\\
\widehat{F}_{(2)}^a&=&d\widehat{A}^a_{(1)}-\widehat{A}^a_{m}\widehat{F}^m_{(2)},\nonumber\\
\widehat{F}_{(1)m}^a&=&d\widehat{A}^a_{m}.
\end{eqnarray}

Using the field redefinitions
\begin{eqnarray*}
\widehat{C}_{(2)}&=&\widehat{B}_{(2)}-\frac{1}{2}B_{(1)m}\wedge \widehat{A}^m_{(1)},\nonumber\\
\widehat{C}_{(1)m}&=&\widehat{B}_{(1)m}-\frac{1}{2}\delta_{ab}\widehat{A}^a_{m}\widehat{A}^b_{(1)},\nonumber\\
\widehat{C}_{mn}&=&B_{mn}+\frac{1}{2}\delta_{ab}\widehat{A}_{m}{}^a\widehat{A}_{n}{}^b,
\end{eqnarray*}
we see that the reduced theory has gauge group $U(1)^{24}$ and may be written in a manifestly $O(4,20)$-invariant way:
\begin{eqnarray}\label{doubledobblewobble}
\mathscr{L}_{6}=e^{-\widehat{\phi}}\left(\widehat{R}*1+*d\widehat{\phi}\wedge d\widehat{\phi}+\frac{1}{4}Tr\left(d\widehat{{\cal M}}\wedge
*d\widehat{{\cal M}}^{-1}\right)-\frac{1}{2}\widehat{\mathcal{H}}_{(3)}\wedge *\widehat{\mathcal{H}}_{(3)}-\frac{1}{2}\widehat{{\cal
M}}_{IJ}\widehat{\mathcal{F}}^I_{(2)}\wedge*\widehat{\mathcal{F}}^J_{(2)}\right)
\end{eqnarray}
where
$$
\widehat{\mathcal{H}}_{(3)}=d\widehat{C}_{(2)}-\frac{1}{2}L_{IJ}\widehat{{\cal A}}_{(1)}^I\wedge d\widehat{{\cal A}}^J_{(1)},  \qquad
\widehat{{\cal F}}^I_{(2)}=d\widehat{\cal A}^I_{(1)}.
$$
Also $I,J=1,2,..24$ and
$$
\widehat{\cal M}^{IJ}= \left(\begin{array}{ccc}
g^{mn} & -B_{np}g^{pm} & -g^{mn}\widehat{A}_{n}{}^a \\
-B_{mp}g^{np} & g_{mn}+g^{pq}B_{mp}B_{nq}+\delta_{ab}\widehat{A}_{m}{}^a\widehat{A}_{n}{}^b & \widehat{A}_{m}{}^a+B_{mp}g^{pn}\widehat{A}_{n}{}^a \\
-\widehat{A}_{n}{}^ag^{mn} & \widehat{A}_{m}{}^a+\widehat{A}_{n}{}^ag^{np}B_{mp} &
\delta^{ab}+\widehat{A}_{m}{}^ag^{mn}\widehat{A}_{n}{}^b
\end{array}\right),
$$
where the $O(4,20)$ vector $\widehat{{\cal A}}^I$ and corresponding field strength $\widehat{{\cal F}}^I$ are
\begin{equation*}\widehat{{\cal A}}^I= \left(\begin{array}{ccc}
\widehat{A}_{(1)}{}^m \\ \widehat{B}_{(1)m} \\ \widehat{A}_{(1)}{}^a
\end{array}\right),  \qquad  \widehat{{\cal F}}^I= \left(\begin{array}{ccc}
\widehat{F}_{(2)}^m \\ \widehat{H}_{(2)m} - B_{mn}\widehat{F}_{(2)}^n-\delta_{ab}\widehat{A}^a_{(1)}\wedge d \widehat{A}^b_{(1)} \\
\widehat{F}_{(2)}^a,
\end{array}\right)
\end{equation*}
and the $O(4,20)$ invariant is
\begin{equation*} L_{IJ}=\left(\begin{array}{ccc}
0 & \bid_{4} & 0 \\ \bid_{4} & 0 & 0 \\ 0 & 0 & \bid_{16}.
\end{array}\right) \end{equation*}

\section{Kaluza-Klein Compactification of IIA Supergravity on $K3$}
The $O(4,20)$-invariant six-dimensional half-maximal theory (\ref{doubledobblewobble}) can also be obtained from a compactification of the maximal IIA
supergravity on $K3$. We define $*\Omega^A\equiv H^A{}_B\Omega^B$, and it is not hard to show that it satisfies $H^A{}_CH^C{}_B=\delta^A{}_B$ and
$\eta_{[A|C}H^C{}_{|B]}=0$, so that
$$
\fdh{A}{B} \in\frac{SO(3,19;\mathbb{R})}{SO(3,\mathbb{R})\times SO(19,\mathbb{R})}.
$$
$H^A{}_B$ therefore parametrizes 57 of the metric moduli coming from the two-forms, the last one is given by an overall size modulus. The moduli
$H^A{}_B$ and $\rho$, as defined, do not depend on the six non-compact directions. It is useful to introduce $\widetilde{H}^A{}_B$ and
$\widetilde{\rho}$ which, as we shall see in the next section, do depend explicitly on the remaining six directions.

\subsection{Dimensional Reduction}
The Lagrangian of Type IIA supergravity is
\begin{eqnarray}
 \mathscr{L}^{IIA}_{10} &=&   e^{-\Phi}\left( \mathscr{R}*1 + d\Phi\wedge *d\Phi + \frac{1}{2}d\mathscr{B}_{(2)}\wedge\ast d\mathscr{B}_{(2)}
+\frac{1}{2}d\mathscr{A}_{(1)}\wedge\ast d\mathscr{A}_{(1)} \right. \nonumber\\
&&+ \left.\frac{1}{2}(d\mathscr{C}_{(3)} - \mathscr{A}_{(1)}\wedge d\mathscr{B}_{(2)})\wedge\ast (d\mathscr{C}_{(3)}- \mathscr{A}_{(1)}\wedge
d\mathscr{B}_{(2)}) -\frac{1}{2} \mathscr{B}_{(2)}\wedge d\mathscr{C}_{(3)}\wedge d\mathscr{C}_{(3)}\right),\nonumber
\end{eqnarray}
where $\Phi$ is the dilaton, $\mathscr{B}$ is the Kalb-Ramond field and $\mathscr{A}_{(1)}$ and $\mathscr{C}_{(3)}$ are Ramond-Ramond fields. We
consider a Kaluza-Klein reduction of IIA supergravity on $K3$. The Kaluza-Klein reduction ansatz is
\begin{eqnarray*}
\mathscr{A}_{(1)}&=&\widetilde{A}_{(1)},\nonumber\\
\mathscr{B}_{(2)}&=&\widetilde{B}_{(2)}+\widetilde{b}_{A}\Omega^A,\nonumber\\
\mathscr{C}_{(3)}&=&\widetilde{C}_{(3)}+\widetilde{C}_{(1)A}\wedge\Omega^A.
\end{eqnarray*}

The resulting six-dimensional theory is best written in terms of an $O(4,20)$ matrix $\widetilde{{\cal M}}_{IJ}$ which takes values in the
coset $O(4,20)/(O(4)\times O(20))$ and is given by
\begin{equation*}
\widetilde{{\cal M}}_{IJ}=\left(
    \begin{array}{ccc}
 e^{-\widetilde{\rho}} + \widetilde{H}^{AB}\widetilde{b}_A\widetilde{b}_B + e^{\widetilde{\rho}} \widetilde{C}^2
        & e^{\widetilde{\rho}} \widetilde{C}
        & -\widetilde{H}^{C}{}_{B}\widetilde{b}_C - e^{\widetilde{\rho}} \widetilde{b}_B\widetilde{C}  \\
 e^\rho \widetilde{C}
        & e^{\widetilde{\rho}}
        & - e^{\widetilde{\rho}} \widetilde{b}_B  \\
 -\widetilde{H}^{B}{}_{A}\widetilde{b}_B - e^{\widetilde{\rho}} \widetilde{b}_A\widetilde{C}
        & - e^{\widetilde{\rho}} \widetilde{b}_A
        & \eta_{AC}\widetilde{H}^{C}{}_{B} + e^{\widetilde{\rho}} \widetilde{b}_A\widetilde{b}_B
    \end{array}\right),
\end{equation*}
where $\widetilde{C}=\frac{1}{2}\eta^{AB}\widetilde{b}_A\widetilde{b}_B$ and $\eta_{AB}$ is the intersection matrix for $K3$. The symmetric matrix of
scalars satisfy $\widetilde{{\cal M}}_{IK}L^{KL}\widetilde{{\cal M}}_{LJ}=L_{IJ}$ with $L_{IJ}$, the invariant of $O(4,20)$, given by
\begin{equation*}
L_{IJ}=\left(\begin{array}{ccc}
0   & -1    & 0 \\
-1  & 0 & 0 \\
0   & 0 & \eta_{AB} \end{array}\right).\nonumber
\end{equation*}
In the theory are also a metric, a dilaton $\widetilde{\phi}$ and a two-form field $\widetilde{B}$, and 24 gauge fields. Of these gauge fields, one comes
from the ten-dimensional gauge field, 22 from the expansion of the three-form field in the two-forms of $K3$, and the last one is the dual of
the three-form field in six dimensions. The six-dimensional supergravity Lagrangian (this can be found, for example, in \cite{DLM}) is given by
\begin{eqnarray*}
\mathscr{L}_{6}^{IIA}&=&e^{-\widetilde{\phi}}\left(\widetilde{R}*1+*d\widetilde{\phi}\wedge d\widetilde{\phi}+\frac{1}{4}d\widetilde{{\cal M}}_{IJ}\wedge
*d\widetilde{{\cal M}}^{IJ}-\frac{1}{2}\widetilde{\mathcal{H}}_{(3)}\wedge *\widetilde{\mathcal{H}}_{(3)}-\frac{1}{2}\widetilde{{\cal
M}}_{IJ}\widetilde{\mathcal{F}}^I_{(2)}\wedge*\widetilde{\mathcal{F}}^J_{(2)}\right)\nonumber\\
&&-\frac{1}{2}L_{IJ}\widetilde{\mathcal{B}}_{(2)}\wedge \widetilde{\mathcal{F}}_{(2)}^I\wedge \widetilde{\mathcal{F}}_{(2)}^J,\nonumber
\end{eqnarray*}
where the field strengths are
$$
\widetilde{\mathcal{H}}_{(3)}=d\widetilde{\mathcal{B}}_{(2)},  \qquad  \widetilde{\mathcal{F}}_{(2)}^I=d\widetilde{\mathcal{A}}^I_{(1)}.
$$

\subsection{Gauge Algebra}
The gauge algebra of this theory is $U(1)^{24}$, where $U(1)^{22}\subset U(1)^{24}$ is generated by antisymmetric tensor transformation with parameters
$\lambda^A$ associated to each of the harmonic two cycles of the $K3$. A further $U(1)$ is inherited directly from ten dimensions as the Abelian
gauge transformation of the Ramond field $\mathscr{A}$. We denote the generator of this transformation by $J$. In six dimensions the three form
part of the Ramond field $\mathscr{C}_{(3)}$ is dual to a one form $\tilde{C}_{(1)}$. A final $U(1)$ comes from the Abelian gauge transformations of
this field, generated by $\tilde{J}$. These generators can be written as an $O(4,20)$ vector $T_I$, with algebra $[T_I,T_J]=0$ where
\begin{eqnarray*}
T_I=\left(%
\begin{array}{c}
  J \\
  \tilde{J} \\
  T_A \\
\end{array}%
\right).
\end{eqnarray*}

\section{Duality-Twist Reductions Over $T^2$}
We have seen that the Heterotic theory compactified in $T^4$ and the IIA theory compactified on $K3$ give the same supergravity with effective
Lagrangian in six dimensions given by
\begin{eqnarray*}
\mathscr{L}_{6}=e^{-\widehat{\phi}}\left(\widehat{R}*1+*d\widehat{\phi}\wedge d\widehat{\phi}+\frac{1}{4}d\widehat{{\cal M}}_{IJ}\wedge
*d\widehat{{\cal M}}^{IJ}-\frac{1}{2}\widehat{\mathcal{H}}_{(3)}\wedge *\widehat{\mathcal{H}}_{(3)}-\frac{1}{2}\widehat{{\cal
M}}_{IJ}\widehat{F}^I_{(2)}\wedge*\widehat{F}^J_{(2)}\right).\nonumber
\end{eqnarray*}
As noted in \cite{Nishino:1986dc}, the theory has $SL(2)\times O(4,20)$ rigid symmetry, a discrete subgroup of which lifts to a duality symmetry of the full
string theory \cite{Hull and Townsend Unity of superstring dualities}. In this Appendix we present we consider a further reduction on $T^2$, twisting with
two commuting elements of $O(4,20)$ over the two cycles of the $T^2$, to give an effective theory in four dimensions.

\subsection{Dimensional Reduction}

Let $y^i$, $i=1,2$ be the $T^2$ coordinates. The reduction ans\"atze are
\begin{eqnarray*}
ds^2_{6}&=&ds^2_{4}+g_{ij}\nu^i\otimes\nu^j,   \nonumber\\
\widehat{\cal A}_{(1)}^I(x,y)&=&(e^{N\cdot y})^I{}_J\left({\cal A}_{(1)}^J(x)+{\cal A}^J_{j}(x)\nu^j\right),\nonumber\\
\widehat{\cal M}^{IJ}(x,y)&=&(e^{N\cdot y})^I{}_K{\cal M}^{KL}(x)(e^{N^T\cdot y})^I{}_L, \nonumber\\
 \widehat{C}_{(2)}(x,y)&=&B_{(2)}(x)+B_{(1)i}(x)\wedge\nu^i+\frac{1}{2}B_{ij}(x)\nu^i\wedge\nu^j,
\nonumber
\end{eqnarray*}
where
\begin{equation*}
\nu^i=dy^i-V^i_{(1)},
\end{equation*}
and the twist matrix is
\begin{equation*}
(e^{N\cdot y})^I{}_J=\exp(N_{iJ}{}^Iy^i).
\end{equation*}
The structure constants $N_{iJ}{}^I$ encode the monodromy around the $i=1,2$ directions:
\begin{eqnarray*}
N_{iJ}{}^I=\left(%
\begin{array}{cc}
 \alpha_J{}^I &
  \beta_J{}^I
\end{array}%
\right),
\end{eqnarray*}
where $e^{\alpha}$ is the $SO(4,20)$ monodromy around the $y^1\sim y^1+1$ direction and $e^{\beta}$ is that around the $y^2\sim y^2+1$
direction where $[\alpha,\beta]=0$. The condition that the two twists commute is
$[\alpha,\beta]_I{}^J=\alpha_I{}^K\beta_K{}^J-\beta_I{}^K\alpha_K{}^J=2N_{I[i|}{}^KN_{|j]K}{}^J=0$. This is equivalent to the Bianchi identity
\begin{equation}
d^2(e^{N\cdot y})_I{}^J=(e^{N\cdot y})_I{}^LN_{Li}{}^KN_{jK}{}^Jdy^i\wedge dy^j=0
\end{equation}
which states that the second cocycle is trivial.

The field strength reductions are
\begin{eqnarray*}
\widehat{\cal F}_{(2)}^I(x,y)&=&\left(e^{N\cdot y}\right)^I{}_J\left\{\left(f^J_{(2)}+{\cal A}_{i}^JG_{(2)}^i\right)+f_{(1)i}^J\wedge
\nu^i\right\},
\end{eqnarray*}
where
\begin{eqnarray*}
f^I_{(2)}&=&d{\cal A}_{(1)}^I-N_{iJ}{}^IV_{(1)}^i\wedge {\cal A}^J_{(1)},\nonumber\\
f^I_{(1)i}&=&d{\cal A}_{i}^I-N_{jJ}{}^IV_{(1)}^j {\cal A}^J_{i}-N_{iJ}{}^I{\cal A}^J_{(1)},\nonumber\\
G_{(2)}^i&=&dV^i_{(1)}.
\end{eqnarray*}
It is useful to make the following field redefinitions:
\begin{eqnarray*}
C_{(1)i}=dB_{(1)i}-\frac{1}{2}L_{IJ}{\cal A}_{i}^I{\cal A}^J_{(1)},\qquad C_{ij}=B_{ij}+\frac{1}{2}L_{IJ}{\cal A}_{i}^I{\cal
A}_{j}^J.
\end{eqnarray*}
These potentials and field strengths can be combined into the $O(6,22)$ multiplets
\begin{eqnarray*}
\mathcal{C}=\left(%
\begin{array}{c}
  V^i_{(1)} \\
  C_{(1)i} \\
  {\cal A}_{(1)}^I \\
\end{array}%
\right),\qquad\mathcal{F}^M=\left(%
\begin{array}{c}
  G^i \\
  H_i \\
  f^I \\
\end{array}%
\right),
\end{eqnarray*}
where
\begin{eqnarray*}
H_{(2)i}=dC_{(1)i}+\frac{1}{2}N_{iIJ}{\cal A}_{(1)}^I\wedge {\cal A}_{(1)}^J,
\end{eqnarray*}
and $N_{iIJ}=-N_{iJI}=L_{IK}N_{iJ}{}^K$. The reduced Lagrangian may be written as
\begin{eqnarray*}
\mathscr{L}_4&=&e^{-\phi}\left(R*1+*d\phi\wedge d\phi+\frac{1}{2}*\mathcal{H}_{(3)}\wedge \mathcal{H}_{(3)}+\frac{1}{4}*D{\cal M}_{MN}\wedge
D{\cal M}^{MN}\right. \nonumber\\ &&- \left.\frac{1}{2}{\cal M}_{MN}*\mathcal{F}_{(2)}^M\wedge\mathcal{F}_{(2)}^N+V*1\right),
\end{eqnarray*}
where $L_{MQ}t_{NP}{}^Q=t_{MNP}$ and the scalar potential is given by
\begin{eqnarray*}
V=- \frac{1}{12}{\cal M}^{MQ}{\cal M}^{NT}{\cal M}^{PS}t_{MNP}t_{QTS}+ \frac{1}{4}{\cal M}^{MQ}L^{NT}L^{PS}t_{MNP}t_{QTS}.
\end{eqnarray*}

The scalar fields ${\cal M}_{MN}$ span the coset $O(6,22)/O(6)\times O(22)$ where
\begin{equation} {\cal M}_{MN}= \left(\begin{array}{ccc}
g_{ij}+{\cal M}_{IJ}{\cal A}^I_{i}{\cal A}^J_{j}+g^{kl}C_{ik}C_{jl} & g^{ik}C_{jk} & g^{jk}C_{ij}L_{IK}{\cal A}^K_{k}+{\cal M}_{IK}{\cal A}^K_{i} \\
g^{ik}C_{jk} & g^{ij} & g^{ij}L_{IK}{\cal A}^K_{j}  \\
 g^{jk}C_{ij}L_{JK}{\cal A}^K_{k}+{\cal M}_{JK}{\cal A}^K_{i}   &  g^{ij}L_{IK}{\cal A}^K_{j} & {\cal
 M}_{IJ}+g^{ij}L_{IK}L_{JL}{\cal A}^K_{i}{\cal A}^L_{j}
\end{array}\right).\nonumber
\end{equation}
The $O(6,22)$ invariant is
\begin{equation*} L_{MN}=\left(\begin{array}{ccccc}
0 & \bid_2 & 0 \\ \bid_2 & 0 & 0 \\ 0 & 0 & L_{IJ}.
\end{array}\right) \end{equation*}

\subsection{Gauge Symmetry}
In ten dimensions the theory has the antisymmetric tensor transformation symmetry
\begin{equation*}
\mathscr{B}\rightarrow \mathscr{B}+d\Lambda_{(1)}.
\end{equation*}
The reduction ansatz for the parameter $\Lambda_{(1)}$ on $T^4$ is $\Lambda_{(1)}=\widehat{\lambda}_{(1)}+\widehat{\lambda}_{m}\nu^m$. The
remainder of the $U(1)^{24}$ gauge symmetry comes from the four $U(1)$ isometries of the $T^4$, $z^m\rightarrow z^m+\omega^m$, under which
$\delta \widehat{A}^m_{(1)}=d\omega^m$ and the $U(1)^{16}$ gauge transformations $\delta \widehat{A}^a_{(1)}=d\epsilon^a$. In six dimensions
this $U(1)^{24}$ gauge symmetry acts on the fields as
$$
\delta_T\widehat{\cal A}^I_{(1)}=d\widehat{\lambda}^I,   \qquad
\delta_T\widehat{C}_{(2)}=d\widehat{\lambda}_{(1)}+\frac{1}{2}L_{IJ}\widehat{\lambda}^I\widehat{{\cal F}}^J_{(2)}.
$$
where we have defined
\begin{eqnarray*}
\widehat{\lambda}^I=\left(%
\begin{array}{c}
  \omega^m \\
  \lambda_{m} \\
\epsilon^a
\end{array}%
\right).
\end{eqnarray*}

\noindent\textbf{Antisymmetric tensor transformations}

\noindent The duality twist reduction ansatz for the six-dimensional gauge parameters $\widehat{\lambda}_{(1)}$ and $\widehat{\lambda}^A$ is
\begin{equation*}
\widehat{\lambda}^I=\left(e^{N\cdot y}\right)^I{}_J\lambda^J,  \qquad  \widehat{\lambda}_{(1)}=\lambda_{(1)}+\lambda_i\nu^i.
\end{equation*}
We denote the infinitesimal variation of the fields under this transformation by $\delta_T$. It is easy to show, by calculating
$d\widehat{\lambda}^I$, that the four-dimensional fields transform as
\begin{eqnarray}\label{willop}
\delta_T{\cal A}^I_{(1)}&=&d\lambda^I+N_{Ji}{}^I\lambda^JV^i_{(1)},\nonumber\\
\delta_T{\cal A}^I&=&N_{Ji}{}^I\lambda^J,\nonumber\\
\delta_TB_{(2)}&=&d\lambda_{(1)}+\lambda_iF^i_{(2)}-\frac{1}{2}L_{IJ}\lambda^I\left({\cal F}_{(2)}^J-{\cal A}^I_{i}G^i_{(2)}\right),\nonumber\\
\delta_TB_{(1)i}&=&d\lambda_i-\frac{1}{2}L_{IJ}\lambda^I{\cal F}_{(2)}^J.
\end{eqnarray}
Using the field redefinition
\begin{equation*}
C_{(1)i}=B_{(1)i}-\frac{1}{2}L_{IJ}{\cal A}^I_{i}{\cal A}^J_{(1)}
\end{equation*}
we see that $C_{(1)i}$ transforms as a connection:
\begin{equation}\label{wollop}
\delta_TC_{(1)i}=d\lambda_i+L_{JK}N_{Ii}{}^K\lambda^I{\cal A}^J_{(1)}.
\end{equation}

\noindent\textbf{$T^2$ Diffeomorphisms}

\noindent The theory must be invariant under reparametrizations of the circle coordinate
\begin{equation*}
y^i\rightarrow y^i+\omega^i.
\end{equation*}
The matrix $e^{N\cdot y}$ changes as $\left(e^{N\cdot y}\right)^I{}_J\rightarrow \left(e^{N\cdot y}\right)^I{}_K\left(e^{N\cdot
\omega}\right)^K{}_J=\left(e^{N\cdot y}\right)^I{}_K\left(\delta^K{}_J+N_{Ji}{}^K\omega^i+...\right)$. From this is it easy to see how the four
dimensional fields must transform in order for the six-dimensional ansatz to be invariant
\begin{equation}\label{wallop}
\delta_Z{\cal A}^I=-N_{Ji}{}^I{\cal A}^J\omega^i,   \qquad \delta_Z{\cal A}^I_{(1)}=-N_{Ji}{}^I{\cal A}^J_{(1)}\omega^i,  \qquad
\delta_ZV^i_{(1)}=d\omega^i.
\end{equation}

\noindent\textbf{Symmetry Algebra}

\noindent We define
\begin{eqnarray}
\delta_Z=\omega^i Z_i, \qquad  \delta_T=\lambda^IT_I,   \qquad  \delta_X=\lambda_iX^i
\end{eqnarray}
where $Z_i$, $X^i$ and $T_I$ are generators of gauge transformations with parameters $\omega^i$, $\lambda_i$ and $\lambda^I$ respectively. It is not hard to show from (\ref{willop}), (\ref{wollop}) and (\ref{wallop}) above that the
Lie algebra of the gauge group is
\begin{eqnarray*}
\left[Z_i,T_I\right]&=&N_{Ii}{}^JT_J,   \qquad\qquad  \left[T_I,T_J\right]=N_{IJi}X^i,
\end{eqnarray*}
with all other commutators vanishing, where we have defined
\begin{equation*}
N_{IJi}=-N_{JIi}=L_{IK}N_{Ji}{}^K.
\end{equation*}
Note that $N_{(IJ)i}=0$ as we are gauging a subgroup of $O(6,22)$.

\subsection{Rewriting the Four-Dimensional Lagrangian}
The gauged theory we have found by dimensional reduction takes the general form
\begin{eqnarray}
\mathscr{L}_4&=&e^{-\phi}\left(R*1+d\phi\wedge *d\phi-\frac{1}{2}\mathcal{H}_{(3)}\wedge *\mathcal{H}_{(3)}+\frac{1}{4}*D{\cal
M}^{MN}\wedge D{\cal M}_{MN}\right. \nonumber\\ &&- \left.\frac{1}{2}{\cal M}_{MN}*\mathcal{F}^M\wedge\mathcal{F}^N\right)+V*1.
\end{eqnarray}

In four dimensions, we may write this in the Einstein frame using the four-dimensional Weyl rescaling
\begin{equation*}
g_{\mu\nu}(x)\rightarrow e^{\phi(x)}g_{\mu\nu}(x).
\end{equation*}
The individual terms in the string frame action rescale as
\begin{eqnarray*}
e^{-\phi}\sqrt{-g}R&\rightarrow& \sqrt{-g}\left(R+\frac{3}{2}(\partial\phi)^2\right),\nonumber\\
e^{-\phi}\sqrt{-g}g^{\mu\lambda}g^{\nu\sigma}g^{\rho\tau}\mathcal{H}_{\mu\nu\rho}\mathcal{H}_{\lambda\sigma\tau}&\rightarrow&
e^{-2\phi}\sqrt{-g}\mathcal{H}^2.
\end{eqnarray*}
In the Einstein frame the action is written as
\begin{eqnarray*}
\mathscr{L}_4&=&R*1+\frac{1}{2}d\phi\wedge *d\phi-\frac{1}{2}e^{-2\phi}\mathcal{H}_{(3)}\wedge *\mathcal{H}_{(3)}+\frac{1}{4}*D{\cal
M}^{MN}\wedge D{\cal M}_{MN}\nonumber\\ &&-\frac{1}{2}e^{-\phi}{\cal M}_{MN}*\mathcal{F}_{(2)}^M\wedge\mathcal{F}_{(2)}^N+V*1.
\end{eqnarray*}

We now consider the dualization of $C_{(2)}$ to a scalar. Let $G_{(3)}=dC_{(2)}=\mathcal{H}_{(3)}-\Omega_{(3)}$, where $\Omega_{(3)}$ is a Chern-Simons term such that
\begin{equation*}
d\Omega_{(3)}=\frac{1}{2}L_{MN}\mathcal{F}_{(2)}^M\wedge \mathcal{F}_{(2)}^N
\end{equation*}
The Bianchi identity $dG_{(3)}=0$ is imposed by the Lagrange multiplier $\chi$
\begin{eqnarray*}
\mathscr{L}_4&=&R*1+\frac{1}{2}d\phi\wedge *d\phi-\frac{1}{2}e^{-2\phi}\mathcal{H}_{(3)}\wedge *\mathcal{H}_{(3)}+\frac{1}{4}*D{\cal
M}^{MN}\wedge D{\cal M}_{MN}\nonumber\\ &&-\frac{1}{2}e^{-\phi}{\cal M}_{AB}*\mathcal{F}_{(2)}^A\wedge\mathcal{F}_{(2)}^B+V*1+d\chi\wedge
G_{(3)}.
\end{eqnarray*}
The $G_{(3)}$ equation of motion is $G_{(3)}=e^{-2\phi}*d\chi$. Substituting this into the Lagrangian gives the dual formulation
\begin{eqnarray*}
{\cal L}_D&=&R*1+\frac{1}{2}d\phi\wedge *d\phi+\frac{1}{2}e^{2\phi}*d\chi\wedge d\chi+\frac{1}{4}*D{\cal M}^{MN}\wedge D{\cal M}_{MN}\nonumber\\
&&-\frac{1}{2}e^{-\phi}{\cal M}_{MN}*\mathcal{F}_{(2)}^M\wedge\mathcal{F}_{(2)}^N+V*1-d\chi\wedge \Omega_{(3)}.
\end{eqnarray*}
Introducing the axio-dilaton $\tau=\chi+ie^{-\phi}$, the $\phi$ and $\chi$ kinetic terms may be written
\begin{equation*}
\frac{1}{2\Im(\tau)^2}d\tau\wedge*d\bar{\tau}=\frac{1}{2}e^{2\phi}d\chi\wedge*d\chi+\frac{1}{2}d\phi\wedge*d\phi.
\end{equation*}
and the topological terms may be rewritten as
\begin{equation*}
-d\chi\wedge \Omega_{(3)}=\frac{1}{2}\Re(\tau)L_{MN}\mathcal{F}_{(2)}^M\wedge \mathcal{F}_{(2)}^N+...\, ,
\end{equation*}
where the dots denote a total derivative term which can be dropped. The Lagrangian may then be written as
\begin{eqnarray*}
\mathscr{L}_4&=&R*1-\frac{1}{2\Im(\tau)^2}d\tau\wedge*d\bar{\tau}+\frac{1}{4}D{\cal M}_{MN}\wedge *D{\cal M}^{MN}\nonumber\\
&&-\frac{1}{2}\Im(\tau){\cal M}_{MN}*\mathcal{F}_{(2)}^M\wedge\mathcal{F}_{(2)}^N+\frac{1}{2}\Re(\tau)L_{MN}\mathcal{F}_{(2)}^M\wedge\mathcal{F}_{(2)}^N+V*1.
\end{eqnarray*}
Written in this form, it is easier to see that the scalars $(\tau,{\cal M}_{MN})$ parameterize the space
\begin{equation*}
\frac{SL(2)}{SO(2)}\times \frac{O(6,22)}{O(6)\times O(22)}.
\end{equation*}

\end{appendix}

\end{document}